\documentclass[oneside]{article}
\usepackage[a4paper, left=1in, right=1in, top=1in, bottom=1in]{geometry}
\usepackage{graphicx}				
\usepackage{booktabs}
\usepackage{pdfpages}
\usepackage{amssymb}
\usepackage{subcaption}
\usepackage{bm}
\usepackage{amsmath}
\usepackage{xcolor}
\newtheorem{theorem}{Theorem}
\newtheorem{lemma}{Lemma}
\usepackage{multirow}
\usepackage{siunitx}
\title{X-chromosome Multilocus Association Studies for Common and Rare Variants}
\author{Ruilin Bai and Bo Chen$^{*}$ \\
School of Statistics and Data Science, LPMC and KLMDASR, \\
Nankai University, Tianjin, China\\
*Corresponding author: bochen@nankai.edu.cn}
\date{}

\begin{document}
\maketitle
\begin{abstract}
{X-chromosome association study has specific model uncertainty challenges, such as unknown X-chromosome inactivation status and baseline allele, and considering nonadditive and gene–sex interaction effects in the analysis or not. Although these challenges have been answered for single-locus X-chromosome variants, it remains unclear how to properly perform multilocus association studies when above uncertainties are present. We first carefully investigate the inferential consequences of these uncertainties on existing multilocus association analysis methods, and then propose a theoretically justified framework to analyze multilocus X-chromosome variants while all the uncertainty issues are addressed. We provide separate solutions for common and rare variants, and simulation results show that our solutions are overall more powerful than existing multilocus methods which were proposed to analyze autosomal variants. We finally provide supporting evidences of our approach by revisiting some published X-chromosome association studies.
}
\end{abstract}
\section{Introduction}

Genome-wide association studies (GWAS) are widely used to identify genetic variants associated with complex phenotypes in case-control and quantitative trait studies.
In recent decades, thousands of genetic variants have been discovered through these studies \cite{Keur2022}. However, the X chromosome is often excluded due to multiple analytical challenges, and the exclusionary problem had expanded from 2013 \cite{Sun2023} after Wise \cite{Wise2013} brought this problem to attention. Several methods have been proposed to analyze single variants, i.e. single nucleotide polymorphisms (SNPs) from X chromosome. Although multilocus analysis has become a common practice for autosome SNPs \cite{Shao2022}, the X chromosome has received less attention and method for X chromosome multilocus analysis remains unclear. Before proceeding to X chromosome multilocus analysis, we first briefly review the challenges and existing solutions to single locus X chromosome analysis and multilocus autosome analysis.

\subsection{Single Locus Association Studies of the X Chromosome}
Challenges for analyzing X chromosome originate from its distinct inheritance pattern. Females carry two copies, while males have one, which requires careful handling of the genotype. The uncertain status of X chromosome inactivation (XCI) for females, including escape (XCIE), random (XCIR) and skewed (XCIS), leads to the uncertainty of the model. Additionally, recent researches reported sex differences in phenotype, genotype missingness and allele frequency \cite{Khramtsova2023, Chen2024}, which indicate potentially nontrivial sex and sex-genotype interaction effects on the traits of interest.

Earlier X chromosome analysis methods needed to assume fixed XCI status, such as XCIE or XCIR. For instance, Clayton et al. \cite{Clayton2008} proposed additive and dominance models, using tests identical to the conventional Pearson chi-squared test under the XCIR assumption, with modifications tailored for the X chromosome. Alternatively, when adjusting for confounding factors, $\text{FM}_{01}$ and $\text{FM}_{02}$ from XWAS \cite{Gao2015} can be used for regression-based association testing, which assumes random XCIR and XCIE in females, respectively. By default, sex is included as a covariate, ensuring robustness under sex-specific allele frequencies. Additionally, XWAS \cite{Gao2015} introduced a sex-stratified test, $\text{FM}_{f}$ and $\text{FM}_{s}$, which first conducts association tests separately in females and males, then combines the results using Fisher's method \cite{Fisher1925} and Stouffer's Z method \cite{Stouffer1949} to obtain a final sex-stratified significance level.

Recently, more approaches have been developed to account for uncertainty of XCI status. Wang et al. \cite{Wang2014} proposed a method using a likelihood ratio test that includes XCIS as an additional possible XCI state. In this approach, the three possible genotypes in females are coded as 0, $\gamma$, and 2, where $\gamma$ is an unknown parameter. The value of $\gamma$ ranges between 0 and 2, allowing it to quantify the degree of skewness. However, it loses some power when XCI is random or escape. Since p-values in this method are estimated using a permutation-based approach, this method is computationally intensive and time-consuming. Wang et al. \cite{Wang2017} later refined this approach to estimate the degree of skewness and determine the underlying XCI pattern for each tested marker. Su et al. \cite{Su2022} used a similar approach to Wang \cite{Wang2014} but applied a score test based on a regression model. This method also does not require specifying a particular XCI pattern. Additionally, it calculates p-values analytically, making it more computationally efficient. Yang et al. \cite{Yang2022} proposed four test statistics: QXcat, QZmax, QMVXcat, and QMVZmax. The first two methods aim to identify the mean differences of the trait value across genotypes, while the latter two examine differences in both means and variances. QMVXcat and QXcat introduce two indicator variables for females, allowing association testing under all XCI patterns. They then combine the p-values from test statistics for females and males directly. QMVZmax and QZmax, on the other hand, use different weights to combine the test statistics for females and males and determine the final test statistic by maximizing these combined values. Additionally, Chen \cite{Chen2018} used a Bayesian model to average over different XCI patterns.

Considering XCI uncertainty, sex and sex-genotype interaction effects altogether, Chen \cite{Chen2021} proposed a regression- and genotype-based model that includes sex as a covariate, allowing analysis of both continuous and binary traits while adjusting for covariate effects, ensuring validity regardless of the X-chromosome inactivation status. The recommended test has three degrees of freedom, incorporating additive and dominance genetic effects, as well as a gene-sex interaction effect. The power of this test remains robust across different genetic models. However, all above mentioned methods were developed for single variant analysis.

\subsection{Multilocus Association Studies of the Autosomes}

In addition to single variant analysis, multilocus methods, i.e. SNP-set analyses, have also been widely implemented because of their higher statistical power and biologically more meaningful interpretation of the combined effects from multiple SNPs. Shao \cite{Shao2022} gave a comprehensive comparison of 22 multilocus methods focusing on autosomal SNPs. There are mainly three types of summary statistics that can be integrated from multiple SNPs: p-values, test statistics (e.g., $z$-scores, chi-squared values), and effect estimates (e.g., effect sizes, odds ratios). For integrating p-values, Fisher's method \cite{Fisher1925}, Stouffer's Z method \cite{Stouffer1949} and minimal p-value (Min p) test \cite{Tippett1931} combine p-values, with the latter converting them into z-scores then use their mean as the test statistic. Other approaches include harmonic mean p-value method \cite{Wilson2019} and Cauchy combination test \cite{Liu2020}, both designed to handle dependencies among p-values effectively. METAL \cite{Willer2010} implements a sample-size-based approach similar to Stouffer's Z method \cite{Stouffer1949}. It transforms p-values into a half-normal distribution, determines the sign of the half-normal z-score based on estimated effect direction, and computes a test statistic using the weighted sum of z-scores, with weights proportional to sample size.

When integrating test statistics and effect estimates from multiple SNPs, because single variant effects can be heterogeneous, referring to variations in effect sizes and directions across multiple SNPs, the linear tests of fixed effects (FE) which assumes an invariant effect size could be under power. It is common to combine the linear tests with quadratic tests that specifically test the heterogeneity effects between multiple SNPs \cite{Derkach2014}. On the other hand, random effects (RE) approach assumes that the true effect size varies across SNPs. Han \cite{Han2011} proposed a new RE method, simultaneously testing mean effect and variance component, which achieves higher statistical power in the presence of heterogeneity, demonstrating its practical utility for discovering associations in multilocus analysis.

Some multilocus methods particularly focuses on rare variants, as single rare variant tests usually do not have enough power to achieve significance at genome-wide level. To account for their rarity, different weights are assigned to SNPs based on their minor allele frequencies (MAFs). A common heuristic is to use weights derived from a Beta distribution with parameters 1 and 25 \cite{Wu2011}.

There are two main classes of methods for analyzing rare variants: burden tests and variance component tests. Burden tests aggregate the weighted contributions of multiple SNPs into a single covariate, which is then tested for association with the phenotype \cite{Morgenthaler2007, Madsen2009}. Although this approach is effective for detecting the cumulative effect of variants, it may suffer from a substantial loss of power when the directions of the variant effects are not consistent. Variance component tests model the genetic contribution of SNPs as a random effect and assess their association with phenotypes accordingly. Examples include Sequence Kernel Association Test (SKAT) \cite{Wu2011}, SSU statistic \cite{Pan2009}, and C-alpha statistic \cite{Neale2011}. While SSU and C-alpha methods typically apply uniform weights across variants, SKAT incorporates MAF-based weights using a Beta distribution. Variance component tests are generally more powerful than burden tests when the effect directions among variants are mixed. Conversely, when all variants affect the phenotype in the same direction, burden tests tend to perform better. To achieve robust performance across both scenarios, SKAT-O \cite{Lee2012} combines burden test and SKAT, effectively adapting to different underlying genetic architectures.

All the approaches mentioned above focus on autosomes, leaving the X chromosome unexplored.
Ma et al. \cite{Ma2015} extended burden test, SKAT, and SKATO to enable their direct application to the association analysis of low-frequency variants for both binary and quantitative traits. These extended methods have been implemented in 'SKAT' R package \cite{Lee2014}.
However, Ma et al. \cite{Ma2015} did not thoroughly consider the impact of uncertainty in coding schemes, which can also lead to model uncertainty. Although their simulations showed only a small loss in power when XCI state was misspecified, their analysis did not sufficiently explore the full extent of this issue.

\subsection{X chromosome Multilocus Association Studies}
To our best knowledge, currently, there has been no work showing how to extend multilocus analysis methods to X chromosome after considering all the analytical challenges discussed in Section 1.1. The main contribution of this paper is to propose a framework for X-chromosome multilocus association studies. Our framework is invariant to the unknown XCI status across multiple SNPs, and takes the effects of sex and sex-genotype interaction into consideration to achieve both model robustness and improved test power. We consider heterogeneity effects across multiple SNPs, carefully investigate the performance of a few multilocus methods when heterogeneity exists and identify the most powerful methods to use in the X chromosome framework. We also give recommendation of the best approach specifically for common and rare variants. 

The rest of this paper is organized as follows. In Section 2, we review several multilocus analysis methods, and explain only the new random effects method \(S_\text{new}\)~\cite{Han2011} and SKATO \cite{Lee2012} remains most powerful under both homogeneous and heterogeneous situation across multiple SNPs. We next derive the framework of applying multilocus methods to X chromosome SNPs. In Section 3, we propose invariant multilocus tests with theoretical justifications, showing the tests are invariant to arbitrary XCI status across multiple X chromosome SNPs. In Section 4, we investigate the maximum power loss from misspecification of XCI status and genotype effects. Although the maximum loss is bounded for single SNP \cite{Chen2021}, we show it may not be bounded as the number of SNPs increases when they have heterogeneous effects, causing problem to all multilocus analysis methods. We find this problem can be solved using Cauchy combination test (CCT) \cite{Liu2020}. In Section 5, we recommend different X chromosome multilocus methods after applying CCT to common and rare variants, and show significant power improvement by extensive simulation studies compared to directly applying autosome multilocus methods to X chromosome. In Section 6, we apply our methods to two real problems with common and rare variants respectively, and demonstrate our methods can potentially identify novel loci from X chromosome. Finally in Section 7, we discuss the limitations and future extension directions of our work. 

    

\section{Multilocus Analysis Methods Review}
In this section, we review commonly used methods for analyzing multiple autosomal variants, including linear tests, quadratic tests and hybrid tests. These methods have also been implemented on X chromosome assuming additive coding and fixed XCI status. Suppose there are \( k \) variants, and the null hypothesis we want to test is $H_0: \beta_1=...=\beta_k=0$. As we will discuss next, linear, quadratic and hybrid tests focus on different alternative hypotheses.


\subsection{Linear Tests}
The most commonly used linear test is Burden test using score statistic. Given the score statistic \( \bm{G}' (Y - \tilde{Y}) \), \( \bm{G} = (G_{1}, \dots, G_{k}) \) is an \( n \times k \) matrix representing the genotypic matrix  of \( k \) variants, \(\bm{X}\) is the covariate matrix including an intercept term (i.e., a column of ones) 
for \( n \) individuals, and \( {\tilde{Y}} \) is an \( n \)-dimensional vector of predicted trait values under $H_0$. Burden test statistic is given by
\begin{align}
Q_B = \left\| \bm{w'G'} (Y - \tilde{Y}) \right\|^2.
\end{align}
 Under the null hypothesis, statistic \( Q_B \) follows a chi-squared distribution with 1 degree of freedom. The choice of weights \( w \) can vary depending on the study design: a simple option is \( w_i = 1 \) for all \( i \) \cite{Morgenthaler2007}, while in rare variant analyses, weights are often determined by minor allele frequency (MAF), giving higher weights to more rare variants \cite{Madsen2009}.

Burden test assumes each genetic effect $\beta_i=w_i \mu$, and the alternative hypothesis is $H_1: \mu \neq 0$, so it is most powerful when the effects of all variants are in the same direction. Their power may substantially decrease under heterogeneous effect scenarios, i.e., when both positive and negative values of \( \beta_{i} \) are present. In such cases, we may want to use quadratic tests by assuming a variance component for each $\beta_i$.

\subsection{Quadratic Tests}
Many quadratic tests are formulated as a quadratic form of the score statistic. For example, the Sequence Kernel Association Test (SKAT) is given by
\begin{align}
Q_S =  {(Y - \tilde{Y})'\bm{G W G'} (Y - \tilde{Y})},
\end{align}
where \( \bm{W} = \mathrm{diag}(w_1^2, \dots, w_k^2) \) is a diagonal weight matrix, and the weights \( w_i \) are chosen as functions of MAF, often using a Beta function to upweight rare variants \cite{Wu2011}. For the other quadratic tests, SSU and C-alpha tests are the special case where $\boldsymbol{W}$ reduce to the identity matrix $\boldsymbol{I}$ \cite{Derkach2014}.

Under the null hypothesis, \( Q_S \) asymptotically follows a weighted sum of chi-squared distributions:
\[
Q_S \sim \sum_{i=1}^{k} \lambda_i \chi^2_{1,i},
\]
where \( \chi^2_{1,i} \) are independent chi-squared random variables with 1 degree of freedom, and \( \lambda_i \) are the eigenvalues of the matrix
\(
\bm{\Lambda}=\bm{W ^{1/2} G' (I - X (X' X)^{-1} X') G W^{1/2}} .
\) The null distribution of \( Q_S \) can be accurately and efficiently approximated using the methods proposed by Davies or Liu \cite{Davies1980, Liu2009}.

SKAT assumes independent random effects with zero mean and a variance component $\tau$ from each genetic variant under the alternative hypothesis, i.e., $\beta_i \sim N(0, w_i^2\tau^2)$ for $i=1,...,k$. Then the null hypothesis is equivalent to $H_0: \tau = 0$, and the alternative hypothesis is $H_1: \tau>0$. So SKAT is the most powerful when $\tau>0$. When $\mu \neq 0$ but $\tau=0$, the test is underpowered compared with burden tests. 

Similar to SKAT, Hotelling's \( T^2 \) test also utilizes a quadratic form of the score statistic. The primary distinction lies in the choice of the weight matrix: Hotelling's test uses \( \bm{W = \Sigma_0^{-1}} \), the inverse covariance matrix of the score statistics. The null distribution follows a chi-squared distribution with \( k \) degrees of freedom. Hotelling's test does not have any specific assumption about the alternative hypothesis, where the genetic effects can be either fixed or random. However, due to the essence of the quadratic test form, its power performance is similar to SKAT, i.e., most powerful when $\tau>0$ but less powerful when $\tau=0$. We show the power comparison results in Supplementary Appendix C. 

P-value combination methods in general has similar power performance with quadratic tests. This is because most genetic effects tests are two-sided, and using p-value as the test statistic will not consider the direction of each $\beta_i$. We choose Fisher's method \cite{Fisher1925} as a representation of p-value combination methods. The combined test statistic is defined as
\[
\chi^2_{2k}=-2\sum_{i=1}^k \log(p_i),
\]
which follows a chi-squared distribution with $2k$ degrees of freedom under null hypothesis. It also has no constraint of the alternative hypothesis. Under various alternative hypotheses, our simulation results in Supplementary Appendix C confirm that it has similar power to SKAT and Hotelling's \( T^2 \).

\subsection{Hybrid Tests}
When the alternative hypothesis is unknown, it is desired to have a test which remains powerful under the unconstrained alternative hypothesis $H_1: \mu \neq 0$ and/or $\tau>0$. The new random effects test $S_{\text{new}}$ \cite{Han2011} is a likelihood ratio test specifically designed to test such alternative hypothesis. $S_{\text{new}}$ assumes independent random effects $\beta_i \sim N(\mu, \tau^2)$ under $H_1$ for $i=1,...,k$. Under the null and alternative hypothesis, the corresponding likelihood functions are:
\begin{align}
L_0 = \prod_{i=1}^{k} \frac{1}{\sqrt{2\pi V_i}} \exp\left(-\frac{\hat{\beta}_{i}^2}{2V_i}\right), \quad
L_1 = \prod_{i=1}^{k} \frac{1}{\sqrt{2\pi (V_i + \tau^2)}} \exp\left(-\frac{(\hat{\beta}_{i} - \mu)^2}{2(V_i + \tau^2)}\right),
\end{align}
$\hat{\beta}_{i}$ denotes regression estimation of ${\beta}_{i}$, with variance \(V_i = \text{Var}(\hat{\beta}_{i})\). The likelihood ratio test statistic is then computed as:
\begin{align}
S_{\text{new}} = -2\log\left(\frac{L_0}{L_1}\right) = \sum_{i=1}^{k} \log\left(\frac{V_i}{V_i + \hat{\tau}^2}\right) + \sum_{i=1}^{k} \frac{\hat{\beta}_{i}^2}{V_i} - \sum_{i=1}^{k} \frac{(\hat{\beta}_{i} - \hat{\mu})^2}{V_i + \hat{\tau}^2}.
\end{align}
Estimates of ${\mu}$ and ${\tau}^2$ are obtained by an iterative procedure proposed by Hardy and Thompson \cite{Hardy1996}.
Asymptotically, the test statistic follows a 1:1 mixture of chi-squared distributions with 1 and 2 degrees of freedom \cite{Self1987}. To improve p-value accuracy when $k$ is not large, Han et al. \cite{Han2011} provided tabulated p-value tables for $k$ from 2 to 50.

An alternative approach to $S_\text{new}$ is to balance the strengths of linear and quadratic tests. As discussed above, burden test tends to be more powerful when the majority of variant effects are in the same direction (either all protective or all deleterious), whereas SKAT is more effective when the effects of causal variants are bidirectional (i.e., a mix of protective and deleterious). SKATO \cite{Lee2012} test was developed to combine burden and SKAT statistics, achieving robust power across a broad range of genetic effect patterns. SKATO test statistic is defined as
\begin{align}
Q_{\rho} = \rho Q_B + (1 - \rho) Q_S, \quad 0 \leq \rho \leq 1,\label{SKATO-Q}
\end{align}
which represents a weighted combination of burden test statistic \( Q_B \) and SKAT statistic \( Q_S \). This formulation allows SKATO to be reduced to SKAT when \( \rho = 0 \), and to the burden test when \( \rho = 1 \).

In practice, the optimal value of \( \rho \) is unknown and must be inferred from the data to maximize statistical power. The final SKATO test statistic is obtained by selecting the value of \( \rho \) that yields the smallest p-value:
\begin{align}
Q_{\text{optimal}} = \min_{0 \leq \rho \leq 1} p_{\rho},
\end{align}
where \( p_{\rho} \) is the \(p\)-value corresponding to each \( Q_{\rho} \).

Similar to $S_\text{new}$, SKATO also assumes random genetic effects, and has no constraint about the alternative hypothesis. The difference is that SKATO can incorporate the weights defined in Burden and SKAT tests to assign higher weights to rare variants. Specifically, the random effects and alternative hypothesis are $\beta_i \sim N(w_i\mu, w_i^2\tau^2)$ for $i=1,...,k$ and $H_1: \mu \neq 0$ and/or $\tau>0$.
\subsection{Brief Summary}
\begin{table}[htbp]
    \centering
    \begin{tabular}{c c c c c c c}
        \toprule& \multicolumn{6}{c}{\textbf{Statistics: }} \\\midrule
        \textbf{Testing $H_1$:}  & Burden Test & SKAT & Hotelling's $T^2$ & Fisher's Method & $S_{\text{new}}$ & SKATO \\ \hline
        $\mu \neq 0, \tau=0$& $\surd$ & $\times$ & $\times$ & $\times$ & $\surd$ & $\surd$ \\ \hline
        $\mu=0, \tau > 0$& $\times$ & $\surd$ & $\surd$ & $\surd$ & $\surd$ & $\surd$ \\ \hline
        $\mu\neq 0$, $\tau > 0$& $\times$ & $\times$ & $\times$ & $\times$ & $\surd$ & $\surd$ \\ \bottomrule
    \end{tabular}
    \caption{Summary of methods for testing genetic effects. A check-mark ($\surd$) indicates that the method has sufficient power under the corresponding alternative hypothesis, while a cross ($\times$) indicates the test is underpowered.}
    \label{tab:methods_review1}
\end{table}

We summarize the ability of different methods to test the parameters $\mu$ and $\tau$ under different alternative hypotheses in Table \ref{tab:methods_review1}. We conclude that only the hybrid tests, i.e., $S_{\text{new}}$ and SKATO are capable of simultaneously testing for mean and variance components. This conclusion is also validated by our simulation results in Supplementary Appendix C, where we compare the power of all six tests under the alternative hypothesis of $\mu \neq 0, \tau=0$ and $\mu=0, \tau>0$ for $k=10$ and 50. In the remaining sections, we will primarily focus on the theoretical developments and application of $S_{\text{new}}$ and SKATO to X chromosome multilocus analysis problem. However, our developed test framework is not restricted to hybrid tests and can also be applied to other linear and quadratic tests if desired.

\section{Multilocus Invariant Testing Framework on X Chromosome}
Association testing problem on the X chromosome is more complex than autosome.  Models that account only for additive effects may fail to capture the full extent of the underlying biological mechanisms. Chen \cite{Chen2021} discussed 8 challenges, including unknown X-chromosome inactivation (XCI) status and reference allele, and decisions about whether to include, sex, gene-sex interaction and dominant effects in the model. For single variant, they proposed a regression-based method with three degrees of freedom (3-df) under generalized linear model:
\begin{align}
    \mathcal{M}: g(E(Y))=\beta_0 +\beta_S{S} + \beta_A G_A + \beta_D G_D + \beta_{GS} GS\label{3dflm_chen}
\end{align} and the corresponding 3 df test, jointly testing: 
\(
H_0: \beta_A = \beta_D = \beta_{GS} = 0,
\) which successfully addresses these challenges. Here \(Y\) denotes continuous or discrete trait vector with link function $g$, \( G_A \), \( G_D \) and \( GS \) denote the additive genotype coding, dominant genotype coding, and gene–sex interaction coding vectors, respectively. All those vectors are with length of sample size \(n\). The corresponding effect size vectors are denoted by \( \beta_A \), \( \beta_D \), and \( \beta_{GS} \), respectively. The coding schemes are summarized in Table \ref{tab:codingschemes}.
\begin{table}[htbp]
    \centering
    \begin{tabular}{c c c c c c c}
        \toprule&   & \multicolumn{5}{c}{ Coding Schemes}\\  \midrule
        Effect  &XCI &\multicolumn{3}{c}{Females~~}&\multicolumn{2}{c}{males~~}\\ 
        Interpretation  &  Status  & \textit{rr}&\textit{rR}&\textit{RR}&\textit{r}&\textit{R} \\\hline
        Additive & Yes &0&0.5&1&0&1\\
        $G_A$  & No &0&1&2&0&1\\\hline
        Interaction  $GS = G \times {S}$ & Either &0&0&0&0&1\\ \hline
        Dominant $G_D$& Either&0&1&0&0&0\\\hline
        $S$ &Either&0&0&0&1&1\\\bottomrule
    \end{tabular}
    \caption{Genotype coding notation. \textit{r} represents the reference allele.}
    \label{tab:codingschemes}
\end{table}

With $k$ variants, the null hypothesis becomes $H_0: \beta_{A,1}=\dots=\beta_{A,k}=\beta_{D,1}=\dots=\beta_{D,k}=\beta_{GS,1}=\dots=\beta_{GS,k}=0$. Our target is to find an X-chromosome multilocus testing framework which could address the above-mentioned challenges. The naive approach is to test additive effects, dominant effects and gene-sex interaction effects separately with the multilocus testing methods discussed in Section 2. However, this approach has two problems. Firstly, the additive test results will not be invariant to XCI status of each variant. For multilocus test with $k$ variants, there are $2^k$ possible testing results considering each variant could be inactivated or not. How to choose or average from such a large number of candidate results remains unclear. Secondly, it is clear that the codings $G_A$, $G_D$ and $GS$ are not independent of each other. After achieving the multilocus test statistics for each coding, we may also have interest with the overall test combining three tests, but the distribution of the sum of these three statistics is unclear.

In this section, we propose a three-step transformation framework of the genotype codings, which are Step 1: Sex Stratification, Step 2: Genotype Reparametrization, and Step 3: Genotype Standardization. After transformation, the codings become invariant to XCI status, so no matter what the XCI status is at each variant, there will be only one invariant multilocus test result. Moreover, the new codings after transformation become independent of each other, so the overall global multilocus test statistic is simply the sum of three multilocus test statistics. To make the transformation framework valid, the principle is that the test under new codings should be equivalent to the old ones for testing $H_0: \beta_A = \beta_D = \beta_{GS} = 0$ for each variant. We will prove that the 3 df statistics of this test are invariant after each transformation step, i.e., our transformation will not affect single variant test results.

\subsection{Sex Stratification}


Sex stratification is not a new idea in X-chromosome association analysis \cite{Gao2015} to eliminate the confounding effects of sex and gene-sex interaction effects. However, previous applications only stratified with additive codings and there was a lack of theoretical justification. In this paper, our contribution is that we show the regression tests are invariant before and after sex stratification, and the invariance property holds for multiple codings stratification. Therefore, the combination test of females and males after stratification is equivalent to the 3 df test of $H_0: \beta_A = \beta_D = \beta_{GS} = 0$.

For clarity, we provide the following notation under sex stratification. For females, the additive genotype is represented by \( G_f =(0,1,2)'\), under the assumption XCIE, and \( G_f =(0,0.5,1)'\), under the assumption XCI. The dominant genotype by \( G_d = (0,1,0)'\).
For males, the genotype coding is denoted by \( G_m = (0,1)'\).
The corresponding effect are \( \beta_f \), \( \beta_d \), and \( \beta_m \), respectively.

The full model for females is
\begin{align}\mathcal{M}_f: g(E(Y_f)) = {\beta_{0,f}} + {G_{f}\beta_{f}}  + {{G_{d}}\beta_d} ,\label{model:female}\end{align}
and for males is 
\begin{align}\mathcal{M}_m: g(E(Y_m)) = {\beta_{0,m}} + {G_{m}\beta_{m}},\label{model:male}\end{align}
where \(Y_f\) and \(Y_m\) represent the trait values in females and males, respectively. Since dominant effects are only defined for females, the male model does not include \(G_d\).  

The relationships among the initial parameters \( \beta_A \), \( \beta_D \), and \( \beta_{GS} \), and the parameters after sex stratification (with females taken as the reference group) are summarized as follows:
\begin{align}
    \beta_{A} &=\beta_{f}\\
    \beta_{D} &= \beta_d\\
    \beta_{GS}&=\beta_{m} -\beta_{f}
\end{align} 

Next, we show the stratified test is equivalent to the individual-level full model \(\mathcal{M}\) with the Wald, Score, and Likelihood Ratio Test (LRT) statistics.

\begin{theorem}
    Let $Y = \left(\begin{array}{c}
                   Y_f \\
                   Y_m \\
\end{array}\right)$,
where $Y_f$ and $Y_m$ are the response vectors with length $n_f$ and $n_m$. 
$X = \left(\begin{array}{c}
                   X_f \\
                   X_m \\
\end{array}\right)$
where $X_f$ and $X_m$ are the covariates matrices with dimension $n_f \times p$ and $n_m \times p$.
$S = \left(\begin{array}{c}
                   0_{n_f} \\
                   1_{n_m} \\
\end{array}\right)$
equals 0 for females and 1 for males. Under generalized linear model, the full regression model is $\mathcal{M}: g(E(Y))=\beta_0+\beta_S S+X\boldsymbol{\beta_1}+XS\boldsymbol{\beta_2}$, where 
$XS = \left(\begin{array}{c}
                   0_{n_f \times p} \\
                   X_m \\
\end{array}\right)$,
$\boldsymbol{\beta_1}$ and $\boldsymbol{\beta_2}$ are parameter vectors with length $p$. The sex stratified models are $\mathcal{M}_f: g(E(Y_f))=\beta_{0,f}+X_f\boldsymbol{\beta_f}$ and $\mathcal{M}_m: g(E(Y_m))=\beta_{0,m}+X_m\boldsymbol{\beta_m}$. Then the Wald, Score and likelihood ratio test (LRT) for testing $H_0: \boldsymbol{\beta_1}=\boldsymbol{\beta_2}=0$ and $H_0: \boldsymbol{\beta_f}=\boldsymbol{\beta_m}=0$ are identical.
\end{theorem}

We present the proof of Theorem 1 in Supplementary Appendix A. We apply the theorem with $p=2$, $X=(G_A,G_D)$ and $\boldsymbol{\beta_1}=(\beta_A,\beta_D)'$. Note that $G_D \times S=0$, so $XS=(GS,0_{n_f+n_m})$ where $GS=G_A \times S$, and $\boldsymbol{\beta_2}=(\beta_{GS},0)'$. Stratifying by sex, we get $X_f=(G_f,G_d)$ with $\boldsymbol{\beta_f}=(\beta_f,\beta_d)'$, and $X_m=(G_m,0_{n_m})$ with $\boldsymbol{\beta_m}=(\beta_m,0)'$. Removing the covariate with all zeros, the theorem shows that testing $H_0: \beta_A=\beta_D=\beta_{GS}=0$ under the full model $\mathcal{M}$ is equivalent to testing $H_0: \beta_f=\beta_d=\beta_m=0$ under sex stratified models \(\mathcal{M}_f\) and \(\mathcal{M}_m\).

\subsection{Genotype Reparametrization}


It is obvious that the sex-stratified test of females and males are independent. However, in the female population, the additive component \( G_f \) and the dominant component \( G_d \) may not be independent. The main purpose of step 2 is to reparametrize $G_f$ and $G_d$ so they are uncorrelated and invariant to unknown XCI status. Assuming that the genotype frequencies for \textit{rr}, \textit{rR}, and \textit{RR} are \(p_1\), \(p_2\), and \(p_3\) (with \(p_1 + p_2 + p_3 = 1\)) respectively, we define \(G_f^*\) and \(G_d^*\) as:
\[
    G_f^* = (-1, 0, 1)', \quad G_d^* = \left(-p_3, \frac{2p_1p_3}{p_2}, -p_1\right)',
\]
The genotype frequencies are usually unknown but can be easily estimated in practice. Because males only have one coding, reparameterization is not necessary, and \(G_m^* = G_m\). After transformation, it is easy to verify that
\begin{align}
    \text{Cov}(G_f^*,G_d^*)  
    &= \text{E}(G_f^*G_d^*) - \text{E}(G_f^*)\text{E}(G_d^*) \notag\\
    &=(p_1p_3-p_1p_3) - (p_3-p_1)(-p_1p_3+2p_1p_3-p_1p_3) \notag\\
    &=0 \notag
\end{align}

The relationships between the reparametrized parameters and the original $\beta_f$ and $\beta_d$ depends on the codings of $G_f$. With XCIE coding such that $G_f=(0,1,2)'$, 
\[
\beta_f^* = \beta_f - \frac{(p_3 - p_1)p_2}{4p_1p_3 + p_1p_2 + p_2p_3} \beta_d;
\]
with XCI coding such that $G_f=(0,0.5,1)'$,
\[
\beta_f^* = \frac{\beta_f}{2} - \frac{(p_3 - p_1)p_2}{4p_1p_3 + p_1p_2 + p_2p_3} \beta_d.
\]
$\beta_d^* = \frac{2p_2}{4p_1p_3 + p_1p_2 + p_2p_3} \beta_d$ regardless of XCI status. $\beta_m^* = \beta_m$ which is not reparametrized.

Depending on $G_f$, the design matrix with transformed codings is a linear transformation of the original codings. If $G_f=(0,1,2)'$,
\[
(1_3,G_f^*,G_d^*)=(1_3,G_f,G_d)\left(\begin{matrix}
    1&-1&-p_3 \\
    0&1&\frac{p_3-p_1}{2}\\
    0&0&\frac{4p_1p_3+p_1p_2+p_2p_3}{2p_2}
\end{matrix}\right),
\]
where $1_3=(1,1,1)'$. Alternatively, if $G_f=(0,0.5,1)'$,
\[
(1_3,G_f^*,G_d^*)=(1_3,G_f,G_d)\left(\begin{matrix}
    1&-1&-p_3 \\
    0&2&p_3-p_1\\
    0&0&\frac{4p_1p_3+p_1p_2+p_2p_3}{2p_2}
\end{matrix}\right).
\]
Chen \cite{Chen2021} proposed Theorem 1 in their paper saying that the Wald, Score, and LRT statistics are invariant, if the full design matrix is a linear transformation, and the design matrix under $H_0$ (which is $1_3$ here) is also a linear transformation between new and old codings. With the existence of the transformation matrices, the theorem implies that the test statistics for testing $H_0 : \beta_f^* = \beta_d^* = 0$ are equivalent to those for $H_0 : \beta_f = \beta_d = 0$ under the original codings. Because male codings are not changed, Step 2 has no change to the test statistic of the 3 df test $H_0: \beta_f=\beta_d=\beta_m=0$ for each variant.

\subsection{Genotype Standardization}
The multilocus test for each of $G_f^*$, $G_d^*$ and $G_m^*$ has become invariant to XCI status after genotype reparametrization. However, the Burden, SKAT and SKATO test statistics directly involve the genotypes, and thus sensitive the scale of genotype codings. Originally not a problem for each single test, it may cause a problem when combining three test statistics because different scales of each coding implicitly assume different weights, and thus leads to different test statistics for the overall test. In practice, genotype scales are not unique. For example, some people use $(0,2)'$ coding for male instead of $(0,1)'$. Our reparameterization $G_f^*$ and $G_d^*$ also appears somewhat ad hoc, and it is troublesome to find that rescaling $G_f^*$ and $G_d^*$ leads to a different overall test result.


To address these issues, we standardize the reparametrized genotype codings. Specifically, we define
\[
G_f^\dagger = \frac{G_f^*}{\text{sd}(G_f^*)}, \quad G_d^\dagger = \frac{G_d^*}{\text{sd}(G_d^*)}, \quad G_m^\dagger = \frac{G_m^*}{\text{sd}(G_m^*)},
\]
where \(\text{sd}(G_f^*)\), \(\text{sd}(G_d^*)\), and \(\text{sd}(G_m^*)\) denote the standard deviations of the reparametrized genotypes \(G_f^*\), \(G_d^*\), and \(G_m^*\), respectively. After standardization, the multilocus test statistic for each of $G_f^\dagger$, $G_d^\dagger$ and $G_m^\dagger$ become invariant to genotype coding scales for all 6 test statistics summarized in Table 1. So the overall tests integrating $G_f^\dagger$, $G_d^\dagger$ and $G_m^\dagger$ are also invariant to coding scales.

After standardization, the corresponding effects are 
\[
\beta_f^\dagger=\beta_f^* \text{sd}(G_f^*), \quad \beta_d^\dagger=\beta_d^* \text{sd}(G_d^*) \quad \beta_m^\dagger=\beta_m^* \text{sd}(G_m^*).
\]
It is clear that $(G_f^\dagger,G_d^\dagger,G_m^\dagger)$ is a linear transformation of $(G_f^*,G_d^*,G_m^*)$. By applying Theorem 1 from Chen \cite{Chen2021} again, the Wald, Score, and LRT statistics remain invariant for testing $H_0: \beta_f^\dagger=\beta_d^\dagger=\beta_m^\dagger=0$ vs. $H_0: \beta_f^*=\beta_d^*=\beta_m^*=0$.

Therefore, we conclude that after the three step transformation, the 3 df test of $H_0: \beta_f^\dagger=\beta_d^\dagger=\beta_m^\dagger=0$ is equivalent to $H_0: \beta_A=\beta_D=\beta_{GS}=0$ with the original codings for each variant. This provides theoretical justifications for our transformation framework.



\subsection{Applying Transformation Framework to Hybrid Tests}
After explaining the transformation Steps 1-3, we now illustrate how the transformation framework can be incorporated into the multilocus hybrid tests discussed in Section 2, i.e., $S_\text{new}$ and SKATO. It needs to be noted that transformation is not required for Hotelling's \( T^2 \) and Fisher's method, because it is easy to show that those test results are invariant to XCI status even without the transformation. Nevertheless, the three step transformation is still valid with Hotelling's \( T^2 \) and Fisher's method because it makes no change to the original test results. 

For SKATO, weights are implicitly added in the standardization step, because the score statistics are directly computed from transformed genotypes. Most genotype codings for rare variants are zero, so $\text{sd}(G_f^*)$, $\text{sd}(G_d^*)$ and $\text{sd}(G_m^*)$ are all smaller compared to common variants, which gives higher weights to the standardized codings $G_f^\dagger$, $G_d^\dagger$ and $G_m^\dagger$ of rare variants. The reason we give higher weights to rare variants was that otherwise rare variants with same effect size to common variants would be less significant. Intuitively speaking, after standardization the score statistic ${G_i^\dagger}' (Y - \tilde{Y})$ for each variant becomes equally important regardless of common and rare, so including a secondary weight function is not necessary. For this reason, we eliminate the notation of weights when applying SKATO methods to X chromosome after transformation. 

It needs to be noted that the implicit weights given by standardization only have effects to score function, and thus only apply to Burden, SKAT and SKATO methods. Regarding $S_\text{new}$, both $\hat{\beta_i}$ and its variance $V_i$ are standardized, which indicates that no weights are adjusted for any variant after standardization.

After transformation, the testing model becomes
\[\mathcal{M}_{f} : g(E(Y_f)) = \beta_{0,f}^\dagger + \bm{G_f^\dagger\beta_f^\dagger}+ \bm{G_d^\dagger\beta_d^\dagger}, \quad
\mathcal{M}_{m} : g(E(Y_m)) = \beta_{0,m}^\dagger + \bm{G_m^\dagger\beta_m^\dagger}, \]
where $\boldsymbol{\beta_f^\dagger}=(\beta_{f,1}^\dagger,...,\beta_{f,k}^\dagger)'$, $\boldsymbol{\beta_d^\dagger}=(\beta_{d,1}^\dagger,...,\beta_{d,k}^\dagger)'$ and $\boldsymbol{\beta_m^\dagger}=(\beta_{m,1}^\dagger,...,\beta_{m,k}^\dagger)'$. $\bm{G_f^\dagger} = (G_{f,1}^\dagger, \dots, G_{f,k}^\dagger)$, $\bm{G_d^\dagger} = (G_{d,1}^\dagger, \dots, G_{d,k}^\dagger)$ and $\bm{G_m^\dagger}=(G_{m,1}^\dagger, \dots, G_{m,k}^\dagger)$ are $n \times k$ matrices with transformed genotypes. For multilocus testing, we consider $\beta_{f,i}^\dagger$, $\beta_{d,i}^\dagger$ and $\beta_{m,i}^\dagger$ as independent random effects for $i=1,...,k$. Specifically, after we eliminate the weights, we assume
\[
\beta_{f,i}^\dagger \sim {N}(\mu_f, \tau_f^2), \quad \beta_{d,i}^\dagger \sim {N}(\mu_d, \tau_d^2), \quad \beta_{m,i}^\dagger \sim {N}(\mu_m, \tau_m^2).
\]
Clearly, the null hypothesis $H_0: \beta_{A,1}=\dots=\beta_{A,k}=\beta_{D,1}=\dots=\beta_{D,k}=\beta_{GS,1}=\dots=\beta_{GS,k}=0$ is 
equivalent to $H_0: \mu_f = \mu_d =\mu_m = \tau_f = \tau_d = \tau_m = 0$. Next, we discuss how to test this hypothesis using hybrid tests.

\subsubsection{New Random Effects Test $S_\text{new}$}

We run regression models separately for females and males after three step transformation, and achieve the estimators \( \hat{\beta}_{f,i}^\dagger \), \( \hat{\beta}_{d,i}^\dagger \), and \( \hat{\beta}_{m,i}^\dagger \) representing  LSEs or MLEs of the corresponding parameters.
Under the null hypothesis, these estimators follow normal distributions with zero mean and variances \( V_{f,i}^\dagger \), \( V_{d,i}^\dagger \), and \( V_{m,i}^\dagger \), respectively. These variances are unknown in practice, so we use their estimated values \( \hat{V}_{f,i}^\dagger \), \( \hat{V}_{d,i}^\dagger \), and \( \hat{V}_{m,i}^\dagger \) instead. We apply an iterative procedure described in~\cite{Hardy1996} to estimate the parameters \( (\mu_f, \tau_f^2) \), \( (\mu_d, \tau_d^2) \), and \( (\mu_m, \tau_m^2) \).

The overall test statistic \( S_\text{new}^\dagger \) consists of three independent components : 
\begin{align}
S_\text{new}^\dagger = S_f + S_d + S_m,
\end{align}
where
\[S_f = \sum_{i=1}^k \log\left(\frac{\hat{V}_{f,i}^\dagger}{\hat{V}_{f,i}^\dagger + \hat{\tau}_{f}^2}\right) + \sum_{i=1}^k \frac{(\hat{\beta}_{{f,i}}^\dagger)^2}{\hat{V}_{f,i}^\dagger} - \sum_{i=1}^k \frac{(\hat{\beta}_{f,i}^\dagger - \hat{\mu}_f)^2}{\hat{V}_{f,i}^\dagger + \hat{\tau}_{f}^2} \]
\[S_d = \sum_{i=1}^k \log\left(\frac{\hat{V}_{d,i}^\dagger}{\hat{V}_{d,i}^\dagger + \hat{\tau}_{d}^2}\right) + \sum_{i=1}^k \frac{(\hat{\beta}_{{d,i}}^\dagger)^2}{\hat{V}_{d,i}^\dagger} - \sum_{i=1}^k \frac{(\hat{\beta}_{d,i}^\dagger - \hat{\mu}_d)^2}{\hat{V}_{d,i}^\dagger + \hat{\tau}_{d}^2} \]
\[S_m = \sum_{i=1}^k \log\left(\frac{\hat{V}_{m,i}^\dagger}{\hat{V}_{m,i}^\dagger + \hat{\tau}_{m}^2}\right) + \sum_{i=1}^k \frac{(\hat{\beta}_{{m,i}}^\dagger)^2}{\hat{V}_{m,i}^\dagger} - \sum_{i=1}^k \frac{(\hat{\beta}_{m,i}^\dagger - \hat{\mu}_m)^2}{\hat{V}_{m,i}^\dagger + \hat{\tau}_{m}^2} \]

To assess the significance of \( S_\text{new}^\dagger \), an efficient approach is to use the asymptotic distribution. Under $H_0$, the statistic asymptotically follows a mixture of \( 3 \), \( 4 \), \( 5 \), and \( 6 \) df chi-squared distributions with a ratio of 1:3:3:1. Self and Liang \cite{Self1987} discussed more details about how to derive the asymptotic distribution. However, the asymptotic result is valid only when $k$ is large. For smaller $k$, the asymptotic p-value may be overly conservative \cite{Han2011}.
To address this, we provide tabulated values. For \(k = 2\) to \(50\), we generate \( 10^7 \) null statistics to construct p-value tables that offer reasonable precision up to \( 10^{-5} \). For p-values more significant than \( 10^{-5} \), we apply the asymptotic p-value, adjusted by the ratio between the asymptotic p-value and the true p-value estimated at \( 10^{-5} \).

\subsubsection{SKATO}
For $k$ variants, the score statistics after transformation are: 
\[
\bm{{G_f^\dagger}'} (Y_f - \tilde{Y}_f), \quad \bm{{G_d^\dagger}'} (Y_f - \tilde{Y}_f), \quad \bm{{G_m^\dagger}'} (Y_m - \tilde{Y}_m),
\]
where \(\tilde{Y}_f, \tilde{Y}_d~\text{and}~\tilde{Y}_m \) are the fitted values under the null hypothesis.

The overall Burden test is the squared linear composite of the score statistics, since three components are independent, and the scales are comparable after the transformation and eliminating the weights:
\begin{align}
    Q_B^\dagger =Q_B^f + Q_B^d + Q_B^m= \left\| \bm{ {G_f^\dagger}'} (Y_f - \tilde{Y}_f) \right\|^2 + \left\| \bm{ {G_d^\dagger}'} (Y_f - \tilde{Y}_f)\right\|^2 + \left\| \bm{ {G_m^\dagger}'} (Y_m - \tilde{Y}_m) \right\|^2,
\end{align}
$Q_B^\dagger$ follows 3 df chi-squared distribution under the null hypothesis. 

The overall SKAT test is analogous to the Burden test:
\begin{align}
    Q_S^\dagger &= Q_S^f + Q_S^d + Q_S^m,
\end{align}
where
\[Q_S^f = (Y_f - \tilde{Y}_f)'\bm{G_f^\dagger W{G_f^\dagger}'} (Y_f - \tilde{Y}_f),\]
\[Q_S^d = (Y_f - \tilde{Y}_f)'\bm{G_d^\dagger W{G_d^\dagger}'} (Y_f - \tilde{Y}_f),\]
\[Q_S^m = (Y_m - \tilde{Y}_m)'\bm{G_m^\dagger W{G_m^\dagger}'} (Y_m - \tilde{Y}_m),\]
and $\boldsymbol{W}$ is the identity matrix without weights. Under the null hypothesis, \( Q_S^\dagger \) asymptotically follows the weighted sum of $3k$ chi-squared distributions with 1 df:
\[
\sum_{i=1}^k \lambda_{f,i} \chi^2_{1,f,i} + \sum_{i=1}^k \lambda_{d,i} \chi^2_{1,d,i} + \sum_{i=1}^k \lambda_{m,i} \chi^2_{1,m,i}.
\]
\( \lambda_{f,i} \), \( \lambda_{d,i} \), and \( \lambda_{m,i} \) are the eigenvalues of the matrices
\[\bm{\Lambda_f=W^{1/2} {G_f^\dagger}'(I - X_f(X_f'X_f)^{-1}X_f')G_f^\dagger W^{1/2}}, \]
\[\bm{\Lambda_d=W^{1/2} {G_d^\dagger}'(I - X_d(X_d'X_d)^{-1}X_d')G_d^\dagger W^{1/2}}, \]
\[\bm{\Lambda_m=W^{1/2} {G_m^\dagger}'(I - X_m(X_m'X_m)^{-1}X_m')G_m^\dagger W^{1/2}}. \]
Given the value of \( Q_S^\dagger \) and its eigenvalues, we use the method proposed by Liu \cite{Liu2009} to approximate its statistical significance.



The computation of the SKATO \(p\)-value follows a procedure similar to that proposed in SKAT, with the primary difference being $\boldsymbol{W}$ replaced by \(\bm{W_\rho}= [(1 - \rho)\bm{I} + \rho\bm{1}\bm{1'}]\bm{W}\). \( \bm{I} \) is the \( k \times k \) identity matrix, and \( \bm{1} \) is a length-\( k \) vector with all elements equal to 1. For each test component, we first estimate the optimal $\rho_f$, $\rho_d$ and $\rho_m$ separately using
\begin{align}
    \rho = \text{ratio}_1^2 (2\text{ratio}_2 - 1)^2,
    \label{estimated_rho}
\end{align}
where \(\text{ratio}_1\) is the proportion of non-zero $\beta_k$'s, and \(\text{ratio}_2\) is the proportion of positive $\beta_k$'s among the non-zero $\beta_k$'s \cite{Lee2012.1}.  
We then compute the \(Q\)-statistics \(Q_\rho^f\), \(Q_\rho^d\), and \(Q_\rho^m\) for the female additive, female dominant, and male additive components, respectively, using formula~(\ref{SKATO-Q}). The computation involves decomposing the matrices 
\[\bm{\Lambda_{\rho,f}=W_{\rho,f}^{1/2} {G_f^\dagger}'(I - X_f(X_f'X_f)^{-1}X_f')G_f^\dagger W_{\rho,f}^{1/2}}, \]
\[\bm{\Lambda_{\rho,d}=W_{\rho,d}^{1/2} {G_d^\dagger}'(I - X_d(X_d'X_d)^{-1}X_d')G_d^\dagger W_{\rho,d}^{1/2}}, \]
\[\bm{\Lambda_{\rho,m}=W_{\rho,m}^{1/2} {G_m^\dagger}'(I - X_m(X_m'X_m)^{-1}X_m')G_m^\dagger W_{\rho,m}^{1/2}}, \]
with $\rho$ in $\boldsymbol{W_\rho}$ replaced by $\rho_f$, $\rho_d$ and $\rho_m$, respectively.  The eigenvalues of $\boldsymbol{\Lambda_{\rho,f}}$, $\boldsymbol{\Lambda_{\rho,d}}$ and $\boldsymbol{\Lambda_{\rho,m}}$ are derived correspondingly as \(\lambda_{\rho,i}^f\), \(\lambda_{\rho,i}^d\), and \(\lambda_{\rho,i}^m\).


The overall SKATO test statistic is defined as the sum of the individual components, 
\begin{align}
    Q_\rho^\dagger &= Q_\rho^f + Q_\rho^d + Q_\rho^m.
\end{align}
Under the null hypothesis, $Q_\rho^\dagger$ follows the weighted sum of $3k$ chi-squared distributions with 1 df when $\rho \neq 1$, where the weights are the eigenvalues \(\lambda_{\rho,i}^f\), \(\lambda_{\rho,i}^d\), and \(\lambda_{\rho,i}^m\). The \(p\)-value can then be computed numerically using the Liu method~\cite{Liu2009}.

For rare variants (minor allele frequency, MAF \(<\) 0.01\(\sim \)0.05)~\cite{Lee2012}, genotypic matrices \(\bm{G_f}\), \(\bm{G_m}\), and \(\bm{G_d}\) tend to be sparse. In particular, certain genotypic codings may be identical across all individual, or for certain variants, the additive and dominant codings may become identical, for example, when only \(rr\) and \(rR\) genotypes are observed in females. In such cases, testing for a dominant effect is meaningless. Therefore, we recommend using the two-degrees-of-freedom (2-df) model, which computes the female and male additive effect \(Q\)-statistics and their corresponding eigenvalues separately. The combined \(Q\)-statistic is obtained by summing the two individual \(Q\)-statistics, and the \(p\)-value can be calculated using the combined set of eigenvalues.




\section{Power Loss Accumulation Problem and Our Solution} \label{Power Loss}
\subsection{The Problem with Multiple Variants}
The primary difference between our proposed multilocus X-chromosome tests and classic multilocus tests is that the degree of freedom of our tests is $3\times$ the classic ones, where the df of our tests can be as large as $3k$ with $k$ genetic variants. The natural question to ask is that when the classic testing model is correct, i.e., the genetic effects are truely additive with no gene-sex interaction and XCI status for each variant are correctly specified, will the extra degree of freedom reduce the power of our tests and make our framework less appealing? This question consists of two problems. First, incorrect specification of the XCI status (through erroneous coding) may result in a loss of power in the classic multilocus tests based on additive model. Second, incorrectly assuming the existence of gene-sex interaction and dominant effects may also lead to a power loss.

For the first problem, Ma \cite{Ma2015} presented some simulation results for Burden, SKAT and SKATO showing that the power loss due to misspecified XCI status was usually not large. However, they only considered a few specific parameter settings. The power loss depends on both effect sizes of each variant and type I error level $\alpha$. We find the power loss could be much bigger than what they found in many other settings. For the second problem, Chen \cite{Chen2021} noted that in a single variant model, jointly testing the null hypothesis $H_0: \beta_A = \beta_D = \beta_{GS} = 0$ leads to a maximum power loss of 18.8\% compared to testing $H_0: \beta_A = 0$ using the 1-df additive model alone. However, for multiple variants, we find the maximum power loss may be accumulated and not capped by 18.8\%. In this section, we compute the maximum possible power loss due to XCI status misspecification and incorrectly assuming gene-sex interaction and dominant effects. We find the maximum power loss is accumulated and can go up to 1 as $k$ increases. 

Formally, we assume the true model only include \(\bm{G_A} = (G_{A,1}, \dots, G_{A,k})\) that there are no dominant or interaction effects and that the XCI status is known. To compute the maximum power loss, we consider the extreme situation where the XCI status for all $k$ variants are misspeciffied. Without loss of generality, we assume all $k$ variants are inactivated. The additive model is
\[\mathcal{M}_A : g(E(Y)) = \beta_0 + \beta_SS + \bm{G_{A}\beta_A}\]
where $G_{A,i}=(0,0.5,1,0,1)'$ with the true XCI status and $G_{A,i}=(0,1,2,0,1)'$ with XCI status misspefication. The dominant and interaction effects \(\bm{G_D} = (G_{D, 1}, \dots, G_{D, k})\) and \(\bm{GS} = (GS_{1}, \dots, GS_{k})\) are included in the full model:
\[\mathcal{M}_{F} : g(E(Y)) = \beta_0 + \beta_SS + \bm{G_A\beta_A}+ \bm{G_D\beta_D} + \bm{GS\beta_{GS}}\]
The full model is equivalent to sex-stratified model after the transformation proposed in Section 3:
\[\mathcal{M}_{f} : g(E(Y_f)) = \beta_{0,f}^\dagger + \bm{G_f^\dagger\beta_f^\dagger}+ \bm{G_d^\dagger\beta_d^\dagger}, \quad
\mathcal{M}_{m} : g(E(Y_m)) = \beta_{0,m}^\dagger + \bm{G_m^\dagger\beta_m^\dagger}, \]
both of which may have reduced power if $\boldsymbol{\beta_D}=\boldsymbol{\beta_{GS}}=0$.

We first derive the maximum power loss from XCI status misspecification of $k$ variants. We start from $k=1$, and derive the distribution of the test statistic for testing $\beta_A=0$ from model $\mathcal{M}_A$ under the alternative hypothesis:

\begin{lemma}
    Let $W_A$ be the test statistic for testing $\beta_A=0$ following $\chi^2(1)$ distribution under $H_0$. Then under $H_1$, it will follow non-central chi-squared distribution with non-centrality parameter $ncp_{A,I}$ and $ncp_{A,N}$ under XCI and no XCI genotype codings respectively, where $ncp_{A,N}<ncp_{A,I}$ and both ncps $\propto \beta_A^2$.
\end{lemma}

Following Lemma 1, we can define $P(x,k,ncp)$ as the CDF of noncentral $\chi^2$ distribution. For \(k=1\) variant, $ncp \propto \beta_A^2$, and we let $ncp_A=c_1\beta_A^2$ under correctly specified XCI status, and $ncp_A=c_2\beta_A^2$ if XCI status is misspecified so that $c_1>c_2>0$. Let \(z_{\alpha}\) and $\chi^2_{k,\alpha}$ be the \((1-\alpha)\)-quantiles of $N(0,1)$ and $\chi^2(k)$ respectively. Then Theorem 2 shows that there exists $\beta_A$-dependent $\alpha$ such that power loss due to XCI status misspecification increases to 1 as $|\beta_A|  \to \infty$: 

\begin{theorem}
For any $a \in (c_2,c_1)$, if $z_{\alpha/2}=\sqrt{a}|\beta_A|$, then $P(\chi^2_{1,\alpha},1,c_2\beta_A^2)-P(\chi^2_{1,\alpha},1,c_1\beta_A^2) \to 1$ as $|\beta_A| \to \infty$.
\end{theorem}

Theorem 2 indicates that as $|\beta_A|$ increases, we can always find smaller $\alpha$ which leads to bigger power loss for any single variant. Because the power loss will accumulate as $k$ increases, it immediately implies that the maximum power loss due to misspecified XCI status is not bounded for any $k$. Therefore, we conclude that the classic multilocus tests based on 1df additive model may suffer severe power loss problem when applying to X-Chromosome SNPs due to misspecified XCI status.

Next, we investigate the maximum power loss due to incorrectly assuming gene-sex interaction and dominant effects. For $k$ variants, it may not be capped by $18.8\%$, and ideally we would like to derive the maximum theoretical power loss specifically for new random effects test \( S_\text{new} \) and SKATO, as well as Hotelling's \(T^2\) and Fisher's method for compassion. This is generally a hard problem, because most of the theoretical distributions of these test statistics under alternative hypothesis are not clear. The special case is Hotelling's \(T^2\), which is known to follow a non-central chi-squared distribution under $H_1$. We show in Theorem 3 that its maximum power loss due to XCI misspecification and misspecifying the full model can go up to 1 as $k \to \infty$. As discussed in Section 2, other quadratic tests such as SKAT and Fisher's method are expected to have similar power performance.

To compute the power loss for Hotelling's \(T^2\) test, we first derive its distribution under the alternative hypothesis by Lemma 2:

\begin{lemma}
    Let $W_A$ and $W_F$ be Hotelling's \(T^2\) Statistics following $\chi^2(k)$ and $\chi^2(3k)$ distribution respectively under $H_0$. Then under $H_1$, they will follow non-central chi-squared distribution: $W_A \sim \chi^2(k, {ncp}_A)$ and $W_F \sim \chi^2(3k, {ncp}_F)$, and the non-centrality parameters ${ncp}_A$ and ${ncp}_F$ has the order $O(k)$.
\end{lemma}
    
When comparing the additive model with the full model, the non-centrality parameter from the full model can not be less than the additive model. Furthermore, the power loss is maximized when $\boldsymbol{\beta_D}=\boldsymbol{\beta_{GS}}=0$, which implies ${ncp}_A={ncp}_F=ck$. Then Theorem 3 shows that power loss due to misspecification of the full model also increases to 1 as $k \to \infty$:

\begin{theorem}
Let ${ncp}_A={ncp}_F=ck$, for any $b \in (\sqrt{2c+\sqrt{6}}-\sqrt{6},\sqrt{2c+\sqrt{2}}-\sqrt{2})$, if $z_\alpha=b\sqrt{k}$ then $P(\chi^2_{3k,\alpha},3k,ck)-P(\chi^2_{k,\alpha},k,ck) \to 1$ as $k \to \infty$.
\end{theorem}
Theorem 3 suggests that as $k$ increases, the maximum power loss is reached at smaller type I error level $\alpha$. The proof of Lemma 1, Theorem 2, Lemma 2 and Theorem 3 are provided in Supplementary Appendix B.

The maximum power loss of \(S_{\text{new}}\) and SKATO due to misspecification of the full model are evaluated through simulation studies. In the simulations, we generate continuous traits using the following formula: 
\[
    y = 0.5S + \bm{G_A \beta_A}+ \epsilon.
\]
\(\bm{G_A} = (G_{A,1}, \dots, G_{A,k})\) is the genotypic matrix for \(k\) variants and 2000 individuals, including 1000 females and 1000 males. All variants are assumed to be inactivated for females. MAFs of each variant are simulated from $Unif(0.1,0.5)$ separately for females and males. For common variants, weights are generally not required, so we simulate \( \beta_{A,i} \sim N(\mu_A,\tau_A^2),\) for each variant under the generative model. The random error term is denoted as \( \epsilon \sim N(0, 1) \). Empirical power is estimated by replicating $1\times10^5$ times and calculating the ratio of p-values smaller than a critical value \( \alpha \).

\begin{figure}[htbp]
    \centering
    \includegraphics[width=0.9\linewidth]{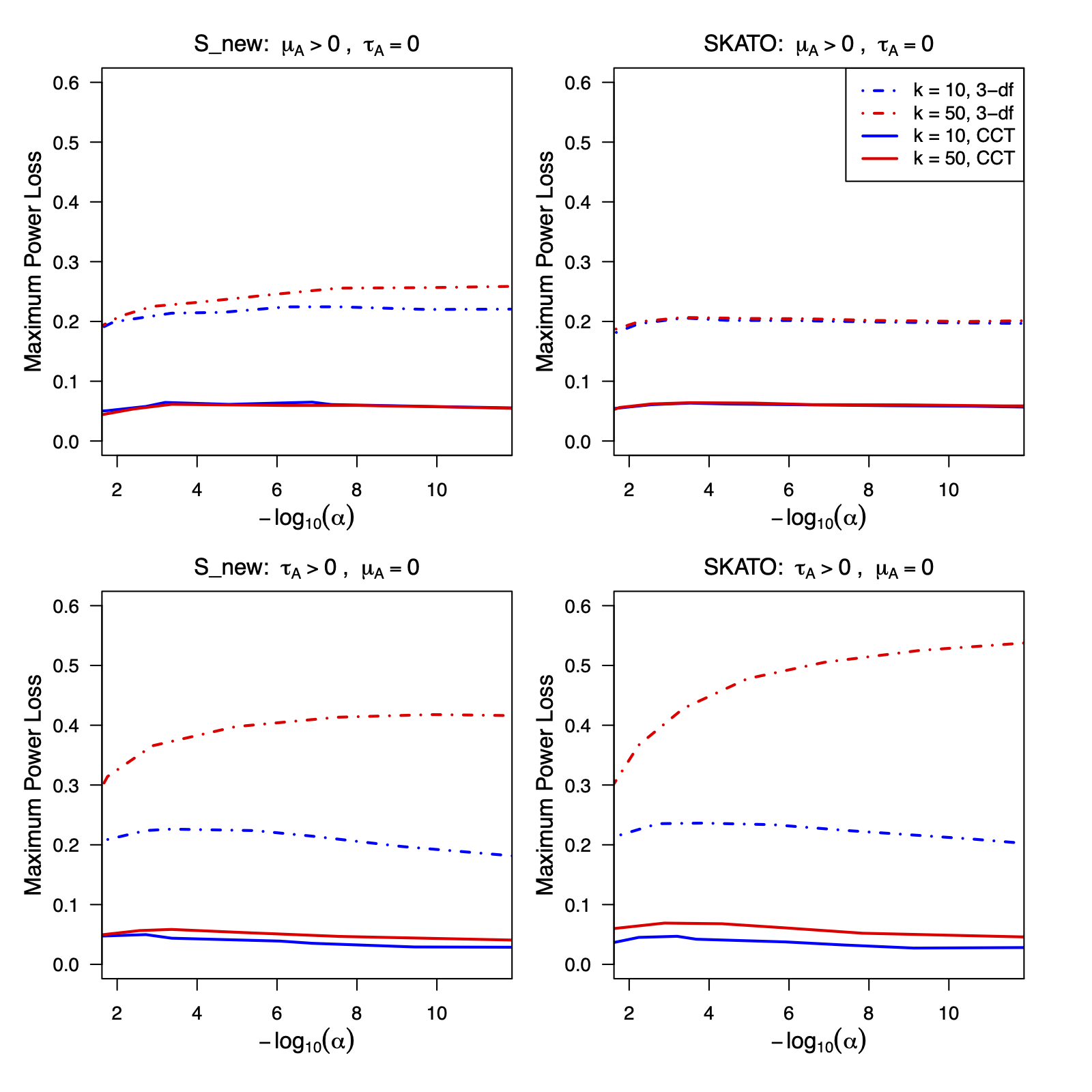}
    \caption{Maximum power loss at various type I error level $-\log_{10} \alpha$ from 1 to 12, compared to the test under true additive 1-df XCI model when gene-sex interaction and dominant effects are not present. First row: $k$ variants have fixed effects $\mu_A>0, \tau_A=0$. Second row: $k$ variants have random effects $\mu_A=0, \tau_A>0$. Left panels: $S_\text{new}$. Right panels: SKATO. Blue curve: $k=10$. Red curve: $k=50$. Dash curve: power loss from 3-df full model. Solid curve: power loss from CCT.}
    \label{fig:PowerLoss_Curves_Hybrid}
\end{figure}

The maximum power losses are estimated comparing the true additive model under XCI status for all $k$ variants, with the full model after transformation. We denote the comparison as 3-vs 1-df because the full model is based on the 3 df model for each variant. It needs to be noted that XCI status does not affect the power under the full model. The power loss depends on both type I error level $\alpha$ and the effect sizes. We consider most realistic $\alpha$ such at $1<-\log_{10}\alpha<12$. For effect sizes, we consider both situation of $\mu_A>0$ and $\tau_A>0$. Given a fixed type I error level $\alpha$, we search different $\mu_A$ and $\tau_A$ to find which parameter values lead to the maximum power loss. We then record the maximum power loss curve as a function of $\alpha$. To show if the maximum power loss increases with $k$, we compare $k=10$ with $k=50$. Results of maximum power loss of \(S_{\text{new}}\) and SKATO as a function of $\alpha$ are summarized in Figure \ref{fig:PowerLoss_Curves_Hybrid}. The full results for \(S_{\text{new}}\), SKATO, Hotelling's \(T^2\) and Fisher's method, showing the relationship between maximum power loss to $\mu_A$ or $\tau_A$ and the corresponding value of $-\text{log}_{10}(\alpha)$ at which the maximum occurs for each test are presented in Supplementary Appendix C.

When \(\mu_A > 0, \tau_A = 0\), simulation suggests that power loss of SKATO does not increase with the number of variants \(k\). This is because SKATO is simply the Burden test in this setting, leading to a reduction in degrees of freedom—from $k$ to 1 for the additive model, and from $3k$ to 3 for the full model. As a result, the power comparison becomes analogous to comparing a 3-df chi-squared statistic with a 1-df chi-squared statistic. Consequently, the maximum power loss does not accumulate as the number of variants \(k\) increases, and is close to the theoretical bound of 0.188. However, other tests, \(S_\text{new}\), Hotelling`s and Fisher`s method exhibit varying degrees of power loss accumulation: the accumulation is slight for \(S_\text{new}\) whereas it is substantial for the quadratic test. 

When \(\tau_A > 0, \mu_A=0\), indicating the presence of random effects, we observe significant power loss accumulation across all methods, where the maximum power loss could reach 0.6 when $k=50$. It suggests that direct application of the multilocus X-chromosome tests to the full model could be underpower if gene-sex interaction and dominant effects are not present.


\subsection{Controlling Maximum Power Loss by Cauchy Combination Test}
As we have identified the power loss accumulation problem, it is not recommended to directly apply any of the multilocus tests to either additive or full model. Because we do not know the true model and XCI status in practice, our solution is to simultaneously consider the additive model with XCI and no XCI codings as well as the full model. Clearly the test statistics from these 3 models are not independent. We propose using Cauchy Combination Test (CCT)\cite{Liu2020}, which can combine \(p\)-values derived from arbitrarily dependent tests, to take a weighted average of the additive XCI, additive no-XCI, and the full model. Specifically, we choose the weight of (0.25, 0.25, 0.5) for each test, and the CCT statistic is defined as follows:
\[
\text{CCT} = 0.25 \cdot \tan\left\{(0.5 - p_\text{xci})\pi\right\} 
           + 0.25 \cdot \tan\left\{(0.5 - p_\text{noxci})\pi\right\} 
           + 0.5 \cdot \tan\left\{(0.5 - p_F)\pi\right\},
\]
Under the null hypothesis, the tail probability of CCT converges to standard Cauchy distribution.


We apply CCT to combine three \(p\)-values derived from \(S_\text{new}\) and SKATO respectively. Simulation results in Figure \ref{fig:PowerLoss_Curves_Hybrid} and Supplementary Appendix C show that CCT is effective in controlling power loss. Compared to the 1-df XCI model, which is the true model, the power loss of CCT remains small with the increasing number of genetic variants, and the maximum power loss is bounded below 0.1. Therefore, we recommend using CCT statistic for multilocus X-chromosome test, which is invariant to XCI status and solves power loss accumulation problem with misspecified gene-sex interaction and dominant effects.


\section{Power Improvements by Our Framework}\label{Power Gain}

We have shown in Section 4.2 that the power loss of CCT compared to additive test is relatively small and bounded with increased $k$ for both \( S_{\text{new}} \) and \(\text{SKATO}\) when the additive model is correct. In this section, we show that when additive model is not correct, i.e., with the presence of dominant and gene-sex interaction effects, our proposed tests have significant power improvements compared to the additive tests. As we discussed in Section 3.4, \( S_{\text{new}} \) imposes no weights, but SKATO imposes higher weights to rare variants after our transformation framework. For this reason, we recommend using \( S_{\text{new}} \) for common variants analysis where it is reasonable to assume equal weight for each variant, and SKATO for rare variants analysis where equal importance of each variant is achieved after the standardization of genotype codings.



\subsection{Common Variants}
Using \( S_{\text{new}} \) for multilocus test, we compare the powers for 1-df additive model and our proposed CCT model which combines additive XCI, additive no XCI the full model by intensive simulation studies. We separately simulate continuous traits of female and male using the following models:
\begin{align}
    y_f = \bm{G_f^\dagger\beta_f^\dagger} + \bm{G_d^\dagger\beta_d^\dagger} + \epsilon\notag, \quad 
    y_m = \bm{G_m^\dagger\beta_m^\dagger} + \epsilon \notag
\end{align}
The sample sizes are 1,000 each for female and male. The random genetic effects are generated from
\[
\beta_{f,i}^\dagger \sim {N}(\mu_f, \tau_f^2), \quad \beta_{d,i}^\dagger \sim {N}(\mu_d, \tau_d^2), \quad \beta_{m,i}^\dagger \sim {N}(\mu_m, \tau_m^2).
\]
For each variant, MAFs are sampled from \(Unif \sim (0,0.5)\) separately for female and male to mimic the common variants situation where female and male MAFs are not necessarily equal. We generate genotypes $(rr,RR,RR,r,R)$ using the sampled female and male MAFs assuming HWE, and then 
apply the transformation framework in Section 3 to compute $\boldsymbol{G_f^\dagger}$, $\boldsymbol{G_d^\dagger}$ and $\boldsymbol{G_m^\dagger}$ for all variants. These transformed genotype codings are invariant to XCI status, which implies specifying XCI status is not required in the generative model. Regarding the testing models, we have explained in Section 4.2 that change of XCI status does not affect the CCT power. However, computing the power under additive model does require specified XCI status. Without loss of generality, we assume random half of the $k$ variants are inactivated and the other half are not inactivated, and the 1-df additive test power is subsequently computed with the genotype codings corresponding to the XCI status of each variant.

\begin{figure}[htbp]
    \centering
    \includegraphics[width=\linewidth]{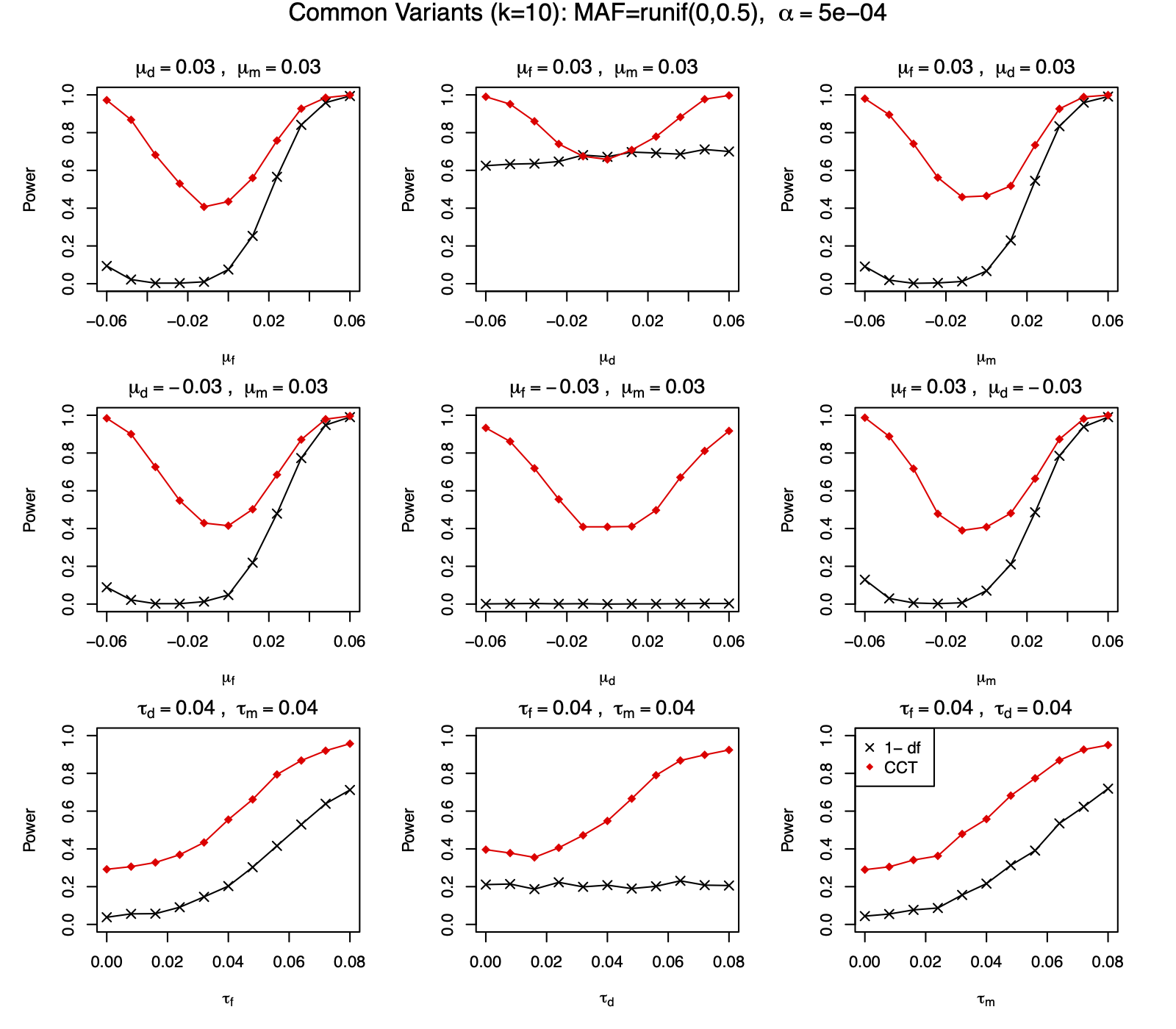}
    \caption{$S_\text{new}$ test power for common variants. First row: two of $\mu_f$, $\mu_d$ and $\mu_m$ are held fixed with same direction while the third is varied, and all \(\tau\) values are set to 0. Second row: two of $\mu_f$, $\mu_d$ and $\mu_m$ are held fixed with opposite directions while the third is varied, and all \(\tau\) values are set to 0. Third row: two of $\tau_f$, $\tau_d$ and $\tau_m$ are held fixed while the third is varied, and all \(\mu\) values are set to 0. Red curve: CCT based multilocus \( S_{\text{new}} \) test power. Black curve: 1-df additive multilocus \( S_{\text{new}} \) test power.
}
    \label{fig:Common Vairants Power Gain}
\end{figure}

We show the power comparison results with various $\mu$ and $\tau$ values. The first two rows in Figure \ref{fig:Common Vairants Power Gain} corresponds to $\mu \neq 0, \tau=0$, and the last row corresponds to $\mu=0, \tau>0$ for all $\mu$'s and $\tau$'s. Power estimations are based on 1,000 replications. The number of common variants are chosen to be \(k=10\). Results when $k=50$ are similar and presented in Supplementary Appendix C.

Figure \ref{fig:Common Vairants Power Gain} shows an overall power improvement using the CCT method under almost all alternative hypotheses. In particular, when $\mu>0, \tau=0$, and the directions of $\mu_f$ and $\mu_m$ are opposite, CCT exhibits a substantial power gain compared to the 1-df additive model. In this scenario, the opposing effects give rise to a gene-sex interaction, which the 1-df model is largely unable to detect. Moreover, when varying \(\mu_d\), the power of the 1-df model remains nearly unchanged, while CCT's power increases with the magnitude of \(\mu_d\). 


When $\mu=0, \tau>0$, CCT outperforms the 1-df model across all three scenarios. Increasing \(\tau_f\) and \(\tau_m\) leads to higher power for both methods. However, while the 1-df model is relatively insensitive to changes in \(\tau_d\), CCT exhibits notable power gains as \(\tau_d\) increases. 

Compared to the maximum power loss shown in Figure \ref{fig:PowerLoss_Curves_Hybrid} which is less than 0.1, the illustrated power improvement using CCT is frequently greater than 0.4 and could sometimes be close to 1 across all alternative hypotheses. Therefore for common variants, we recommend using CCT based \( S_{\text{new}} \) test instead of the classic additive \( S_{\text{new}} \) test for the multilocus testing problem in practice.

\subsection{Rare Variants}

Rare variants may play a significant role in the etiology of complex traits and help explain the missing heritability not accounted for by common variants \cite{Lee2012}. These rare variants, often characterized by low minor allele frequencies (MAF), tend to have stronger associations with phenotypes. Since SKATO incorporates standardized genotype codings to prioritize lower MAF variants, we recommend its use in rare variants association testing.

As we explained in Section 3.4.2 that $\boldsymbol{G_f}$ and $\boldsymbol{G_d}$ are usually not distinguishable for rare variants, we compare the power of the 1-df additive model with that of our proposed 2-df CCT method. We simulate continuous traits separately for females and males using the following models:
\[
y_f = \bm{G_f\beta_f}+ \epsilon, \quad 
y_m = \bm{G_m\beta_m} + \epsilon
\]
Unlike common variants, MAFs are sampled from a \(\text{Beta}(1, 40)\) distribution to mimic typical rare variant scenarios, separately for female and male. Then the true genetic effects are generated from \(|\beta_{f,i}| = c_f |\log_{10}(\text{MAF}_i)|\) and \(|\beta_{m,i}| = c_m |\log_{10}(\text{MAF}_i)|\) for each variant, which gives higher weights to rare variants. We fix \(c_m\), thereby fixing the magnitude of \(\bm{\beta_m}\), and vary \(c_f\) to adjust the magnitude of \(
\bm{\beta_f}\). For the directions of $\beta_{f,i}$ and $\beta_{m,i}$, we consider $\boldsymbol{\beta_m}$ to be all positive, randomly positive or negative and 0. The power patterns observed when $\boldsymbol{\beta_m}$ are all negative are symmetric to all positive $\boldsymbol{\beta_m}$ and are therefore omitted. Then the direction of $\boldsymbol{\beta_f}$ is set to be all same to $\boldsymbol{\beta_m}$, all opposite to $\boldsymbol{\beta_m}$ and random to $\boldsymbol{\beta_m}$. For the genotypes, because weights have been imposed with the genetic effects, we use the unstandardized $\boldsymbol{G_f}$ and $\boldsymbol{G_m}$ instead of $\boldsymbol{G_f^\dagger}$ and $\boldsymbol{G_m^\dagger}$ in the generative model. Because $\boldsymbol{G_f}$ is not invariant to XCI status, similar to common variants, we assume random half of the variants are inactivated. 

Regarding the testing models, the 2-df test is computed from standardized genotype codings $\boldsymbol{G_f^\dagger}$ and $\boldsymbol{G_m^\dagger}$, so no further weights are required to compute the SKATO statistic. However, the classic SKATO test based on 1-df additive model uses the original codings, so we still need to impose a weight for rare variants. We choose the standard weight Beta$(MAF_i,1,25)$ suggested by Wu \cite{Wu2011}. To guarantee best power performance of the classic SKATO test, the testing model is based on correctly specified XCI status of each variant.

\begin{figure}[htbp]
    \centering
    \includegraphics[width=\linewidth]{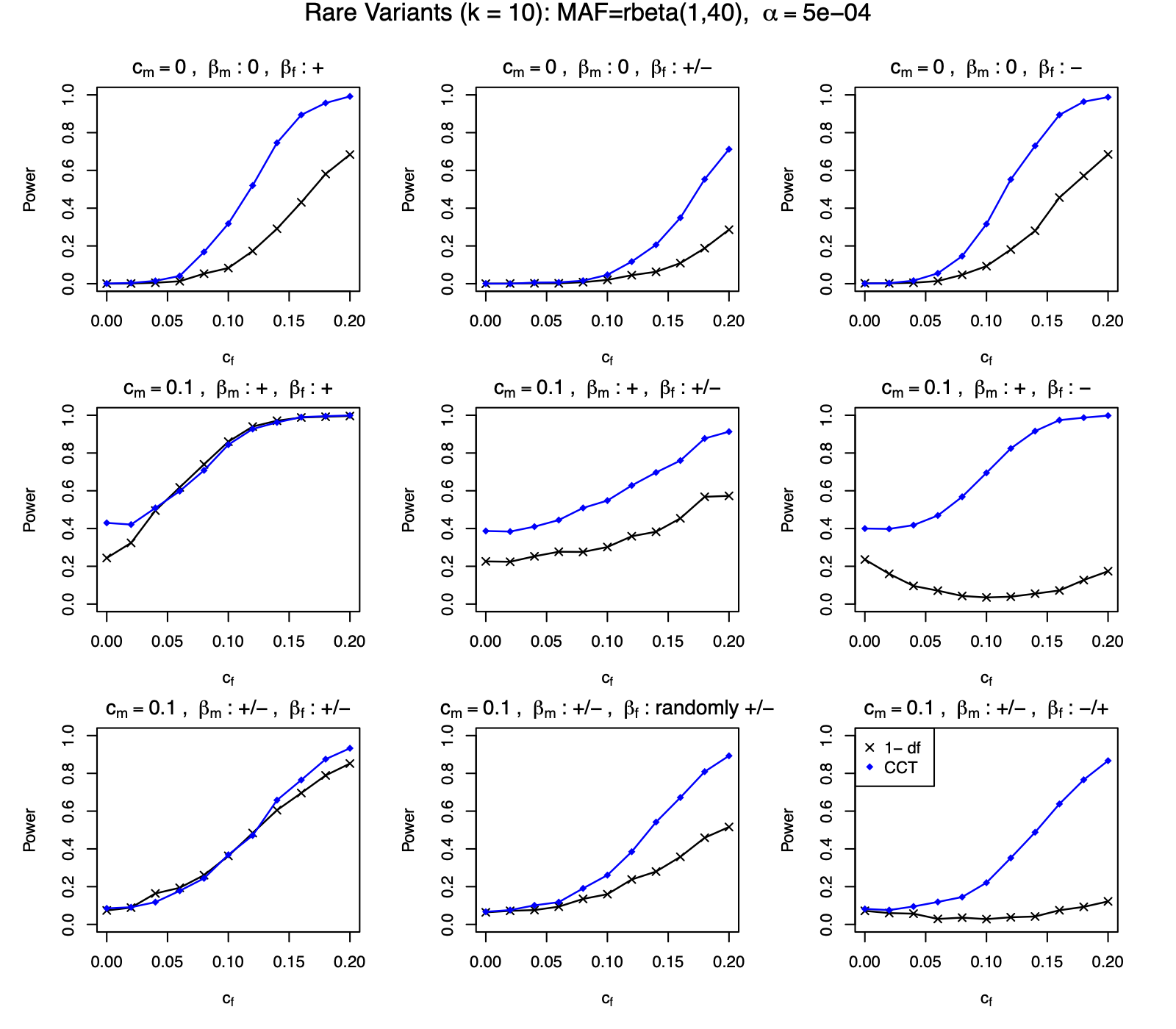}
    \caption{
SKATO test power for rare variants. First row: \(\bm{\beta_m = 0}\), and \(\bm{\beta_f}\) takes on values with positive, 50\% positive / 50\% negative, and negative signs. Second row: \(\bm{\beta_m > 0}\), and the directions of \(\bm{\beta_f}\) and \(\bm{\beta_m}\) vary as follows: 100\% concordant, 50\% concordant / 50\% discordant, and 100\% discordant. Third row: the half part of \(\bm{\beta_m}\) is randomly assigned positive values and the other half negative. Correspondingly, \(\bm{\beta_f}\) direction is set as 100\% concordant, randomly half positive, or 100\% discordant. Blue curve: CCT based multilocus SKATO test power. Black curve: 1-df additive multilocus SKATO test power.
}
    \label{fig:Rare Vairants Power Gain}
\end{figure}

In Figure~\ref{fig:Rare Vairants Power Gain}, we show power comparisons between 1-df additive model and 2-df CCT method, under varying configurations of the directions of $\boldsymbol{\beta_f}$ and $\boldsymbol{\beta_m}$. Similar to common variants, the CCT method also shows a significant improvement of power in general, and we observe power improvement under almost all configurations. Specifically, when \(\bm{\beta_m = 0}\), increasing the magnitude of \(\bm{\beta_f}\) (indicating the presence of a gene-sex interaction effect) leads to greater power for CCT compared to the 1-df additive model. When \(\bm{\beta_m > 0}\), the power gap between the two methods widens as the level of discordance between \(\bm{\beta_f}\) and \(\bm{\beta_m}\) increases. In more complex scenarios where \(\bm{\beta_m}\) values are randomly assigned to be positive or negative, the power of the 1-df additive model becomes sensitive to the direction of \(\bm{\beta_f}\), while CCT remains powerful.
Based on these results, we recommend using CCT based SKATO test for multilocus rare variant analysis.

\section{Application to Real Data}
\subsection{Common Variants Application: Cystic Fibrosis Lung Disease}
We first apply our proposed method to real data from common variants, and revisit the Cystic Fibrosis (CF) Lung Disease study by Soave et al.~\cite{Soave2015}. In medical diagnostics, meconium ileus—an intestinal obstruction present at birth—occurs in approximately 15\% of CF cases. The goal is to identify loci on the X chromosome that are associated with meconium ileus. The dataset includes \(n = 3199\) individuals, consisting of \(n_f = 1477\) females and \(n_m = 1722\) males, and contains \(k = 14{,}280\) SNPs on the X chromosome. Among these, 13,168 are common variants with minor allele frequency (MAF) greater than 3\%. In the study by Soave et al.~\cite{Soave2015}, single variant analysis was performed under additive genetic effect assumption, and three X-chromosomal SNPs in the gene \textit{SLC6A14}, rs12839137 (MAF = 0.24), rs5905283 (MAF = 0.49), and rs3788766 (MAF = 0.40), were identified as significantly associated with lung disease in cystic fibrosis. 

For multilocus variant analysis, we implement our proposed CCT based new random effects test \(S_\text{new}\), which jointly tests the additive, dominant and gene-sex interaction effects. A moving-window approach can be employed by analyzing one genomic window at a time. We use a window size of 8 SNPs and a step size of 1 SNP. The significance threshold, adjusted for multiple testing, is \(\alpha = 3.37 \times 10^{-6}\).

In table \ref{tab:Real Rata Common Variant}, three moving windows located on different genes are identified as significantly associated with meconium ileus using the CCT based \(S_\text{new}\) test. The first gene \textit{SLC6A14} includes the SNPs: {rs12839137}, {rs5905283} and {rs3788766}, which confirms results from the previous study of Soave et al.~\cite{Soave2015}. Additionally, our proposed method identifies moving windows located on two other X chromomse genes: \textit{CLCN4} and \textit{PCDH19} which were not previously reported to be associated with meconium ileus. In contrast, the 1 df additive multilocus tests are not able to identify these genes. It suggests that our proposed method has stronger ability to find novel genes, which could be potentially ignored by classic single variant X chromosome analysis.

\begin{table}[htbp]
\centering

\setlength{\tabcolsep}{3pt} 
\begin{tabular}{ccccc}
\toprule
\textbf{Gene} & \textbf{1 d.f. Test(xci)} & \textbf{1 d.f. Test(noxci)} & \textbf{3 d.f. Test} & \textbf{CCT} \\
\midrule
SLC6A14 & $9.27\times10^{-29}$ & $8.89\times10^{-25}$ &$1.35\times10^{-27}$ & $3.26\times10^{-28}$ \\
CLCN4 & $1.92\times10^{-3}$ & $1.56\times10^{-1}$ & $2.24\times10^{-8}$ & $4.48\times10^{-8}$ \\
PCDH19 & $4.88\times10^{-4}$ & $7.23\times10^{-8}$ & $6.59\times10^{-12}$ & $1.32\times10^{-11}$ \\
\bottomrule
\end{tabular}
\caption{CF patients X chromosome association analysis with meconium ileus. Multilocus test with locus size 8 using \(S_\text{new}\) and significant level ~\(\alpha = 3.37 \times 10^{-6}\).}
\label{tab:Real Rata Common Variant}
\end{table}

\subsection{Rare Variants Application: FT4 and TSH}
 
To examine our improved SKATO and its CCT method in rare variants application, we use the dataset from the genome-wide association analysis (GWAS) meta-analysis study of thyroid dysfunction \cite{Teumer2018}. In daily clinical practice, thyroid function is assessed by measuring circulating free thyroxine (FT4) and thyrotropin (TSH) levels. In the original study, they performed single variant tests on 8 million common genetic variants across the whole genome in 72,167 individuals to identify their associations to FT4 and TSH. In contrast, we particularly focus on rare variants from X-chromosome. The GWAS on FT4 and TSH, including 
3846 and 5787 rare variants respectively, were calculated using a linear model after inverse-normal transformation of the thyroid hormone measurements (outcome), providing summary statistics.
Our focus is on assessing the association between FT4 or TSH levels and rare variants, with the null hypothesis
\(
H_0: \beta_f^\dagger = \beta_m^\dagger = 0.
\)
We first use the sex stratified statistic with following expression:
\[
Q_\rho = (1-\rho){\sum_{i=1}^{k}}w_i^2{\hat{\beta}_{i}^\dagger{^2}} + \rho({\sum_{i=1}^{k}}w_i{\hat{\beta}_{i}^\dagger})^2,
\]
where \( w_i \) is a function of the minor allele frequency (MAF) for the \( i \)th variant, and \( \rho \) is estimated separately for each sex using the formula in Equation~(\ref{estimated_rho}). The eigenvalues are obtained by decomposing the matrix
\(
\bm{W} \left[ (1 - \rho)\bm{I} + \rho \bm{1}\bm{1}' \right] \bm{W},
\)
where \( \bm{I} \) is the \( k \times k \) identity matrix, and \( \bm{1} \) is a length-\( k \) vector with all elements equal to 1. Similar to the full data analysis, we aggregate \( Q_\rho \) and use the sets of eigenvalues obtained separately from the female and male strata to numerically compute the corresponding \( p \)-values with Davies method~\cite{Davies1980}.

We apply a moving-window approach that combines consecutive locus, and examine each window where the proportion of rare variants (RVs) is no less than 20\%. For multiple testing correction, we conservatively assume that the number of tests equals the number of rare variants. 
We find three new loci in table \ref{tab:Real Data Rare Variant FT4}, which are associated with FT4 but not found in the original study. The MAOA and MAOB are known to encode enzymes that break down neurotransmitters like serotonin, dopamine, and norepinephrine \cite{Whibley2010}. The CCT method identifies both genes, While the classic 1 df SKATO test fails to detect their significance. JPX and FTX are long noncoding RNA (lncRNA) genes that regulate X chromosome inactivation in mammals \cite{Rosspopoff2023}. CCT p-value coincides with 1 df SKATO but is more significant. The new findings suggest that our proposed CCT is more powerful to identify multilocus rare variants compared to classic SKATO method. 
\begin{table}[htbp]
\centering
\begin{tabular}{cccccc}
\toprule
\textbf{Gene} & \textbf{Window Size} &\textbf{No. of RVs}& \textbf{1 d.f. Test} & \textbf{2 d.f. Test}&  \textbf{CCT}\\
\midrule
MAOA & 59 & 14 & $8.42 \times 10^{-1}$ & $1.10 \times 10^{-7}$ & $2.20 \times 10^{-7}$ \\
MAOB & 41 & 28 &$9.24 \times 10^{-1}$ & $4.10 \times 10^{-6}$ & $8.21 \times 10^{-6}$\\
JPX/FTX & 17 & 6 & $1.52 \times 10^{-6}$ & $1.28 \times 10^{-7}$ & $2.37 \times 10^{-7}$ \\
\bottomrule
\end{tabular}
\caption{Novel X chromosome rare variants (MAF$<3\%$) loci associated with free thyroxine (FT4) as a measurement of thyroid function:
Multilocus variants combination with $\text{SKATO}$, \(\alpha = 1.30 \times 10^{-5}\).}
\label{tab:Real Data Rare Variant FT4}
\end{table}



\section{Discussion} 
In this paper, we have proposed a universal framework for X-chromosomal multilocus variant association studies. Our contributions are fourfold. First, the proposed framework addresses two key challenges in X-chromosomal multilocus association analysis: XCI uncertainty in genetic coding schemes and selection between additive vs. full model. Second, we have addressed the issue of power loss accumulation caused by genetic coding or model misspecification, making the power loss bounded with increasing number of combined variants. Third, we design separate procedures for common and rare variants. Simulation results show that our proposed methods, the CCT method, are more powerful than the conventional 1-df multilocus test. Finally, we apply our framework to the Cystic Fibrosis (CF) dataset \cite{Soave2015} and the Thyroid Dysfunction dataset \cite{Teumer2018}. In addition to replicating previously reported findings in CF, our method also identifies a novel locus on the X chromosome in both datasets.

There has been a long debate choosing between the 1 df additive model and 3 df full model when performing single variant X chromosome analysis. Although the benefits of choosing 3 df model is clear \cite{Chen2021}, power loss of the 3 df model is the major concern when additive model is correct. In this paper, we have shown the CCT method combining the 1 df and 3 df model has superior power performance to each of the single model under the multilocus testing framework, which gives a clear answer to end the debate.

It needs to be noted that our testing framework can also be implemented when the original data is difficult to access and only summary statistics are available. In new random effects test, we only need to know the estimated effects and their corresponding variance estimates of each single variant; in SKATO, only score functions are required to compute Burden and SKAT test statistics. These information are more likely to be available in meta-analysis compared to the original data, e.g., in the Thyroid Dysfunction dataset \cite{Teumer2018}.



Linkage disequilibrium (LD) usually needs to be taken into account in multilocus association analysis. Shao \cite{Shao2022} reviewed how existing autosome multilocus analysis methods could deal with LD. The primary focus of this paper is to propose a novel X-chromosome multilocus analysis method. When LD exists, it is straightforward to include an LD matrix when computing the SKATO statistic, and our proposed framework for analyzing X-chromosome variants remains the same. Alternatively, LD-free p-value combination methods, such as harmonic mean p-value \cite{Wilson2019} and CCT \cite{Liu2020} may be used to combine single variant test results, which are robust against correlation between SNPs. Depending on different correlation structures, these p-value combination methods could be underpower or not. Further work is required to give a thorough investigation of their power performance in the context of multiple X chromosome SNPs with LD.


\bibliographystyle{unsrt}  
\bibliography{references}   

@article{Keur2022,
	author = {Keur, Nick and Rica{\~n}o-Ponce, Isis and Kumar, Vinod and Matzaraki, Vasiliki},
	doi = {10.1093/bib/bbac287},
	eprint = {https://academic.oup.com/bib/article-pdf/23/5/bbac287/48174005/bbac287.pdf},
	issn = {1477-4054},
	journal = {Briefings in Bioinformatics},
	month = {07},
	number = {5},
	pages = {bbac287},
	title = {A systematic review of analytical methods used in genetic association analysis of the X-chromosome},
	url = {https://doi.org/10.1093/bib/bbac287},
	volume = {23},
	year = {2022}}

@article{Sun2023,
	author = {Sun, Lei and Wang, Zhong and Lu, Tianyuan and Manolio, Teri A. and Paterson, Andrew D.},
	date = {2023/06/01},
	doi = {10.1016/j.ajhg.2023.04.009},
	isbn = {0002-9297},
	journal = {The American Journal of Human Genetics},
	journal1 = {The American Journal of Human Genetics},
	number = {6},
	pages = {903--912},
	publisher = {Elsevier},
	title = {eXclusionarY: 10 years later, where are the sex chromosomes in GWASs?},
	type = {doi: 10.1016/j.ajhg.2023.04.009},
	url = {https://doi.org/10.1016/j.ajhg.2023.04.009},
	volume = {110},
	year = {2023},
	year1 = {2023}}

@article{Wise2013,
  author    = {Wise, Andrea L. and Gyi, Laramie and Manolio, Teri A.},
  title     = {eXclusion: Toward Integrating the X Chromosome in Genome-wide Association Analyses},
  journal   = {American Journal of Human Genetics},
  year      = {2013},
  volume    = {92},
  number    = {5},
  pages     = {643--647},
  doi       = {10.1016/j.ajhg.2013.03.017},
  pmid      = {23643377},
  pmcid     = {PMC3644627}
}

@article{Khramtsova2023,
  author    = {Khramtsova, Ekaterina A. and Wilson, Melissa A. and Martin, Joanna and others},
  title     = {Quality control and analytic best practices for testing genetic models of sex differences in large populations},
  journal   = {Cell},
  year      = {2023},
  volume    = {186},
  number    = {10},
  pages     = {2044--2061},
  doi       = {10.1016/j.cell.2023.04.014}
}

@article{Chen2024,
  author    = {Chen, D. Z. and Roshandel, D. and Wang, Z. and Sun, L. and Paterson, A. D.},
  title     = {Comprehensive whole-genome analyses of the UK Biobank reveal significant sex differences in both genotype missingness and allele frequency on the X chromosome},
  journal   = {Human Molecular Genetics},
  year      = {2024},
  volume    = {33},
  number    = {6},
  pages     = {543--551},
  doi       = {10.1093/hmg/ddad201}
}

@article{Clayton2008,
	author = {Clayton, David},
	doi = {10.1093/biostatistics/kxn007},
	eprint = {https://academic.oup.com/biostatistics/article-pdf/9/4/593/17736312/kxn007.pdf},
	issn = {1465-4644},
	journal = {Biostatistics},
	month = {04},
	number = {4},
	pages = {593-600},
	title = {Testing for association on the X chromosome},
	url = {https://doi.org/10.1093/biostatistics/kxn007},
	volume = {9},
	year = {2008}}

@article{Gao2015,
	author = {Gao, Feng and Chang, Diana and Biddanda, Arjun and Ma, Li and Guo, Yingjie and Zhou, Zilu and Keinan, Alon},
	doi = {10.1093/jhered/esv059},
	eprint = {https://academic.oup.com/jhered/article-pdf/106/5/666/10007277/esv059.pdf},
	issn = {0022-1503},
	journal = {Journal of Heredity},
	month = {08},
	number = {5},
	pages = {666-671},
	title = {XWAS: A Software Toolset for Genetic Data Analysis and Association Studies of the X Chromosome},
	url = {https://doi.org/10.1093/jhered/esv059},
	volume = {106},
	year = {2015}}

@book{Fisher1925,
  title={Statistical Methods for Research Workers},
  author={Fisher, R.A.},
  year={1925},
  publisher={Oliver and Boyd},
  address={Edinburgh},
  isbn={0-05-002170-2}
}

@book{Stouffer1949,
  title={The American Soldier: Adjustment during Army Life},
  author={Stouffer, S.A. and Suchman, E.A. and DeVinney, L.C. and Star, S.A. and Williams, R.M., Jr.},
  year={1949},
  publisher={Princeton University Press},
  address={Princeton, NJ, USA},
  volume={1},
  series={Studies in Social Psychology in World War II}
}

@article{Wang2014,
  author = {Wang, J. and Yu, R. and Shete, S.},
  title = {X-chromosome genetic association test accounting for X-inactivation, skewed X-inactivation, and escape from X-inactivation},
  journal = {Genetic Epidemiology},
  year = {2014},
  volume = {38},
  number = {6},
  pages = {483--493},
  doi = {10.1002/gepi.21814}
}

@article{Wang2017,
author = {Jian Wang and Rajesh Talluri and Sanjay Shete},
title ={Selection of X-chromosome Inactivation Model},
journal = {Cancer Informatics},
volume = {16},
pages = {1176935117747272},
year = {2017},
doi = {10.1177/1176935117747272},
}

@Article{Su2022,
AUTHOR = {Su, Youpeng and Hu, Jing and Yin, Ping and Jiang, Hongwei and Chen, Siyi and Dai, Mengyi and Chen, Ziwei and Wang, Peng},
TITLE = {XCMAX4: A Robust X Chromosomal Genetic Association Test Accounting for Covariates},
JOURNAL = {Genes},
VOLUME = {13},
YEAR = {2022},
NUMBER = {5},
ARTICLE-NUMBER = {847},
URL = {https://www.mdpi.com/2073-4425/13/5/847},
PubMedID = {35627231},
ISSN = {2073-4425},
DOI = {10.3390/genes13050847}
}

@article{Yang2022,
	author = {Yang, Zi-Ying and Liu, Wei and Yuan, Yu-Xin and Kong, Yi-Fan and Zhao, Pei-Zhen and Fung, Wing Kam and Zhou, Ji-Yuan},
	date = {2022/10/01},
	doi = {10.1038/s41437-022-00560-y},
	id = {Yang2022},
	isbn = {1365-2540},
	journal = {Heredity},
	number = {4},
	pages = {244--256},
	title = {Robust association tests for quantitative traits on the X chromosome},
	url = {https://doi.org/10.1038/s41437-022-00560-y},
	volume = {129},
	year = {2022}}

@article{Chen2018,
	author = {Chen, Bo and Craiu, Radu V and Sun, Lei},
	doi = {10.1093/biostatistics/kxy049},
	eprint = {https://academic.oup.com/biostatistics/article-pdf/21/2/319/32914751/kxy049.pdf},
	issn = {1465-4644},
	journal = {Biostatistics},
	month = {09},
	number = {2},
	pages = {319-335},
	title = {Bayesian model averaging for the X-chromosome inactivation dilemma in genetic association study},
	url = {https://doi.org/10.1093/biostatistics/kxy049},
	volume = {21},
	year = {2018}}

@article{Chen2021,
  author = {Chen, B. and Craiu, R. V. and Strug, L. J. and Sun, L.},
  title = {The X factor: A robust and powerful approach to X-chromosome-inclusive whole-genome association studies},
  journal = {Genetic Epidemiology},
  year = {2021},
  volume = {45},
  pages = {694--709},
  doi = {10.1002/gepi.22422}
}

@article{Shao2022,
  author    = {Shao, Zhiyu and Wang, Ting and Qiao, Junwei and others},
  title     = {A comprehensive comparison of multilocus association methods with summary statistics in genome-wide association studies},
  journal   = {BMC Bioinformatics},
  year      = {2022},
  volume    = {23},
  pages     = {359},
  doi       = {10.1186/s12859-022-04897-3}
}

@book{Tippett1931,
  title={Methods of Statistics},
  author={Tippett, L.H.C.},
  year={1931},
  publisher={Williams Norgate},
  address={London, UK}
}

@article{Wilson2019,
  title={The harmonic mean p-value for combining dependent tests},
  author={Wilson, D. J.},
  journal={Proceedings of the National Academy of Sciences of the United States of America},
  year={2019},
  volume={116},
  number={4},
  pages={1195--1200},
  doi={10.1073/pnas.1814092116},
  note={Published correction appears in Proc Natl Acad Sci U S A. 2019 Oct 22;116(43):21948. doi: 10.1073/pnas.1914128116.}
}

@article{Liu2020,
  title={Cauchy combination test: a powerful test with analytic p-value calculation under arbitrary dependency structures},
  author={Liu, Y. and Xie, J.},
  journal={Journal of the American Statistical Association},
  year={2020},
  volume={115},
  number={529},
  pages={393--402},
  doi={10.1080/01621459.2018.1554485}
}

@article{Willer2010,
  title={METAL: fast and efficient meta-analysis of genomewide association scans},
  author={Willer, C.J. and Li, Y. and Abecasis, G.R.},
  journal={Bioinformatics},
  year={2010},
  volume={26},
  number={17},
  pages={2190--2191},
  doi={10.1093/bioinformatics/btq340}
}

@article{Derkach2014,
  author = {Andriy Derkach and Jerry F. Lawless and Lei Sun},
  title = {Pooled Association Tests for Rare Genetic Variants: A Review and Some New Results},
  journal = {Statistical Science},
  volume = {29},
  number = {2},
  pages = {302--321},
  year = {2014},
  month = {May}
}

@article{Han2011,
	author = {Han, Buhm and Eskin, Eleazar},
	date = {2011/05/13},
	doi = {10.1016/j.ajhg.2011.04.014},
	isbn = {0002-9297},
	journal = {The American Journal of Human Genetics},
	journal1 = {The American Journal of Human Genetics},
	number = {5},
	pages = {586--598},
	publisher = {Elsevier},
	title = {Random-Effects Model Aimed at Discovering Associations in Meta-Analysis of Genome-wide Association Studies},
	type = {doi: 10.1016/j.ajhg.2011.04.014},
	url = {https://doi.org/10.1016/j.ajhg.2011.04.014},
	volume = {88},
	year = {2011},
	year1 = {2011}}

@article{Wu2011,
	author = {Wu, Michael C. and Lee, Seunggeun and Cai, Tianxi and Li, Yun and Boehnke, Michael and Lin, Xihong},
	date = {2011/07/15},
	doi = {10.1016/j.ajhg.2011.05.029},
	isbn = {0002-9297},
	journal = {The American Journal of Human Genetics},
	journal1 = {The American Journal of Human Genetics},
	number = {1},
	pages = {82--93},
	publisher = {Elsevier},
	title = {Rare-Variant Association Testing for Sequencing Data with the Sequence Kernel Association Test},
	type = {doi: 10.1016/j.ajhg.2011.05.029},
	url = {https://doi.org/10.1016/j.ajhg.2011.05.029},
	volume = {89},
	year = {2011},
	year1 = {2011}}

@article{Morgenthaler2007,
title = {A strategy to discover genes that carry multi-allelic or mono-allelic risk for common diseases: A cohort allelic sums test (CAST)},
journal = {Mutation Research/Fundamental and Molecular Mechanisms of Mutagenesis},
volume = {615},
number = {1},
pages = {28-56},
year = {2007},
issn = {0027-5107},
doi = {https://doi.org/10.1016/j.mrfmmm.2006.09.003},
url = {https://www.sciencedirect.com/science/article/pii/S0027510706002740},
author = {Stephan Morgenthaler and William G. Thilly},
}

@article{Madsen2009,
  title={A Groupwise Association Test for Rare Mutations Using a Weighted Sum Statistic},
  author={Madsen, B.E. and Browning, S.R.},
  journal={PLoS Genetics},
  volume={5},
  number={2},
  pages={e1000384},
  year={2009},
  doi={10.1371/journal.pgen.1000384}
}

@article{Pan2009,
  author = {Pan, W.},
  title = {Asymptotic tests of association with multiple SNPs in linkage disequilibrium},
  journal = {Genetic Epidemiology},
  year = {2009},
  volume = {33},
  number = {6},
  pages = {497-507},
  doi = {10.1002/gepi.20402},
  pmid = {19170135},
  pmcid = {PMC2732754}
}

@article{Neale2011,
  author = {Neale, B. M. and Rivas, M. A. and Voight, B. F. and Altshuler, D. and Devlin, B. and et al.},
  title = {Testing for an Unusual Distribution of Rare Variants},
  journal = {PLoS Genetics},
  year = {2011},
  volume = {7},
  number = {3},
  pages = {e1001322},
  doi = {10.1371/journal.pgen.1001322}
}

@article{Lee2012,
	author = {Lee, Seunggeun and Emond, Mary J. and Bamshad, Michael J. and Barnes, Kathleen C. and Rieder, Mark J. and Nickerson, Deborah A. and Christiani, David C. and Wurfel, Mark M. and Lin, Xihong},
	date = {2012/08/10},
	doi = {10.1016/j.ajhg.2012.06.007},
	isbn = {0002-9297},
	journal = {The American Journal of Human Genetics},
	number = {2},
	pages = {224--237},
	publisher = {Elsevier},
	title = {Optimal Unified Approach for Rare-Variant Association Testing with Application to Small-Sample Case-Control Whole-Exome Sequencing Studies},
	type = {doi: 10.1016/j.ajhg.2012.06.007},
	url = {https://doi.org/10.1016/j.ajhg.2012.06.007},
	volume = {91},
	year = {2012}}

@article{Ma2015,
author = {Ma, C. and Boehnke, M. and Lee, S.},
title = {Evaluating the Calibration and Power of Three Gene-Based Association Tests of Rare Variants for the X Chromosome},
journal = {Genetic Epidemiology},
volume = {39},
number = {6},
pages = {499-508},
year = {2015},
doi = {10.1002/gepi.21935}
}

@misc{Lee2014,
  author = {Lee, S.},
  year = {2014},
  title = {SKAT: SNP-set (Sequence) Kernel Association Test},
  note = {R package version 1.0.5},
  url = {http://CRAN.R-project.org/package=SKAT}
}

@article{Hardy1996,
  title     = {A likelihood approach to meta-analysis with random effects},
  author    = {Hardy, Rebecca J. and Thompson, Simon G.},
  journal   = {Statistics in Medicine},
  volume    = {15},
  number    = {6},
  pages     = {619--629},
  year      = {1996},
  doi       = {10.1002/(SICI)1097-0258(19960330)15:6<619::AID-SIM188>3.0.CO;2-A}
}

@article{Self1987,
  title={Asymptotic Properties of Maximum Likelihood Estimators and Likelihood Ratio Tests Under Nonstandard Conditions},
  author={Self, Steven G. and Liang, Kung-Yee},
  journal={Journal of the American Statistical Association},
  volume={82},
  number={398},
  pages={605--610},
  year={1987},
  publisher={Taylor \& Francis},
  doi={10.2307/2289471}
}

@article{Davies1980,
  author    = {Davies, R. B.},
  title     = {Algorithm AS 155: The Distribution of a Linear Combination of $\chi^2$ Random Variables},
  journal   = {Journal of the Royal Statistical Society. Series C (Applied Statistics)},
  volume    = {29},
  number    = {3},
  pages     = {323--333},
  year      = {1980},
  doi       = {10.2307/2346911},
  publisher = {Wiley for the Royal Statistical Society}
}

@article{Lee2012.1,
  author    = {Lee, S. and Wu, M. C. and Lin, X.},
  title     = {Optimal tests for rare variant effects in sequencing association studies},
  journal   = {Biostatistics},
  year      = {2012},
  volume    = {13},
  number    = {4},
  pages     = {762--775},
  doi       = {10.1093/biostatistics/kxs014}
}

@article{Soave2015,
  author    = {Soave, D. and Corvol, H. and Panjwani, N. and Gong, J. and Li, W. and Boëlle, P.Y. and Durie, P.R. and Paterson, A.D. and Rommens, J.M. and Strug, L.J. and Sun, L.},
  title     = {A Joint Location-Scale Test Improves Power to Detect Associated SNPs, Gene Sets, and Pathways},
  journal   = {American Journal of Human Genetics},
  year      = {2015},
  volume    = {97},
  number    = {1},
  pages     = {125--138},
  month     = {July},
  doi       = {10.1016/j.ajhg.2015.05.015},
  pmid      = {26140448},
  pmcid     = {PMC4572492}
}

@article{Teumer2018,
  author    = {Teumer, Alexander and Chaker, Layal and Groeneweg, Stefan and Li, Yang and Di Munno, Chiara and Barbieri, Caterina and others},
  title     = {Genome-wide analyses identify a role for SLC17A4 and AADAT in thyroid hormone regulation},
  journal   = {Nature Communications},
  year      = {2018},
  volume    = {9},
  pages     = {4455},
  doi       = {10.1038/s41467-018-06356-1}
}

@book{Muirhead2005,
  author    = {Muirhead, R.},
  title     = {Aspects of Multivariate Statistical Theory (2nd Edition).},
  publisher = {Wiley},
  year      = {2005}
}

@book{Fisher1934,
  author    = {Fisher, R. A.},
  title     = {Statistical Methods for Research Workers},
  publisher = {Oliver and Boyd},
  address   = {Edinburgh, UK},
  year      = {1934}
}

@article{Liu2009,
title = {A new chi-square approximation to the distribution of non-negative definite quadratic forms in non-central normal variables},
journal = {Computational Statistics \& Data Analysis},
volume = {53},
number = {4},
pages = {853-856},
year = {2009},
issn = {0167-9473},
doi = {https://doi.org/10.1016/j.csda.2008.11.025},
url = {https://www.sciencedirect.com/science/article/pii/S0167947308005653},
author = {Huan Liu and Yongqiang Tang and Hao Helen Zhang}
}

@book{Keith2000,
  title     = {Mathematical Statistics},
  author    = {Keith Knight},
  year      = {2000},
  publisher = {Chapman \& Hall/CRC}
}

@article{Whibley2010,
title = {Deletion of MAOA and MAOB in a male patient causes severe developmental delay, intermittent hypotonia and stereotypical hand movements},
journal = {European Journal of Human Genetics},
volume = {18},
pages = {1095-1099},
year = {2010},
doi = {https://doi.org/10.1038/ejhg.2010.41},
author = {Annabel Whibley and Jill Urquhart and Jonathan Dore and Lionel Willatt and Georgina Parkin and Lorraine Gaunt and Graeme Black and Dian Donnai and F Lucy Raymond}
}

@article{Rosspopoff2023,
title = {Species-specific regulation of XIST by the JPX/FTX orthologs},
journal = {Nucleic Acids Research},
volume = {51},
number = {5},
pages = {2177-2194},
year = {2023},
doi = {https://doi.org/10.1093/nar/gkad029},
author = {Olga Rosspopoff and Emmanuel Cazottes and Christophe Huret and Agnese Loda and Amanda J Collier and Miguel Casanova and Peter J Rugg-Gunn and Edith Heard and Jean-François Ouimette and Claire Rougeulle}
}

\end{document}


\maketitle
\thispagestyle{empty}  
\newpage

\setcounter{page}{1}
\section*{Appendix A: Proof of Theorem 1}
\subsection*{Meta-Analysis and Mega-Analysis}
We define 
\[Y_f = (Y_{f,1}, \dots, Y_{f,n_f})',\quad
Y_m = (Y_{m,1}, \dots, Y_{m,n_m})'
\]
Assume that all \(Y_{f,i}\) and \(Y_{m,i}\)  are independent, and they have the same dispersion parameter \(\phi\).

We may combine individual level data from all studies. 
We have the full GLM by combining all individual level data:
\[
\mathcal{M}: g(E(Y)) = \left(\begin{matrix}
    1_{n_f} & X_f & 0_{n_f} & 0_{n_f}\\
    1_{n_m} & X_m & 1_{n_m} & X_m \\
\end{matrix}\right)\left( \begin{matrix}
    \beta_0 \\\boldsymbol{\beta_1} \\\beta_S\\\boldsymbol{\beta_2}\\
\end{matrix}
\right)
\]

where \(g(\cdot)\) is the link function. (e.g. 
\(g(x) = x\) is for linear regression,
\(g(x) = \text{log}(\frac{x}{1-x})\) is for logistic regression and \(g(x) = \text{log}(x)\) is for Poisson regression.).

We may also perform sex-stratified analysis in the study, for the female study, we assume the following generalized linear regression model:
\[
\mathcal{M}_f: g(E(Y_f)) = \beta_{0,f} + X_f\boldsymbol{\beta_f},
\]
and for the male study:
\[
\mathcal{M}_m: g(E(Y_m)) = \beta_{0,m} + X_m\boldsymbol{\beta_m},
\]
Comparing \(\mathcal{M}\) with \(\mathcal{M}_f\) and \(\mathcal{M}_m\), we can find the relationship:
\[
\beta_{0,f} = \beta_{0},~
\beta_{0,m} = \beta_{0}+\beta_{S}, ~ 
\boldsymbol{\beta_f} = \boldsymbol{\beta_1},~\text{and}~\boldsymbol{\beta_m}  = \boldsymbol{\beta_1}+\boldsymbol{\beta_2}.
\]
In the full model, we may estimate genotype effects by testing \(H_0: \boldsymbol{\beta_1} = \boldsymbol{\beta_2}=0,\) which is equivalent to testing \(H_0: \boldsymbol{\beta_f} = \boldsymbol{\beta_m}=0\) in sex stratified study.
\subsection*{Parameters Estimation}
There is no explicit expression of the MLE in the GLM, hence maximum likelihood estimates must generally be obtained using some iterative numerical methods. 
Here we use the Fisher's scoring algorithm. \cite{Keith2000}
We use \(\hat{\cdot}\) and \(\tilde{\cdot}\) to denote unconstrained and constrained (under \(H_0\)) estimators, respectively and we use those notations below. 
\begin{itemize}
    \item For the female model \(\mathcal{M}_f\):
\begin{align}
    \boldsymbol{\theta_f}=&(\theta_{f,1},\dots,\theta_{f,n_f}) = g^{-1}(\beta_{0,f}+X_f \boldsymbol{\beta_f})\notag\\
     V(\theta_{f,i}) =& Var(Y_{f,i})/\phi, \quad V(\boldsymbol{\theta_{f}}) = diag\{V(\theta_{f,1}), \dots, V(\theta_{f,n_f})\}\notag\\
     w(\theta_{f,i}) =& \frac{1}{V(\theta_{f,i})[g'(\theta_{f,i})]^2}, \quad W(\boldsymbol{\theta_{f}}) = diag\{ w(\theta_{f,1}), \dots, w(\theta_{f,n_f})\}\notag\\
     z(\theta_{f,i}) =& g(\theta_{f,i}) + g'(\theta_{f,i})(Y_{f,i}-\theta_{f,i}), \quad z(\boldsymbol{\theta_{f}}) = ( z(\theta_{f,1}), \dots,  z(\theta_{f,n_f}))'.\notag
\end{align}
    \item For the male model \(\mathcal{M}_m\):
\begin{align}
    \boldsymbol{\theta_m}=&(\theta_{m,1},\dots,\theta_{m,n_m}) = g^{-1}(\beta_{0,m}+X_m \boldsymbol{\beta_m})\notag\\
     V(\theta_{m,i}) =& Var(Y_{m,i})/\phi, \quad V(\boldsymbol{\theta_{m}}) = diag\{V(\theta_{m,1}), \dots, V(\theta_{m,n_m})\}\notag\\
     w(\theta_{m,i}) =& \frac{1}{V(\theta_{m,i})[g'(\theta_{m,i})]^2}, \quad W(\boldsymbol{\theta_{m}}) = diag\{ w(\theta_{m,1}), \dots, w(\theta_{m,n_m})\}\notag\\
     z(\theta_{m,i}) =& g(\theta_{m,i}) + g'(\theta_{m,i})(Y_{m,i}-\theta_{m,i}), \quad z(\boldsymbol{\theta_{m}}) = ( z(\theta_{m,1}), \dots,  z(\theta_{m,n_m}))'.\notag
\end{align}
    \item For the full model \(\mathcal{M}\):
\[
\boldsymbol{\theta} = \left(\boldsymbol{\theta_f}',\boldsymbol{\theta_m}'\right)',\quad
V(\boldsymbol{\theta}) = \left(
 \begin{matrix}
      V(\boldsymbol{\theta_f})& 0\\
      0& V(\boldsymbol{\theta_m})
 \end{matrix}
 \right)
\]
\[ 
 W(\boldsymbol{\theta}) = \left(
 \begin{matrix}
      W(\boldsymbol{\theta_f} )& 0\\
      0& W(\boldsymbol{\theta_m})
 \end{matrix}
 \right), \quad
z(\boldsymbol{\theta}) = 
\left( z(\boldsymbol{\theta_f})',z(\boldsymbol{\theta_m})'\right)'.\notag
\]

\end{itemize}

If the dispersion parameter \(\phi\) is unknown,
in this proof, we use
\[
\hat{\phi}_f = \frac{1}{n_f}(Y_f-\boldsymbol{\hat{\theta}_f})'V(\boldsymbol{\hat{\theta}_f})^{-1}(Y_f-\boldsymbol{\hat{\theta}_f}), 
\quad
\hat{\phi}_m = \frac{1}{n_m}(Y_m-\boldsymbol{\hat{\theta}_m})'V(\boldsymbol{\hat{\theta}_m})^{-1}(Y_m-\boldsymbol{\hat{\theta}_m}), \]
\[
\hat{\phi} = \frac{1}{n}(Y-\boldsymbol{\hat{\theta}})'V(\boldsymbol{\hat{\theta}})^{-1}(Y-\boldsymbol{\hat{\theta}}) =  \frac{1}{n}(n_f\hat{\phi}_f + n_m\hat{\phi}_m)  
\]
Likewise, the constrained estimator also satisfies
\[
\tilde{\phi}_f = \frac{1}{n_f}(Y_f-\boldsymbol{\tilde{\theta}_f})'V(\boldsymbol{\tilde{\theta}_f})^{-1}(Y_f-\boldsymbol{\tilde{\theta}_f}), 
\quad
\tilde{\phi}_m = \frac{1}{n_m}(Y_m-\boldsymbol{\tilde{\theta}_m})'V(\boldsymbol{\tilde{\theta}_m})^{-1}(Y_m-\boldsymbol{\tilde{\theta}_m}), \]
\[
\tilde{\phi} = \frac{1}{n}(Y-\boldsymbol{\tilde{\theta}})'V(\boldsymbol{\tilde{\theta}})^{-1}(Y-\boldsymbol{\tilde{\theta}}) =  \frac{1}{n}(n_f\tilde{\phi}_f + n_m\tilde{\phi}_m)  
\]

We begin with \(\boldsymbol{\hat{\beta_f}^{(0)}}=\boldsymbol{\hat{\beta_m}^{(0)}}=\boldsymbol{\hat{\beta_1}^{(0)}}=\boldsymbol{\hat{\beta_2}^{(0)}} = 0\), since there is no prior information that the effect is positive or negative. So choosing 0 as the start point is reasonable.

Under this initialization, for each study:
 \[
 \hat{\beta}_{0,f}^{(0)} = g(\bar{Y}_f),~\bar{Y}_f = \frac{1}{n_f}\sum_{i=1}^{n_f}Y_{f,i},\quad 
  \hat{\beta}_{0,m}^{(0)} = g(\bar{Y}_m),~\bar{Y}_m = \frac{1}{n_m}\sum_{i=1}^{n_m}Y_{m,i}
\]
 
For the full study:
\[
\hat{\beta}_0^{(0)} = g(\bar{Y}_f), \quad  \widehat{\beta_0 + \beta_S}^{(0)} = g(\bar{Y}_m).
\]
By the invariance of MLE, \(\widehat{\beta_0 + \beta_S}^{(0)} = \hat{\beta}_0^{(0)} + \hat{\beta}_S^{(0)},\) hence we have 
\[\hat{\beta}_{0,f}^{(0)} = \hat{\beta}_0^{(0)},~\text{and}~\hat{\beta}_m^{(0)} = \hat{\beta}_0^{(0)}+\hat{\beta}_S^{(0)}.\]
So the fitted values are \[ \boldsymbol{\hat{\theta}_f^{(0)}} = g^{-1}(1_{n_f}\hat{\beta}_{0,f}^{(0)}), \quad 
\boldsymbol{\hat{\theta}_m^{(0)}} = g^{-1}(1_{n_m}\hat{\beta}_{0,m}^{(0)})\]
~\text{and}~
\[
\boldsymbol{\hat{\theta}^{(0)}}  = \left(\begin{matrix}
    g^{-1}(1_{n_f}\hat{\beta}_{0}^{(0)})\\
     g^{-1}(1_{n_m}(\hat{\beta}_0^{(0)}+\hat{\beta}_S^{(0)}))
\end{matrix}\right) = \left(\begin{matrix}
    g^{-1}(1_{n_f}\hat{\beta}_{0,f}^{(0)})\\
     g^{-1}(1_{n_m}\hat{\beta}_{0,m}^{(0)})
\end{matrix}\right) = \left(\begin{matrix}\boldsymbol{\hat{\theta}_f^{(0)}}\\\boldsymbol{\hat{\theta}_m^{(0)}}
\end{matrix}\right)\]   
If we further assume 
\[
\hat{\beta}_{0,f}^{(l)} = \hat{\beta}_0^{(l)},~
\hat{\beta}_{0,m}^{(l)} = \hat{\beta}_0^{(l)}+\hat{\beta}_S^{(l)},~
\boldsymbol{\hat{\beta}_{f}^{(l)}} = \boldsymbol{\hat{\beta}_{1}^{(l)}},~
\boldsymbol{\hat{\beta}_{m}^{(l)}} = \boldsymbol{\hat{\beta}_{1}^{(l)}}+\boldsymbol{\hat{\beta}_{2}^{(l)}}
\]
then it yields 
 \[\boldsymbol{\hat{\theta}^{(l)}} = \left(\begin{matrix}\boldsymbol{\hat{\theta}_f^{(l)}}\\\boldsymbol{\hat{\theta}_m^{(l)}}
\end{matrix}\right), ~
W(\boldsymbol{\hat{\theta}^{(l)}}) = 
\left(\begin{matrix}
W(\boldsymbol{\hat{\theta}_f^{(l)}})& 0\\0 &W(\boldsymbol{\hat{\theta}_m^{(l)}})
\end{matrix}\right) ~\text{and}~ z(\boldsymbol{\hat{\theta}^{(l)}}) = (z(\boldsymbol{\hat{\theta}_f^{(l)}})', z(\boldsymbol{\hat{\theta}_m^{(l)}})')'.\] 

At the \((l+1)\)th iteration, for the stratified model:
\[
\left(\begin{matrix}
    \hat{\beta}_{0,f}^{(l+1)}\\
    \boldsymbol{\hat{\beta}_f^{(l+1)}}
\end{matrix}\right) = (\boldsymbol{X_f}'W(\boldsymbol{\hat{\theta}_f^{(l)}})\boldsymbol{X_f})^{-1}\boldsymbol{X_f}'W(\boldsymbol{\hat{\theta}_f^{(l)}})z(\boldsymbol{\hat{\theta}_f^{(l)}}),~\text{where}~\boldsymbol{X_f} = (1_{n_f}, X_f).
\]
\[
\left(\begin{matrix}
    \hat{\beta}_{0,m}^{(l+1)}\\
    \boldsymbol{\hat{\beta}_m^{(l+1)}}
\end{matrix}\right) = (\boldsymbol{X_m}'W(\boldsymbol{\hat{\theta}_m^{(l)}})\boldsymbol{X_m})^{-1}\boldsymbol{X_m}'W(\boldsymbol{\hat{\theta}_m^{(l)}})z(\boldsymbol{\hat{\theta}_m^{(l)}}),~\text{where}~\boldsymbol{X_m} = (1_{n_m}, X_m).
\]

For the full model, we estimate the fitted values using \( \hat{\beta}_{0,f}^{(l)},~\hat{\beta}_{0,m}^{(l)},~\boldsymbol{\hat{\beta}_{f}^{(l)}}\) and \(\boldsymbol{\hat{\beta}_{m}^{(l)}}\). These parameters are related to \( \hat{\beta}_0^{(l)},~\hat{\beta}_S^{(l)},~\boldsymbol{\hat{\beta}_1^{(l)}} \) and \( \boldsymbol{\hat{\beta}_2^{(l)}} \), allowing us to express the model components consistently across different parameterizations.
\[ g(\boldsymbol{\hat{\theta}^{(l)}}) = \left(\begin{matrix}
    1_{n_f} & X_f & 0_{n_f} & 0_{n_f}\\
    1_{n_m} & X_m & 1_{n_m} & X_m \\
\end{matrix}\right)T^{-1}
T
\left( \begin{matrix}
    \hat{\beta}_0^{(l)} \\\boldsymbol{\hat{\beta}_1^{(l)}} \\ \hat{\beta}_S^{(l)}\\ \boldsymbol{\hat{\beta}_2^{(l)}}
\end{matrix}\right) = 
\left(\begin{matrix}
    \boldsymbol{X_f} & 0 \\
    0 & \boldsymbol{X_m} \\
\end{matrix}\right)
\left( \begin{matrix}
    \hat{\beta}_{0,f}^{(l)}\\
    \boldsymbol{\hat{\beta}_f^{(l)}} \\ \hat{\beta}_{0,m}^{(l)}\\ \boldsymbol{\hat{\beta}_m^{(l)}}
\end{matrix}\right),~\text{here}~T = \left(\begin{matrix}
1 & 0 & 0 & 0\\
0 & 1 & 0 & 0\\
1 & 0 & 1 & 0\\
0 & 1 & 0 & 1\\
\end{matrix}\right)
\]
For the \((l+1)\)th iteration, the parameters estimated with the full model are
\[
\left(\begin{matrix}
     \hat{\beta}_{0,f}^{(l+1)}\\
    \boldsymbol{\hat{\beta}_f^{(l+1)}}\\
         \hat{\beta}_{0,m}^{(l+1)}\\
    \boldsymbol{\hat{\beta}_m^{(l+1)}}\\
\end{matrix}\right) = 
\left(\begin{matrix}
(\boldsymbol{X_f}'W(\boldsymbol{\hat{\theta}_f^{(l)}})\boldsymbol{X_f})^{-1}\boldsymbol{X_f}'W(\boldsymbol{\hat{\theta}_f^{(l)}})z(\boldsymbol{\hat{\theta}_f^{(l)}})\\
(\boldsymbol{X_m}'W(\boldsymbol{\hat{\theta}_m^{(l)}})\boldsymbol{X_m})^{-1}\boldsymbol{X_m}'W(\boldsymbol{\hat{\theta}_m^{(l)}})z(\boldsymbol{\hat{\theta}_m^{(l)}})\\
\end{matrix}\right),
\] they are exactly the estimated parameters under each study.
Hence for the next iteration,
\[
\hat{\beta}_{0,f}^{(l+1)} = \hat{\beta}_0^{(l+1)},~
\hat{\beta}_{0,m}^{(l+1)} = \hat{\beta}_0^{(l+1)}+\hat{\beta}_S^{(l+1)},~
\boldsymbol{\hat{\beta}_{f}^{(l+1)}} = \boldsymbol{\hat{\beta}_{1}^{(l+1)}},~
\boldsymbol{\hat{\beta}_{m}^{(l+1)}} = \boldsymbol{\hat{\beta}_{1}^{(l+1)}}+\boldsymbol{\hat{\beta}_{2}^{(l+1)}}
\]
Therefore, mathematical induction and limiting argument lead to the MLE of those parameters satisfying this relationship:
\[
\hat{\beta}_{0,f} = \hat{\beta}_0,~
\hat{\beta}_{0,m} = \hat{\beta}_0+\hat{\beta}_S,~
\boldsymbol{\hat{\beta}_{f}} = \boldsymbol{\hat{\beta}_{1}},~
\boldsymbol{\hat{\beta}_{m}} = \boldsymbol{\hat{\beta}_{1}}+\boldsymbol{\hat{\beta}_{2}}
\] and the fitted value
\(\boldsymbol{\hat{\theta}} = (\boldsymbol{\hat{\theta}_f}',\boldsymbol{\hat{\theta}_m}')'\)
\subsection*{Equivalence of Test Statistics}
Under the GLM, the sex-stratified and individual-level full models yield equivalent Wald, Score, and LRT statistics.

\subsubsection*{The Wald Statistic}
For \(\mathcal{M}_f\), the  Wald Statistic is :
\[
Wald_f = \frac{1}{\hat{\phi}_f}\{\left(\hat{\beta}_{0,f},\boldsymbol{\hat{\beta}_f}'\right)L_f'[L_f(\boldsymbol{X_f}'W(\boldsymbol{\hat{\theta}_f})\boldsymbol{X_f})^{-1}L_f']^{-1}L_f
\left(\begin{matrix}
   \hat{\beta}_{0,f}\\\boldsymbol{\hat{\beta}_f}
\end{matrix}\right)\},
\] where \(L_f = \left(\begin{matrix}0&I_p\end{matrix}\right)\), and for \(\mathcal{M}_m\) is
\[
Wald_m = \frac{1}{\hat{\phi}_m}\{\left(\hat{\beta}_{0,m},\boldsymbol{\hat{\beta}_m}'\right)L_m'[L_m(\boldsymbol{X_m}'W(\boldsymbol{\hat{\theta}_m})\boldsymbol{X_m})^{-1}L_m']^{-1}L_m
\left(\begin{matrix}
   \hat{\beta}_{0,m}\\\boldsymbol{\hat{\beta}_m}
\end{matrix}\right)\},
\] where \(L_m = \left(\begin{matrix}0&I_p\end{matrix}\right)\).

For the full study \(\mathcal{M}\), the  Wald Statistic is
\begin{align}
Wald =& \frac{1}{\hat{\phi}}\{\left(\hat{\beta}_{0,f},\boldsymbol{\hat{\beta}_f}',\hat{\beta}_{0,m},\boldsymbol{\hat{\beta}_m}'\right)L'[L\left(\begin{matrix}
    (\boldsymbol{X_f}'W(\boldsymbol{\hat{\theta}_f})\boldsymbol{X_f})^{-1}&0 \\
     0&  (\boldsymbol{X_m}'W(\boldsymbol{\hat{\theta}_m})\boldsymbol{X_m})^{-1}
\end{matrix}\right)L']^{-1}L
\left(\begin{matrix}
   \hat{\beta}_{0,f}\\\boldsymbol{\hat{\beta}_f}\\
   \hat{\beta}_{0,m}\\\boldsymbol{\hat{\beta}_m}
\end{matrix}\right)\}\notag
\\
=&\frac{1}{\hat{\phi}}
\{\left(\hat{\beta}_{0,f},\boldsymbol{\hat{\beta}_f}'\right)L_f'[L_f(\boldsymbol{X_f}'W(\boldsymbol{\hat{\theta}_f})\boldsymbol{X_f})^{-1}L_f']^{-1}L_f
\left(\begin{matrix}
   \hat{\beta}_{0,f}\\\boldsymbol{\hat{\beta}_f}
\end{matrix}\right)+\notag\\
&\left(\hat{\beta}_{0,m},\boldsymbol{\hat{\beta}_m}'\right)L_m'[L_m(\boldsymbol{X_m}'W(\boldsymbol{\hat{\theta}_m})\boldsymbol{X_m})^{-1}L_m']^{-1}L_m
\left(\begin{matrix}
   \hat{\beta}_{0,m}\\\boldsymbol{\hat{\beta}_m}
\end{matrix}\right)
\}\notag,\end{align}
where $L = \left(\begin{matrix}
    L_f&0\\0&L_m
\end{matrix}\right)$.
As the sample size becomes sufficiently large, \(\hat{\phi}_f,~\hat{\phi}_m\) and \(\hat{\phi}\to \phi\), then we have
\[Wald = Wald_f + Wald_m\]
Thus, the sex-stratified and full individual-level models produce equivalent Wald statistics.

\subsubsection*{The Score Statistic}
For the female study \(\mathcal{M}_f\), the Score statistic is 
\[
Score_f = \frac{1}{\tilde{\phi}_f}(Y_f-\boldsymbol{\tilde{\theta}_f})'V(\boldsymbol{\tilde{\theta}_f})^{-1/2}W(\boldsymbol{\tilde{\theta}_f})^{1/2}X_f(\tilde{R}_f'W(\boldsymbol{\tilde{\theta}_f})\tilde{R}_f)^{-1}X_f'W(\boldsymbol{\tilde{\theta}_f})^{1/2}V(\boldsymbol{\tilde{\theta}_f})^{-1/2}(Y_f-\boldsymbol{\tilde{\theta}_f}),
\]
where \(\boldsymbol{\tilde{\theta}_f}\) is the fitted value under the null model, and 
\[
\tilde{R}_f = X_f - 1_{n_f}(1_{n_f}'W(\boldsymbol{\tilde{\theta}_f})1_{n_f})^{-1}1_{n_f}'W(\boldsymbol{\tilde{\theta}_f})X_f.
\]
The same notation can be applied to the male model \(\mathcal{M}_m\), with the relevant components replaced by their corresponding counterparts.

For the full study \(\mathcal{M}\), 
\begin{align}
    Score = & \frac{1}{\tilde{\phi}}\left(
(Y_f-\boldsymbol{\tilde{\theta}_f})'V(\boldsymbol{\tilde{\theta}_f})^{-1/2}W(\boldsymbol{\tilde{\theta}_f})^{1/2}X_f,\quad(Y_m-\boldsymbol{\tilde{\theta}_m})'V(\boldsymbol{\tilde{\theta}_m})^{-1/2}W(\boldsymbol{\tilde{\theta}_m})^{1/2}X_m
\right)\notag\\
&(\tilde{R}'W(\boldsymbol{\tilde{\theta}})\tilde{R})^{-1}
\left(\begin{matrix}  X_f'W(\boldsymbol{\tilde{\theta}_f})^{1/2}V(\boldsymbol{\tilde{\theta}_f})^{-1/2}(Y_f-\boldsymbol{\tilde{\theta}_f})\\
X_m'W(\boldsymbol{\tilde{\theta}_m})^{1/2}V(\boldsymbol{\tilde{\theta}_m})^{-1/2}(Y_m-\boldsymbol{\tilde{\theta}_m})
\end{matrix}\right)\notag,
\end{align}
where 
\begin{align}
\tilde{R} =& 
\left(\begin{matrix}
    X_f&0\\0&X_m
\end{matrix}\right) - 
\left(\begin{matrix}
 1_{n_f}(1_{n_f}'W(\boldsymbol{\tilde{\theta}_f})1_{n_f})^{-1} 1_{n_f}'W(\boldsymbol{\tilde{\theta}_f})X_f&0\\
0&1_{n_m}(1_{n_m}'W(\boldsymbol{\tilde{\theta}_m})1_{n_m})^{-1}1_{n_m}'W(\boldsymbol{\tilde{\theta}_m})X_m
\end{matrix}\right)\notag\\
=& \left(\begin{matrix}
    \tilde{R}_f&0\\0&\tilde{R}_m
\end{matrix}\right)\notag
\end{align}
As the sample size becomes sufficiently large, \(\tilde{\phi}_f,~ \tilde{\phi}_m\) and \(\tilde{\phi}\to \phi\). Hence we have 
\[
Score = Score_f + Score_m.
\]
\subsubsection*{The LRT statistic}
The likelihood ratio test (LRT) statistic for the female model \(\mathcal{M}_f\) is defined as
\[
LRT_f = 2\sum_{i=1}^{n_f} \left\{ \log\left(\mathcal{L}(Y_{f,i}, \hat{\beta}_{0,f}, \boldsymbol{\hat{\beta}}_f, \hat{\phi}_f)\right) - \log\left(\mathcal{L}(Y_{f,i}, \tilde{\beta}_{0,f}, \boldsymbol{0}, \tilde{\phi}_f)\right) \right\},
\]
and for the male model \(\mathcal{M}_m\) as
\[
LRT_m = 2\sum_{i=1}^{n_m} \left\{ \log\left(\mathcal{L}(Y_{m,i}, \hat{\beta}_{0,m}, \boldsymbol{\hat{\beta}}_m, \hat{\phi}_m)\right) - \log\left(\mathcal{L}(Y_{m,i}, \tilde{\beta}_{0,m}, \boldsymbol{0}, \tilde{\phi}_m)\right) \right\},
\]
where \( \mathcal{L} \) denotes the likelihood function. For \(\mathcal{M}_f\), it is given by
\[
\mathcal{L}(Y_{f,i}, \beta_{0,f}, \boldsymbol{\beta}_f, \phi) = 
\exp\left[
\frac{Y_{f,i} \cdot \boldsymbol{X}_{f,i} (\beta_{0,f}, \boldsymbol{\beta}_f')' - b\left( \boldsymbol{X}_{f,i} (\beta_{0,f}, \boldsymbol{\beta}_f')' \right)}{\phi} + c(Y_{f,i}, \phi)
\right],
\]
and for \(\mathcal{M}_m\):
\[
\mathcal{L}(Y_{m,i}, \beta_{0,m}, \boldsymbol{\beta}_m, \phi) = 
\exp\left[
\frac{Y_{m,i} \cdot \boldsymbol{X}_{m,i} (\beta_{0,m}, \boldsymbol{\beta}_m')' - b\left( \boldsymbol{X}_{m,i} (\beta_{0,m}, \boldsymbol{\beta}_m')' \right)}{\phi} + c(Y_{m,i}, \phi)
\right].
\]

For the full model \(\mathcal{M}\), using the relationship of the MLEs, we have:
\begin{align}
    LRT = &2\sum_{i=1}^{n_1}\{\text{log}(\mathcal{L}(Y_{f,i},\hat{\beta}_0, \boldsymbol{\hat{\beta}_1},\hat{\phi}))-\text{log}(\mathcal{L}(Y_{f,i},\tilde{\beta}_0, \boldsymbol{0},\tilde{\phi}))\}+\notag\\
&2\sum_{i=1}^{n_m}\{\text{log}(\mathcal{L}(Y_{m,i},\hat{\beta}_0+\hat{\beta}_S, \boldsymbol{\hat{\beta}_1}+\boldsymbol{\hat{\beta}_2},\hat{\phi}))-\text{log}(\mathcal{L}(Y_{m,i},\tilde{\beta}_0+\tilde{\beta}_S, \boldsymbol{0},\tilde{\phi}))\}\notag\\
=&2\sum_{i=1}^{n_1}\{\text{log}(\mathcal{L}(Y_{f,i},\hat{\beta}_{0,f}, \boldsymbol{\hat{\beta}_f},\hat{\phi}))-\text{log}(\mathcal{L}(Y_{f,i},\tilde{\beta}_{0,f}, \boldsymbol{0},\tilde{\phi}))\}+\notag\\
&2\sum_{i=1}^{n_m}\{\text{log}(\mathcal{L}(Y_{m,i},\hat{\beta}_{0,m}, \boldsymbol{\hat{\beta}_m},\hat{\phi})-\text{log}(\mathcal{L}(Y_{m,i},\tilde{\beta}_{0,m}, \boldsymbol{0},\tilde{\phi}))\}\notag\\
=&LRT_f + LRT_m\notag.
\end{align}
We thereby show the equivalence of the LRT statistics.

\newpage
\section*{Appendix B}
\subsection*{Proof of Lemma 1} 


We provide proof of linear model.

\(\mathcal{M}_2^*: g(E(Y)) =  \beta_0^* + \beta_S^*S^* +G_A^*\beta_A^* + GS^*\beta_{GS}^*,\) is reparametrized model, of which additive and interaction components are uncorrelated. \(\mathcal{M}_1^*: g(E(Y)) =  \beta_0^* +\beta_S^*S^*+ G_A^*\beta_A^*\).

We already know that \(\mathcal{M}_2^*\) is invariant to jointly test \(H_0: \beta_A^*=\beta_{GS}^*=0\) \cite{Chen2021}. Additionally, non-centrality parameter \(ncp^*\) of statistic \(W_2 \sim \chi^2(2)\) can be decomposed as two uncorrelated parts:
\begin{align}
    ncp^* =  ncp^*_A + ncp^*_{GS}
    \label{formula:decomposed ncp}
\end{align}

To calculate \(ncp^*\) of \(\mathcal{M}_2^*\), we assume \(Y \sim N(X^*\bm{\beta^*},\sigma^2I)\) (in order to avoid the uninteresting case in which \(ncp^* \to \infty\) as n grows, we assume \(\bm{\beta^*} = \frac{\bm{c^*}}{\sqrt{n}}\)
for a fixed vector \(\bm{c}\), so that \(\bm{\beta^*} \stackrel{p}{\to} \bm{0}\) and \(ncp^*\) converges to a finite number as \(n \to \infty\)), where \(X^* = (1_n,~S^*, G_A^*, GS^*) = (X_S^*, G_A^*, GS^*)\) and 
\(\bm{\beta^*} = (\beta_0^*, \beta_S^*, \beta_A^*, \beta_{GS}^*)'\).

Since the estimator of \(\bm{\beta}^*\) is \[\bm{\hat{\beta}^*} = ({X^*}'X^*)^{-1}{X^*}'Y \sim N(\bm{\beta^*}, \sigma^2({X^*}'X^*)^{-1})\]
hence 
\[L\bm{\hat{\beta}^*} \sim N(L\bm{\beta^*}, \sigma^2L({X^*}'X^*)^{-1}L'),\]

where \(L = \begin{bmatrix}
0&0&1&0\\
0&0&0&1\end{bmatrix}.\) Hence $ncp^*$ of \(\mathcal{M}_2^*\) is
\begin{align}
    ncp^* &= \frac{1}{\sigma^2}{\bm{\beta^*}}'L'
(L({X^*}'X^*)^{-1}L')^{-1}L\bm{\beta^*}\notag\\
 &= \frac{1}{\sigma^2}\begin{bmatrix}
     \beta_A^*&\beta_{GS}^*
 \end{bmatrix}
 \begin{bmatrix}
     {G_A^*}'\\
     {GS^*}'
 \end{bmatrix}
( I_n -   X_S^*({X_S^*}'X_S^*)^{-1}{X_S^*}' )\begin{bmatrix}
     {G_A^*}&{GS^*}
 \end{bmatrix}
  \begin{bmatrix}
\beta_A^*\\
\beta_{GS}^*
 \end{bmatrix}
\end{align}
We can decompose \(ncp^*\) into 3 parts such that \(ncp^* = a_1 + a_2 + a_3\), where 
\begin{align}
a_1 =& \frac{1}{\sigma^2} \beta_A^*{G_A^*}'(I_n -   X_S^*({X_S^*}'X_S^*)^{-1}{X_S^*}') {G_A^*}\beta_A^*\notag\\
a_2 =& \frac{2}{\sigma^2} \beta_{GS}^*{GS^*}'(I_n -   X_S^*({X_S^*}'X_S^*)^{-1}{X_S^*}'){G_A^*}\beta_A^*\notag\\
a_3 =& \frac{1}{\sigma^2} \beta_{GS}^*{GS^*}'(I_n -   X_S^*({X_S^*}'X_S^*)^{-1}{X_S^*}'){GS^*}\beta_{GS}^*\notag
\end{align}
Since 
\[
\frac{1}{n}{GS^*}'X_S^* \stackrel{p}{\to} (E(GS^*),E(GS^*)) = (0, 0)
\]
and 
\[
\frac{1}{n}{GS^*}'G_A^* \stackrel{p}{\to} E(GS^*G_A^*) = 0
\]
we have \(a_2 \stackrel{p}{\to} 0, \) 
so \(ncp^* \stackrel{p}{\to} a_1+a_3\), 

To compute \(a_1\)
\begin{align}
    \frac{1}{n}
    \begin{bmatrix}
        {X_S^*}'X_S^*&{X_S^*}'G_A^*\\
        {G_A^*}'X_S^*&{G_A^*}'G_A^*\\
    \end{bmatrix}\stackrel{p}{\to}
\begin{bmatrix}
1 & E(S^*) &E(G_A^*)\\
E(S^*) & E({S^*}^2) & E(S^*G_A^*) \\
E(G_A^*)  &E(G_A^*S^*) & E({G_A^*}^2)\\
\end{bmatrix} = 
\begin{bmatrix} 
P_{11}^*&P_{12}^*\\
P_{21}^*&P_{22}^*
\end{bmatrix},
\label{formula:matrix partition}
\end{align}

where \(P_{22}^* = E({G_A^*}^2)\), \(P_{21}^* = {P_{12}^*}'= (E(G_A^*), E(G_A^*S^*)) \) and \(P_{11}^* = \begin{bmatrix}
    1 & E(S^*)\\ E(S^*) & E({S^*}^2)
\end{bmatrix},\)
which implies 
\[
nL({X_A^*}'X_A^*)^{-1}L' \stackrel{p}{\to} ({P_{22}^* - P_{21}^*{P_{11}^*}^{-1}P_{12}^*})^{-1}. 
\]
Therefore, 
\begin{align}
    a_1 \stackrel{p}{\to} \frac{n}{\sigma^2}(P_{22}^* - P_{21}^*{P_{11}^*}^{-1}P_{12}^*){\beta_A^*}^2
    \label{formula:M2ncp},
\end{align}
Similarly, \(ncp\) of \(\mathcal{M}_1^*\), denoted as \(ncp_A^*\), can be derived by formula \ref{formula:matrix partition} and also converges to the value in formula \ref{formula:M2ncp}, which means \(ncp_A^* \to a_1\).

Next, we provide reparametrized genotype codings for the additive, interaction, and sex effects in the following tables, ensuring \(E(GS^*)=E(GS^*G_A^*)=Cov(GS^*,G_A^*)=0\). Assuming genotypes \((rr, rR, RR, r, R)\) with the following genotype frequencies:
\[
(\frac{(1-f_{female})^2}{2}, f_{female}(1-f_{female}),\frac{f_{female}^2}{2}, \frac{1-f_{male}}{2}, \frac{f_{male}}{2})
\]

Under the XCI assumption:
\begin{center}
\begin{tabular}{c| c | c | c |c | c} 
 \toprule
 & \multicolumn{3}{c}{Female}&\multicolumn{2}{|c}{Male}\\
 \midrule
Coding & rr & rR & RR & r & R \\
 \hline
 ${G^{*}_{A,I}}$ & $-1$ &  $0$  & $1$ & $-1$ & $1$\\ \hline
 ${GS^{*}_{I}}$& $-f_{female}$ & $\frac{1}{2}-f_{female}$ & $1-f_{female}$ &$\frac{f_{female}(1-f_{female})}{2(1-f_{male})}$ & $-\frac{f_{female}(1-f_{female})}{2f_{male}}$\\
 \hline
 $S^*$ & $-1$ & $-1$ & $-1$ & $1$ &  $1$ \\ \bottomrule
\end{tabular}
\end{center}
\begin{align}
&{X^{*}_I} = (1,~S^*, {G^{*}_{A,I}}, {GS^{*}_{I}}) = ({X^{*}_{A,I}}, {GS^{*}_{I}}),~\bm{{\beta^{*}_I}} = ({\beta^{*}_{0,I}}, \beta^{*}_{S,I}, \beta^{*}_{A,I}, \beta^{*}_{GS,I})' = (\bm{{\beta^{*}_{A,I}}'}, \beta^{*}_{GS,I})'\notag\\
&E({G^*_{A,I}}^2) =  f_{female}^2-f_{female}+1
 \quad 
E({G^*_{A,I}}) =  f_{female}+f_{male}-1 \notag\\
&E({G^*_{A,I}}S^*) = f_{male}-f_{female} \quad  E(S^*)=0 \quad E({S^*}^2)=1.\notag
\end{align}

Under the noXCI assumption:
\begin{center}
\begin{tabular}{c| c | c | c |c | c} 
 \toprule
 & \multicolumn{3}{c}{Female}&\multicolumn{2}{|c}{Male}\\
 \midrule
Coding & rr & rR & RR & r & R \\
 \hline
 $G^{*}_{A,N}$ & $-1$ &  $0$  & $1$ & $-1$ & $0$\\ \hline
 ${GS^{*}_N}$& $-f_{female}$ & $\frac{1}{2}-f_{female}$ & $1-f_{female}$ &$\frac{f_{female}(1-f_{female})}{(1-f_{male})}$ & $-\frac{f_{female}(1-f_{female})}{f_{male}}$\\
 \hline
 $S^*$ & $-1$ & $-1$ & $-1$ & $1$ &  $1$ \\ \bottomrule
\end{tabular}
\end{center}
\begin{align}
&{X^{*}_N} = (1,~S^*, {G^{*}_{A,N}}, {GS^{*}_{N}}) = ({X^{*}_{A,N}}, {GS^{*}_{N}}),~\bm{{\beta^{*}_N}} = ({\beta^{*}_{0,N}}, \beta^{*}_{S,N}, \beta^{*}_{A,N}, \beta^{*}_{GS,N})' = (\bm{{\beta^{*}_{A,N}}'}, \beta^{*}_{GS,N})' \notag\\
& E({G^*_{A,N}}^2)  =  
    f_{female}^2-f_{female}+1-\frac{f_{male}}{2}
 \quad 
E({G^*_{A,N}}) =  f_{female}+\frac{f_{male}}{2}-1\notag\\
& E({G^*_{A,N}}S^*) = \frac{f_{male}}{2}-f_{female} \quad  E(S^*)=0 \quad E({S^*}^2)=1.\notag
\end{align}

$X^*_I$, $X^*_N$ are the reparametrized design matrices derived from $X_I$ and $X_N$, respectively.
 \[ 
X^*_I= {X_I}L_1,\quad X^*_N= {X_N}L_2,\quad {X_I}= {X_N}T,\quad
X^*_I= X^*_NL_2^{-1}TL_1,\quad
 \bm{\beta^{*}_{I}} = L_1^{-1}T^{-1}L_2\bm{\beta^{*}_{N}},
 \]
where \[
 L_1 = 
 \begin{bmatrix}
     1 & -1 & -1 & -f_{female}\\
     0 & 2 & 0 & f_{female} + \frac{f_{female}(1-f_{female})}{2(1-f_{male})}\\
     0 & 0 & 2 &1\\
     0 & 0 & 0 & -1-\frac{f_{female}(1-f_{female})}{2f_{male}(1-f_{male})}\\
 \end{bmatrix}, \quad 
 L_2 = 
 \begin{bmatrix}
     1 & -1 & -1 & -f_{female}\\
     0 & 2 & 0 & f_{female} + \frac{f_{female}(1-f_{female})}{1-f_{male}}\\
     0 & 0 & 1 & \frac{1}{2}\\
     0 & 0 & 0 & -\frac{1}{2} -\frac{f_{female}(1-f_{female})}{f_{male}(1-f_{male})}\\
 \end{bmatrix},
 \]
and
\[
 T = 
 \begin{bmatrix}
     1 & 0 & 0 & 0\\
     0 & 1 & 0 & 0\\
     0 & 0 & \frac{1}{2} & 0\\
     0 & 0 & \frac{1}{2}  & 1\\
 \end{bmatrix}.
 \]
Hence we have
\begin{align}
    {\beta^*_{A,I}} =& \frac{f_{male}(1-f_{male})+f_{female}(1-f_{female})}{2f_{male}(1-f_{male})+f_{female}(1-f_{female})} {\beta^*_{A,N}}  + (\frac{f_{male}(1-f_{male})}{2f_{male}(1-f_{male})+f_{female}(1-f_{female})}-\frac{1}{2}){\beta^*_{GS,N}}\notag\\
    {\beta^*_{GS,I}} =&\frac{2f_{male}(1-f_{male})}{2f_{male}(1-f_{male})+f_{female}(1-f_{female})} {\beta^*_{A,N}}  + (\frac{2f_{male}(1-f_{male})+2f_{female}(1-f_{female})}{2f_{male}(1-f_{male})+f_{female}(1-f_{female})}){\beta^*_{GS,N}}\notag
\end{align}
If the XCI assumption holds and there is no interaction effect, then \({\beta^*_{GS,I}}=0\).
Denote 
\begin{align}
    d_1 &=f_{female}(1-f_{female})\notag\\
    d_2 &= f_{male}(1-f_{male})\notag
\end{align}
The relationship between \({\beta^*_{A,I}}~\text{and}~{\beta^*_{A,N}}\) is:
\begin{align}
    {\beta^*_{A,I}} = \frac{2d_1^2+2d_2^2+5d_1d_2}{2d_1^2+4d_2^2+6d_1d_2}{\beta^*_{A,N}} =\frac{2d_1+d_2}{2d_1+2d_2}{\beta^*_{A,N}} 
    \label{formula:ratio_of_beta}
\end{align}
From formulas \ref{formula:M2ncp} and \ref{formula:ratio_of_beta}, the ratio of the non-centrality parameters under the XCI and no-XCI assumptions is a function of the minor allele frequencies in females and males, and $ncp_A^*$ is proportional to \({\beta_A^*}^2\).
\begin{align}
    \frac{ncp^*_{A,I}}{ncp^*_{A,N}} \stackrel{p}{\to} \frac{(2d_1+d_2)^2}{(2d_1+2d_2)^2}\frac{d_1+2d_2}{d_1+\frac{1}{2}d_2}=\frac{2(d_1+d_2)^2+d_1d_2}{2(d_1+d_2)^2}=1+\frac{d_1d_2}{2(d_1+d_2)^2}
    \label{ncp_prop}
\end{align}

From formula \ref{ncp_prop}, we know \(ncp_{A,N}^* < ncp_{A,I}^*\). For model \(\mathcal{M}_1^*\), testing \(H_0: \beta_A^* = 0\) is equivalent to the additive model without reparametrization. Hence, \(ncp_{A,N} < ncp_{A,I}\), and both \(ncp\) values are proportional to \(\beta_A^2\).

\subsection*{Proof of Theorem 2}
Using normal approximations \cite{Muirhead2005}, we estimate the theoretical power loss by the following formula:
\begin{align}
\Phi\left\{ 
	\underbrace{\frac{\chi^2_{1,\alpha}-(1+c_2\beta_A^2)}{\sqrt{2(1+2c_2\beta_A^2)}}}_{\text{Term 1.1}}
\right\}
-
\Phi\left\{
	\underbrace{\frac{\chi^2_{1,\alpha}-(1+c_1\beta_A^2)}{\sqrt{2(1+2c_1\beta_A^2)}}}_{\text{Term 1.2}}
\right\}.
\end{align}
For \(\text{Term 1.1}\) :
\[
\text{Term 1.1} = \frac{(a-c_2)\beta_A^2-1}{\sqrt{2(1+2c_2\beta_A^2)}  }
\]
\(\text{Term 1.2}\) :
\[
\text{Term 1.2} = \frac{(a-c_1)\beta_A^2-1}{\sqrt{2(1+2c_1\beta_A^2)}}
\]
Since \(a-c_2 > 0\) and \(a-c_1 < 0\), as \( |\beta_A| \to \infty \), we have:
\[
\text{Term 1.1} \to +\infty, \quad \text{Term 1.2} \to -\infty.
\]
\noindent
Hence the power loss due to coding misspecification approaches 1 as \( |\beta_A|  \to \infty \), indicating a complete loss of power.
\subsection*{Proof of Lemma 2} 
Notation: 
\[X_S = (1_n, S),~f_{female}=(f_{female,1},\dots,f_{female,k})'~\text{and}~f_{male}=(f_{male,1},\dots,f_{male,k})'.\]
\[\bm{G_A} = (G_{A,1},\dots,G_{A,k}),~\bm{G_D} = (G_{D,1},\dots,G_{D,k}),~\bm{GS} = (GS_{1},\dots,GS_{k}),\]
\[\bm{G_F} = (\bm{G_A}, \bm{G_D},\bm{GS})\]
\[\bm{{\beta}_A} = (\beta_{A,1},\dots,\beta_{A,k})',
~\bm{{\beta}_D} = (\beta_{D,1},\dots,\beta_{D,k})',
~\bm{{\beta}_{GS}} = (\beta_{GS,1},\dots,\beta_{GS,k})',
\]
\[\bm{{\beta}_F} = (\bm{{\beta}_A}',\bm{{\beta}_D}',\bm{{\beta}_{GS}}')'\]
The test statistics for the $k$- and $3k$-df models are denoted as $W_A \sim \chi^2(k, {ncp}_A)$ and $W_F \sim \chi^2(3k, {ncp}_F)$, respectively.
\begin{align}
W_A&=\frac{1}{\sigma^2}\bm{\hat{\beta}_A}'[\bm{G_A'}(I_n-X_S({X_S}'X_S)^{-1}{X_S}')\bm{G_A}]\bm{\hat{\beta}_A}\\
W_F&=\frac{1}{\sigma^2}\boldsymbol{\hat{\beta}_F}'[\bm{G_F'}(I_n-X_S({X_S}'X_S)^{-1}{X_S}')\bm{G_F}]\bm{\hat{\beta}_F},
\end{align}
\[\bm{\hat{\beta}_A}\sim N(\bm{\beta_A},\sigma^2 [\bm{G_A'}(I_n-{X_S}({X_S}'X_S)^{-1}{X_S}')\bm{G_A}]^{-1})\]is the estimated effects under model \(\mathcal{M}_A\),
and \[\bm{\hat{\beta}_F}\sim N(\bm{\beta_F},\sigma^2 [\bm{G_F'}(I_n-{X_S}({X_S}'X_S)^{-1}{X_S}')\bm{G_F}]^{-1}) \] is the estimated effects under model \(\mathcal{M}_F\),

We show how to calculate the non-central parameter of \(W_A\) under \(H_1\). Let \(\bm{X_A} = (X_S,\bm{G_A}) \)~and $P$ be the limit of ${{\bm{X_A}}'\bm{X_A}}/{n}$:
\[\frac{{\bm{X_A}}'\bm{X_A}}{n} \stackrel{p}{\to} P.
\]
\[P = \begin{bmatrix}
1 & E(S) &  E(G_{A,1})&\dots&E(G_{A,k})\\
E(S) & E(S^2) &  E(S\cdot G_{A,1})&\dots&E(S\cdot G_{A,k})\\
E(G_{A,1}) & E(G_{A,1} \cdot S) &  E(G_{A,1}^2)&\dots&E(G_{A,1} \cdot G_{A,k})\\
\vdots&\vdots&\vdots& & \vdots\\
E(G_{A,k}) & E(G_{A,k} \cdot S) &  E(G_{A,k} \cdot G_{A,1})&\dots&E(G_{A,k}^2)
\end{bmatrix}\]
We assume that the populations of females and males are equal, so $E(S) = 0.5$. For the $i$th-SNP, assume that $G_{A,i} = (0, 0.5, 1, 0, 1)$ with the corresponding frequencies given by:  
\[
\left(\frac{(1-f_{\text{female},i})^2}{2}, f_{\text{female},i}(1-f_{\text{female},i}), \frac{f_{\text{female},i}^2}{2}, \frac{1-f_{\text{male},i}}{2}, \frac{f_{\text{male},i}}{2} \right)
\]  
The expected values are:
\begin{align}
E(G_{A,i}) &= \frac{f_{\text{female},i}(1-f_{\text{female},i})}{2} + \frac{f_{\text{female},i}^2}{2} + \frac{f_{\text{male},i}}{2} = \frac{f_{\text{female},i} + f_{\text{male},i}}{2}
\\
E(G_{A,i}\cdot G_{A,j}) &= E(G_{A,i})\cdot E(G_{A,j})\notag\\
&= \frac{(f_{female,i}+f_{male,i})(f_{female,j}+f_{male,j})}{4}\notag\\
&= \frac{f_{female,i}f_{female,j}+f_{female,i}f_{male,j}+f_{male,i}f_{female,j}+f_{male,i}f_{male,j}}{4}\\
E(G_{A,i}^2) &= \frac{f_{female,i}(1-f_{female,i})}{4}+\frac{f_{female,i}^2}{2}+\frac{f_{male,i}}{2}\notag\\
&=\frac{f_{female,i}(1+f_{female,i})}{4}+\frac{f_{male,i}}{2}\\
E(S\cdot G_{A,i}) &= \frac{f_{male,i}}{2}
\end{align}
Corresponding to the split $({X_S},\bm{G_A})$, 
\(P\) is partitioned as \(P = \begin{bmatrix}
	P_{11}&P_{12}\\
	P_{21}&P_{22}
\end{bmatrix}\),
so that
\[P_{11} = \begin{bmatrix}
	1&E(S)\\E(S)&E(S^2)
	\end{bmatrix},~
	P_{12}=P_{21}'= \begin{bmatrix}
		E(G_{A,1})&\dots&E(G_{A,k})\\
		E(S\cdot G_{A,1})&\dots&E(S\cdot G_{A,k})
		\end{bmatrix},\]

\[P_{22} = \begin{bmatrix}
	E(G_{A,1}^2)&\dots&E(G_{A,1} \cdot G_{A,k})\\
	\vdots&&\vdots\\
	E(G_{A,k} \cdot G_{A,1})&\dots&E(G_{A,k}^2)
	\end{bmatrix}.\]

Then the asymptotic value of ${ncp}$ is calculated as:
\[ \frac{n}{\sigma^2}\bm{{\beta}_A}'[P_{22} - P_{21}(P_{11})^{-1}P_{12}]\bm{{\beta}_A},\]
where
\begin{align}
	P_{21}(P_{11})^{-1}P_{12} &= \left(\frac{f_{female}+f_{male}}{2},~\frac{f_{male}}{2}\right)
\left(
\begin{matrix}
	2&-2\\-2&4
\end{matrix}
\right)\left(\frac{f_{female}+f_{male}}{2},~\frac{f_{male}}{2}\right)' \notag\\
&= \frac{f_{female}f_{female}'+ f_{male}f_{male}'}{2},\notag
\end{align}
Each element in $P_{21}(P_{11})^{-1}P_{12}$ is given by:
\[
\frac{f_{\text{female},i} f_{\text{female},j} + f_{\text{male},i} f_{\text{male},j}}{2}.
\]
Hence, the elements in $P_{22} - P_{21}(P_{11})^{-1}P_{12}$ are:
\[
\frac{f_{\text{female},i}(1-f_{\text{female},i}) + 2f_{\text{male},i}(1-f_{\text{male},i})}{4}, \quad i = 1, \dots, k,
\]
and
\[
\frac{(f_{\text{female},i} - f_{\text{male},i})(f_{\text{male},j} - f_{\text{female},j})}{4}, \quad i \neq j.
\]
If $f_{\text{female},i} = f_{\text{male},i} = f_i$, then $P_{22} - P_{21}(P_{11})^{-1}P_{12}$ is a diagonal matrix.
\begin{align}
ncp_A{\to} \frac{n}{\sigma^2} \sum_{i=1}^{k} \frac{3f_i(1-f_i)\beta_{A,i}^2}{4}.
\label{ncp1}
\end{align}
If $f_{\text{female},i} \neq f_{\text{male},i}$, we can apply an orthogonal transformation to $P_{22} - P_{21}(P_{11})^{-1}P_{12}$, defined non-negatively. This gives:
\[
P_{22} - P_{21}(P_{11})^{-1}P_{12} = U' \Lambda U,
\]
and
\[
{\bm{O_A}} = U \bm{\beta_A},
\]
where $\Lambda$ is the diagonal matrix of eigenvalues $\lambda_i(>0)$. Then, $ncp_A$ converges to:
\begin{align}
ncp_A  {\to} \frac{n}{\sigma^2} \sum_{i=1}^{k} \lambda_i O_{A,i}^2,
\label{ncp2}
\end{align}
which increases with the increment of $k$.

We assume that \(\beta_{A,1}, \dots, \beta_{A,k} \stackrel{i.i.d}{\sim} N(\mu_A, \tau_A^2)\). $ncp$ in (\ref{ncp1}) can then be rewritten in terms of the parameters \(\mu_A\) and \(\tau_A\):
\[
E(\beta_{A,i}^2) = \text{Var}(\beta_{A,i}) + [E(\beta_{A,i})]^2 = \tau_A^2 + \mu_A^2.
\]
\begin{align}
    ncp_A {\to} \frac{n}{\sigma^2} \sum_{i=1}^{k} \frac{3f_i(1-f_i)(\tau_A^2 + \mu_A^2)}{4}.
\end{align}
As \(f_{\text{female},i} \neq f_{\text{male},i}\), we partition \(U\) as \(U = (u_1, \dots, u_k)'\), where \(u_i\)s are \(k\)-dimensional orthogonal vectors. Hence, \(O_{A,i} \sim N(\mu_A u_i' 1_k, \tau_A^2)\), for \(i = 1, \dots, k\). We can rewrite (\ref{ncp2}) as:
\begin{align}
    ncp_A{\to} \frac{n}{\sigma^2} \sum_{i=1}^{k} \lambda_i \left(\tau_A^2 + \mu_A^2 u_i' 1_k {1_k}' u_i \right).
\end{align}
Hence \(ncp_A\) has the order \(O(k)\).

To derive non-centrality parameter of \(\mathcal{M}_F\), let \(\bm{X_F} = (X_S, \bm{G_F})\). \(\bm{{X_F}'X_F}/n\) converges to \(P_F\), 
\[
P_F = 
\begin{bmatrix}
1&E(S)&E(\bm{G_F})\\
E(S)&E(S^2)&E(S\cdot\bm{G_F})\\
E(\bm{G_F})&E(\bm{G_F}\cdot{S})&E(\bm{G_F}^2)\\
\end{bmatrix},
\]
similar to above procedures, it is easy to find that \(ncp_F\) also has the order \(O(k)\).
\subsection*{Proof of Theorem 3}
We can approximate the theoretical power loss of model \(\mathcal{M}_F\) compared to model \(\mathcal{M}_A\), when \(\mathcal{M}_A\) is the true model with the following formula \cite{Muirhead2005}:
\begin{align}
\Phi\left\{ 
	\underbrace{\frac{\chi^2_{3k,\alpha}-(3k+ncp)}{\sqrt{2(3k+2ncp)}}}_{\text{Term 2.1}}
\right\}
-
\Phi\left\{
	\underbrace{\frac{\chi^2_{k,\alpha}-(k+ncp)}{\sqrt{2(k+2ncp)}}}_{\text{Term 2.2}}
\right\}.
\end{align}
The quantile of the \(\chi^2\) distribution under the null hypothesis at critical value \(\alpha\) is approximated by a normal distribution~\cite{fisher1934}:
\[
\chi^2_{k,\alpha} \approx \frac{1}{2}\left( z_{\alpha} + \sqrt{2k} \right)^2,
\]

We show that:
\[
\frac{\chi_k^2 - k}{\sqrt{2k}} \stackrel{d}{\rightarrow} N(0, 1) \quad \text{as} \quad k \rightarrow \infty.
\]
Multiplying both numerator and denominator by $2$, we obtain:
\[
\sqrt{2k}\left( \frac{\chi_k^2}{k} - 1 \right) \stackrel{d}{\rightarrow} N(0, 4) \quad \text{as} \quad k \rightarrow \infty.
\]
By delta method, define \( g(x) = \sqrt{x} \), then \( g'(x) = \frac{1}{2\sqrt{x}} \). Applying delta method gives:
\[
\sqrt{2k}\left[ g\left( \frac{\chi_k^2}{k} \right) - g(1) \right] \stackrel{d}{\rightarrow} N\left( 0, 4 [g'(1)]^2 \right) \quad \text{as} \quad k \rightarrow \infty,
\]
that is,
\[
\sqrt{2\chi_k^2} - \sqrt{2k} \stackrel{d}{\rightarrow} N(0, 1) \quad \text{as} \quad k \rightarrow \infty.
\]
Hence, for a given $\alpha$,
\[
\alpha = P\left( \sqrt{2\chi_k^2} - \sqrt{2k} > z_\alpha \right) = P\left( \chi_k^2 > \frac{1}{2} (z_\alpha + \sqrt{2k})^2 \right).
\]
Thus, the approximate upper quantile at critical value $\alpha$ is \(\frac{1}{2}(z_\alpha + \sqrt{2k})^2\).
For any \(b \in (\sqrt{2c+\sqrt{6}}-\sqrt{6},\sqrt{2c+\sqrt{2}}-\sqrt{2})\) and \(z_{\alpha} = b\sqrt{k}\)

First, for \(\text{Term~2.1}:\)
\begin{align}
	\text{Term~2.1}
&= \frac{\frac{1}{2}(z_{\alpha}+\sqrt{6k})^2-3k-ck}{\sqrt{2(3k+2ck)}}
= \frac{(\frac{1}{2}b^2+\sqrt{6}b-c)}{\sqrt{2(3+2c)}}\sqrt{k}\notag
\end{align}
Similarly, for \(\text{Term~2.2}\):
\begin{align}
	\text{Term~2.2}
&= \frac{\frac{1}{2}(z_{\alpha}+\sqrt{2k})^2-k-ck}{\sqrt{2(k+2ck)}}
= \frac{(\frac{1}{2}b^2+\sqrt{2}b-c)}{\sqrt{2(1+2c)}}\sqrt{k}\notag
\end{align}
where \( \frac{1}{2}b^2+\sqrt{6}b-c > 0 \) and \( \frac{1}{2}b^2+\sqrt{2}b-c < 0\) 
Thus, as \(k\to\infty\), \(\text{Term~2.1}\to \infty\)
and \(\text{Term~2.2}\to -\infty\), respectively.

Combining the two results, we conclude that the theoretical power loss is approaching 1 as \(k\to\infty\).

\newpage
\section*{Appendix C: Additional Figures}

\begin{figure}[htbp]
    \centering
    \includegraphics[width=.85\textwidth]{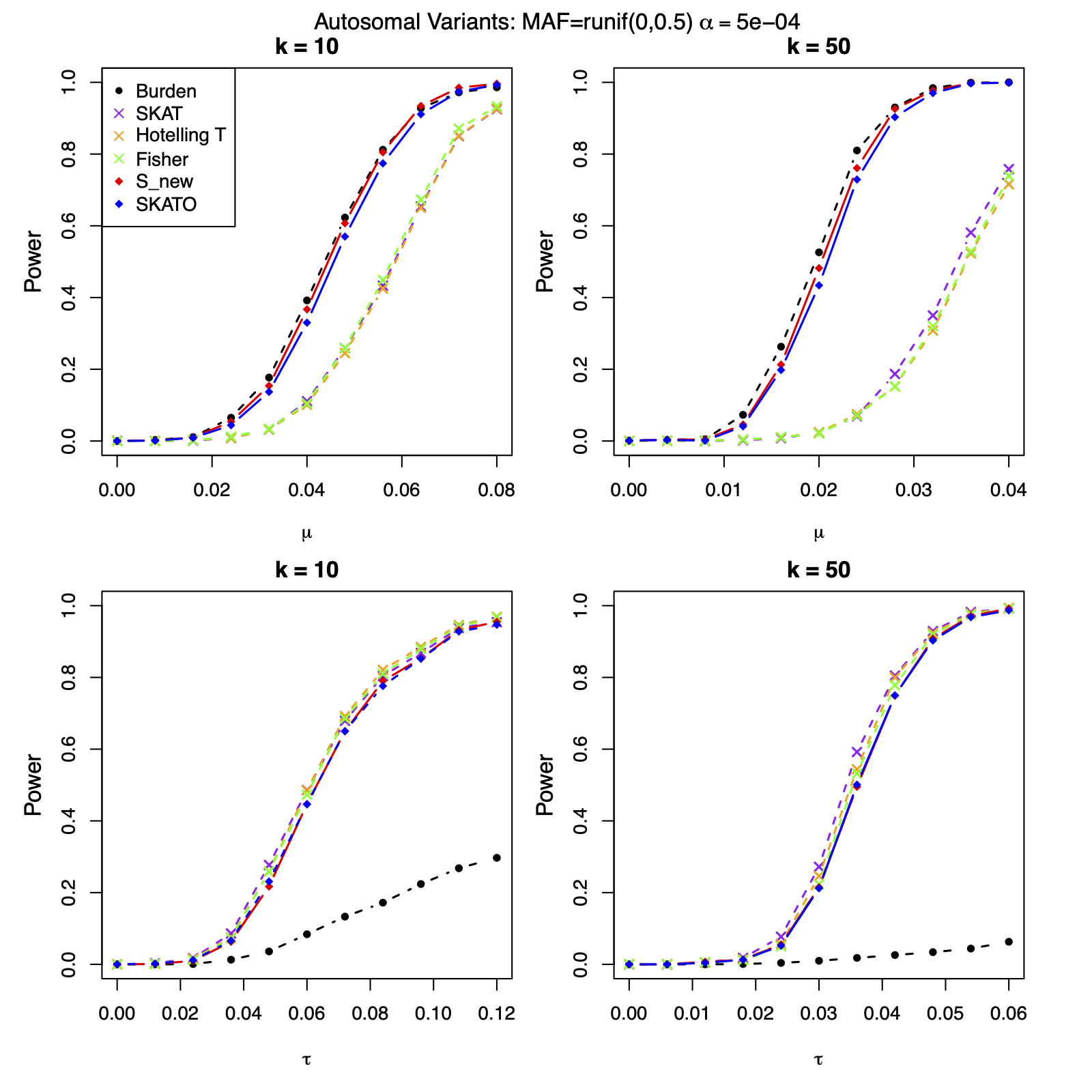}
    \caption{Power comparison settings: In the first row, $k$ variants have fixed effects as $\tau = 0$ and $\mu>0$. In the second row, k variants have random effects as $\tau > 0$ and $\mu=0$. The weights for each variants applied in Burden, SKAT and SKATO are \(w_i = 1, (i = 1,\dots,k)\). The significance level is set at $\alpha = 5 \times 10^{-4}$ and the number of variants is \(k = 10\) and  \(k = 50\).
}
\end{figure}

\begin{figure}[htbp]
    \centering
    \includegraphics[width=0.85\linewidth]{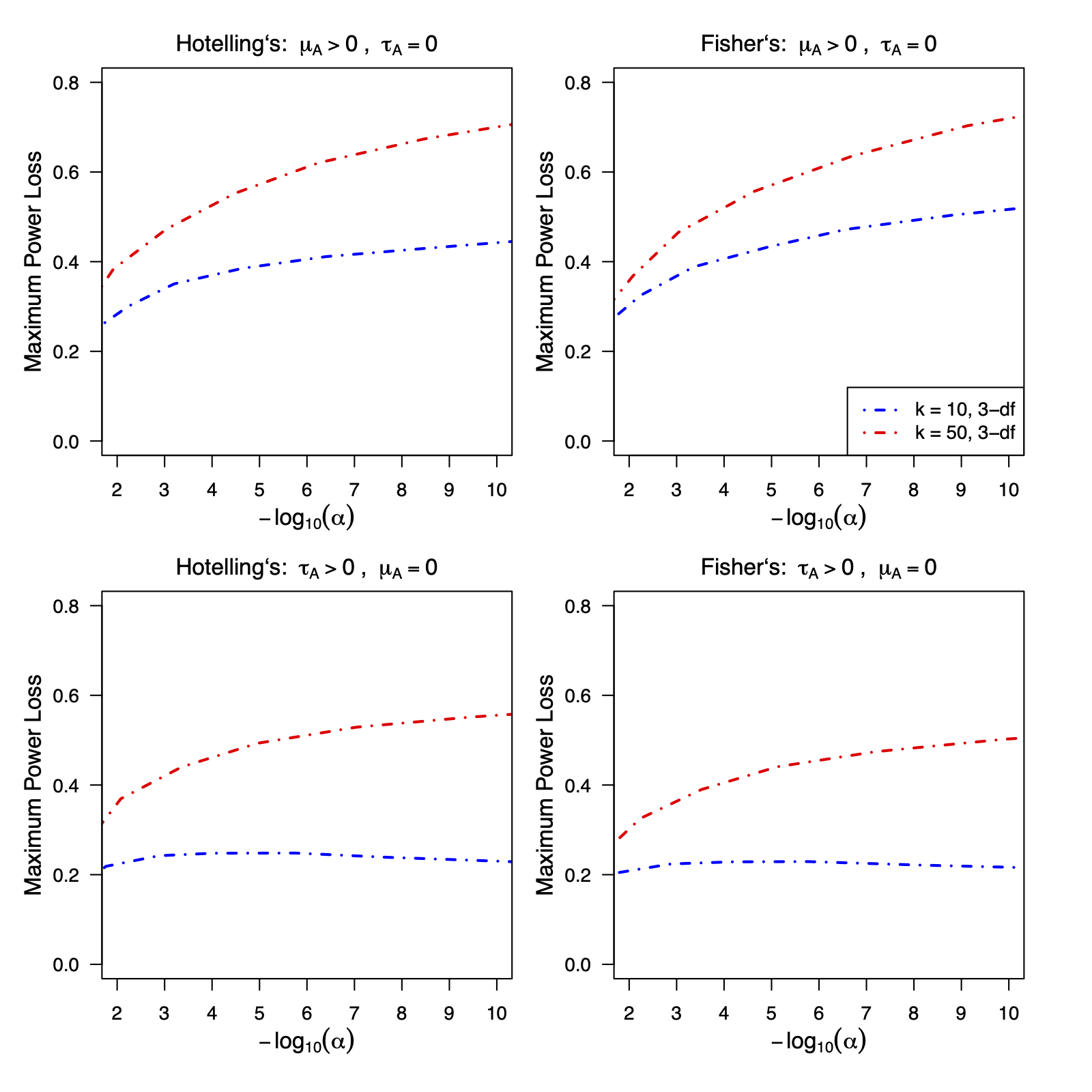}
    \caption{Maximum power loss at various type I error level $-\log_{10} \alpha$ from 2 to 10, compared to the test under true additive 1-df XCI model when gene-sex interaction and dominant effects are not present. First row: $k$ variants have fixed effects $\mu_A>0, \tau_A=0$. Second row: $k$ variants have random effects $\mu_A=0, \tau_A>0$. Left panels: Hotelling`s $T^2$ method. Right panels: Fisher`s method. Blue curve: $k=10$. Red curve: $k=50$. Dash curve: power loss from 3-df full model.}
    \label{PowerLoss_Curves_Quadratic}
\end{figure}

\begin{figure}[htbp]
    \centering
    \includegraphics[width=0.85\linewidth]{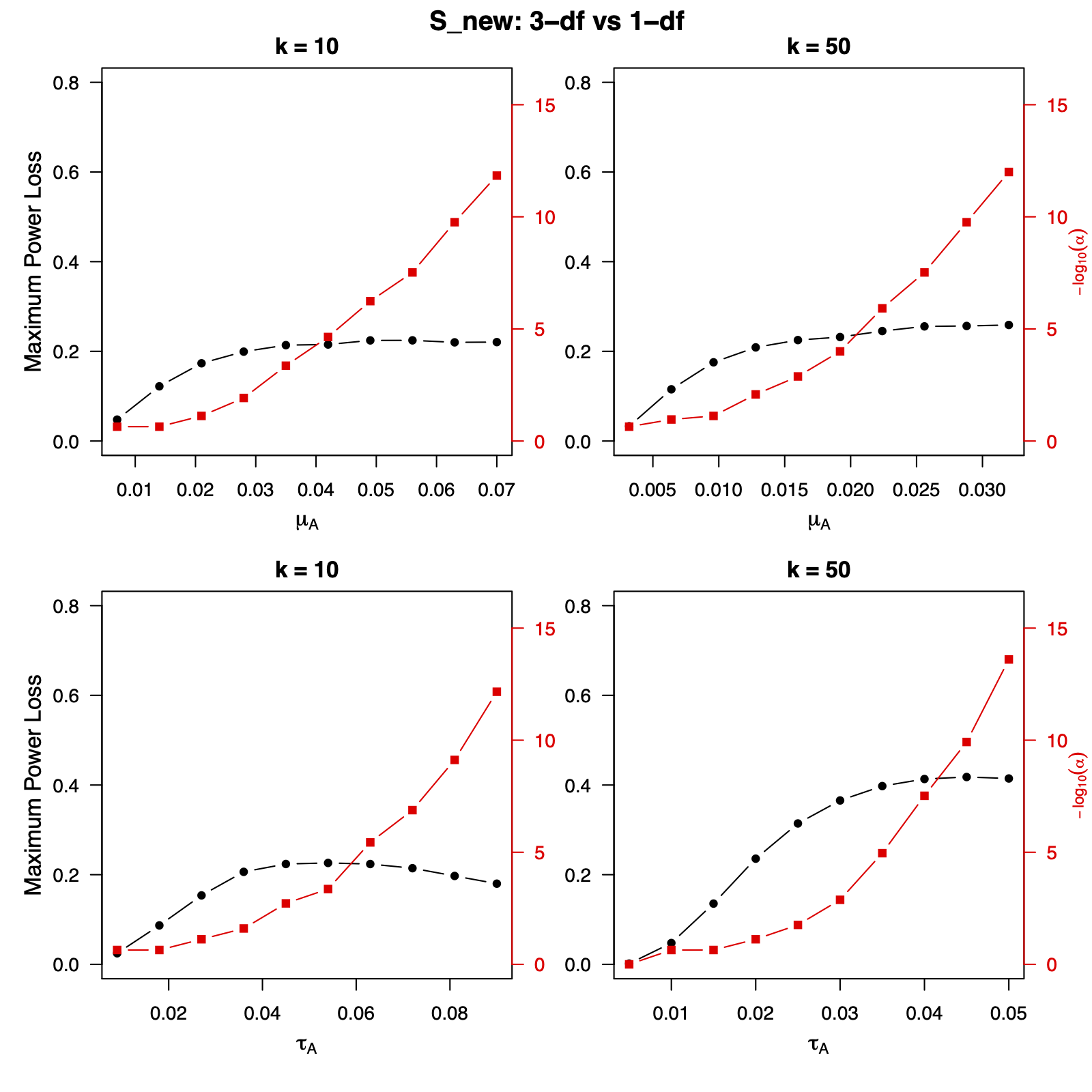}
    \caption{Power Loss Curves of \(S_{\text{new}}\) 3-df vs 1-df. In the first row, $k$ variants have fixed effects, \(\tau_A=0, \mu_A>0\), in the second row $k$ variants have random effects \(\tau_A >0,\) setting \(\mu_A = 0\). We compare the power loss due to model misspecification, 3-df vs 1-df, as 1-df is the correct model. Black Curve: maximum power loss. Red Curve: values of \(-\text{log}_{10}\alpha\) to reach the maximum power loss. Left column: $k=10$, right column: $k=50$.}
    \label{PowerLoss_S_new_3−df vs 1−df}
\end{figure}

\begin{figure}[htbp]
    \centering
    \includegraphics[width=0.85\linewidth]{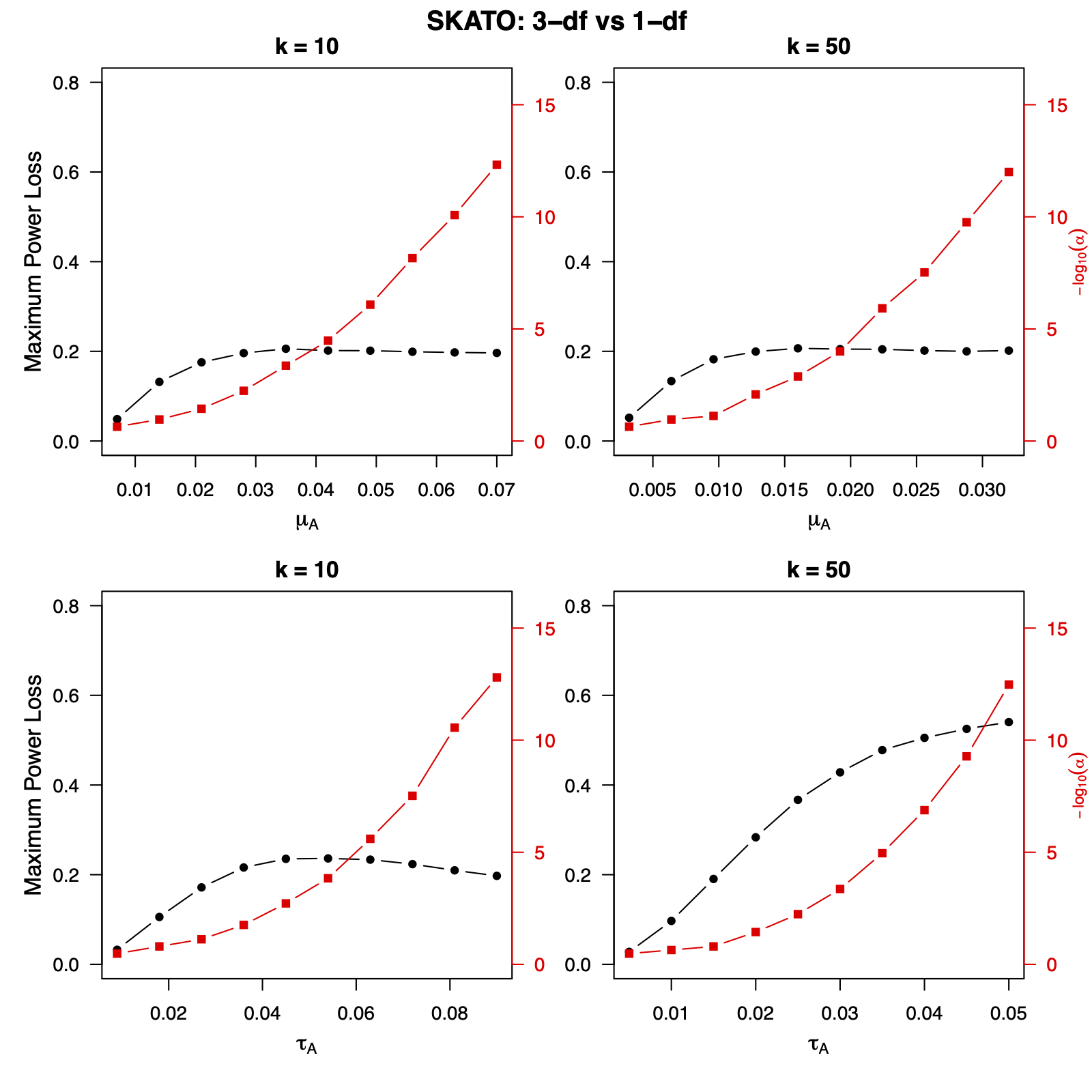}
    \caption{Power Loss Curves of {SKATO} 3-df vs 1-df. In the first row, $k$ variants have fixed effects, \(\tau_A=0, \mu_A>0\), in the second row $k$ variants have random effects \(\tau_A>0,\) setting \(\mu_A = 0\). We compare the power loss due to model misspecification, 3-df vs 1-df, as 1-df is the correct model. Black Curve: maximum power loss. Red Curve: values of \(-\text{log}_{10}\alpha \) to reach the maximum power loss. Left column: $k=10$, right column: $k=50$.}
    \label{PowerLoss_SKATO_3−df vs 1−df}
\end{figure}

\begin{figure}[htbp]
    \centering
    \includegraphics[width=0.85\linewidth]{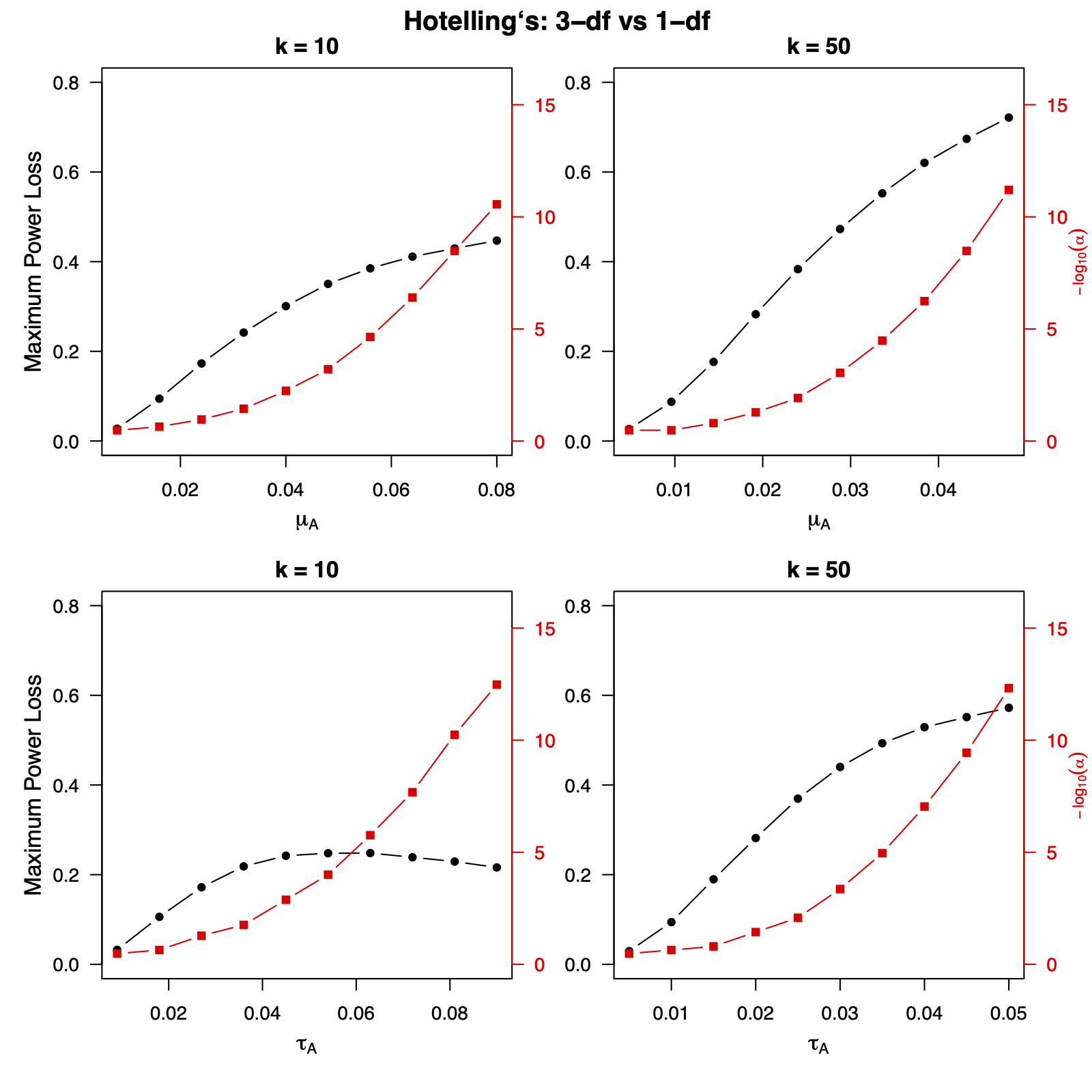}
    \caption{Power Loss Curves of Hotelling`s \(T^2\) method 3-df vs 1-df. In the first row, $k$ variants have fixed effects, \(\tau_A=0, \mu_A>0\), in the second row $k$ variants have random effects \(\tau_A>0,\) setting \(\mu_A = 0\). We compare the power loss due to model misspecification, 3-df vs 1-df, as 1-df is the correct model. Black Curve: maximum power loss. Red Curve: values of \(-\text{log}_{10}\alpha\) to reach the maximum power loss. Left column: $k=10$, right column: $k=50$.}
    \label{PowerLoss_Hotelling_3−df vs 1−df}
\end{figure}

\begin{figure}[htbp]
    \centering
    \includegraphics[width=0.85\linewidth]{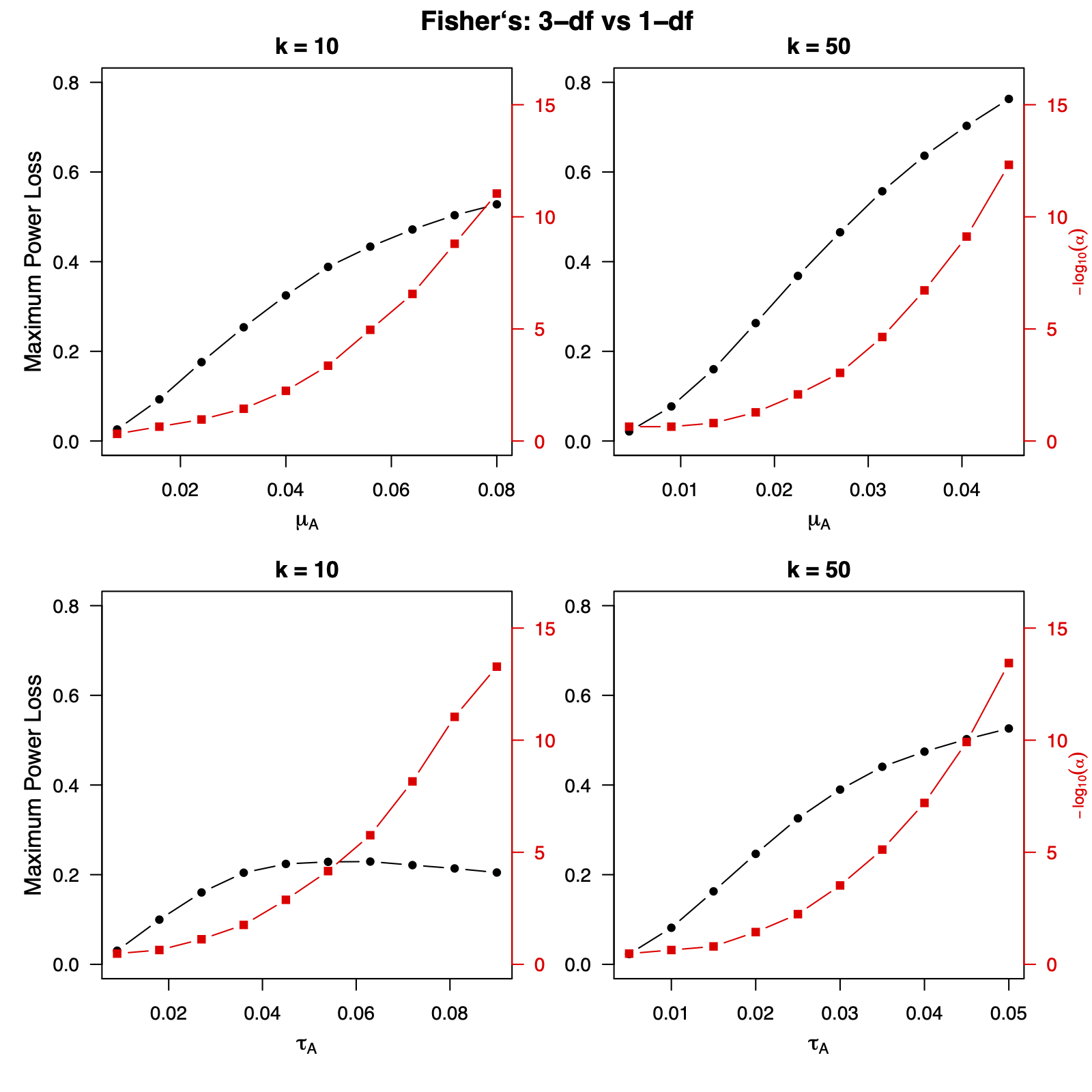}
    \caption{Power Loss Curves of Fisher`s method 3-df vs 1-df. In the first row, $k$ variants have fixed effects, \(\tau_A=0, \mu_A>0\), in the second row $k$ variants have random effects \(\tau_A>0,\) setting \(\mu_A = 0\). We compare the power loss due to model misspecification, 3-df vs 1-df, as 1-df is the correct model. Black Curve: maximum power loss. Red Curve: values of \(-\text{log}_{10}\alpha\) to reach the maximum power loss. Left column: $k=10$, right column: $k=50$.}
    \label{PowerLoss_Fisher_3−df vs 1−df}
\end{figure}

\begin{figure}[htbp]
    \centering
    \includegraphics[width=0.85\linewidth]{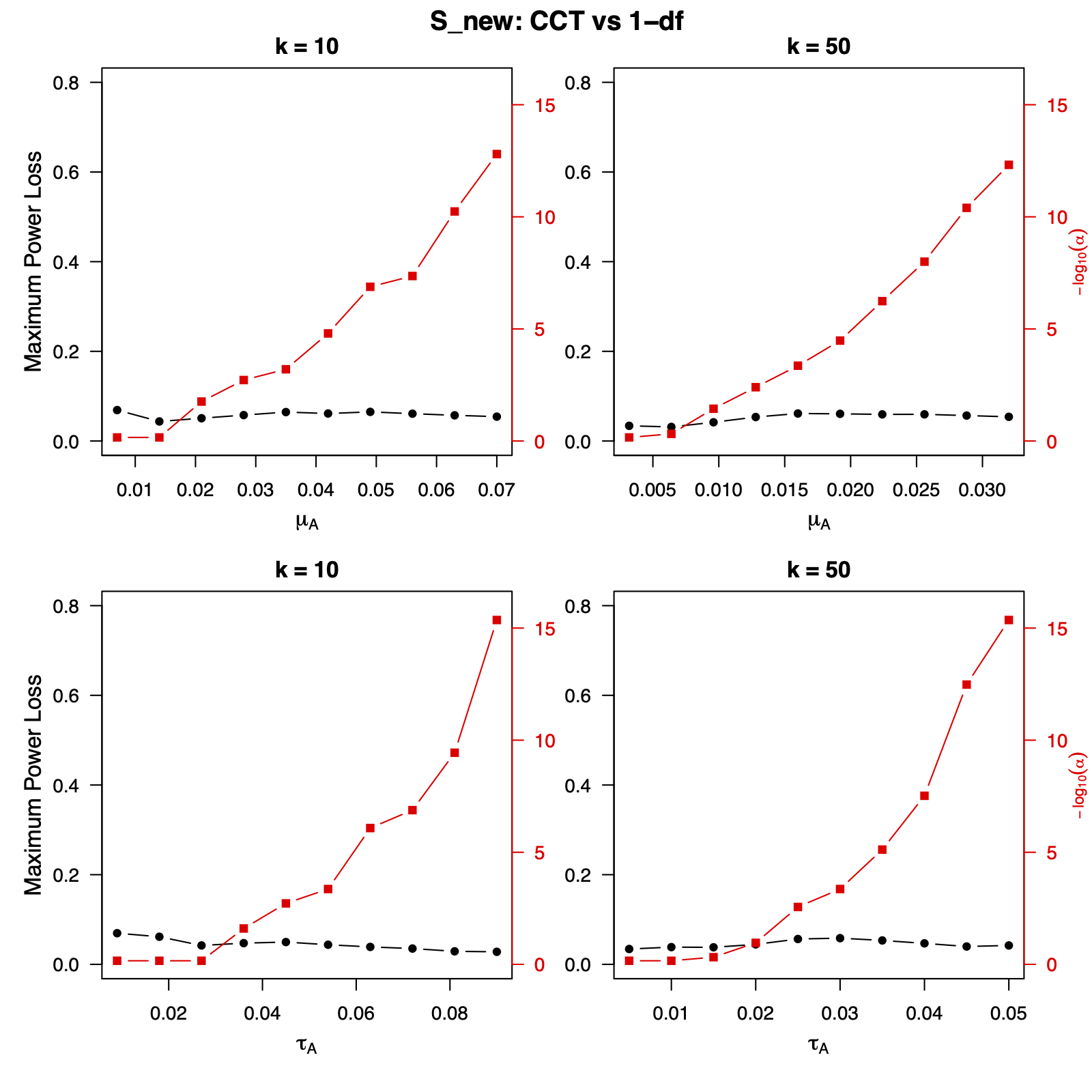}
    \caption{Power Loss Curves of \(S_{\text{new}}\) CCT vs 1-df. In the first row, $k$ variants have fixed effects, \(\tau_A=0, \mu_A>0\), in the second row $k$ variants have random effects \(\tau_A>0,\) setting \(\mu_A = 0\). We compare the power loss due to model misspecification, CCT vs 1-df, as 1-df is the correct model. Black Curve: maximum power loss. Red Curve: values of \(-\text{log}_{10}\alpha\) to reach the maximum power loss. Left column: $k=10$, right column: $k=50$.}
    \label{PowerLoss_S_new_CCT vs 1−df}
\end{figure}
\begin{figure}[htbp]
    \centering
    \includegraphics[width=0.85\linewidth]{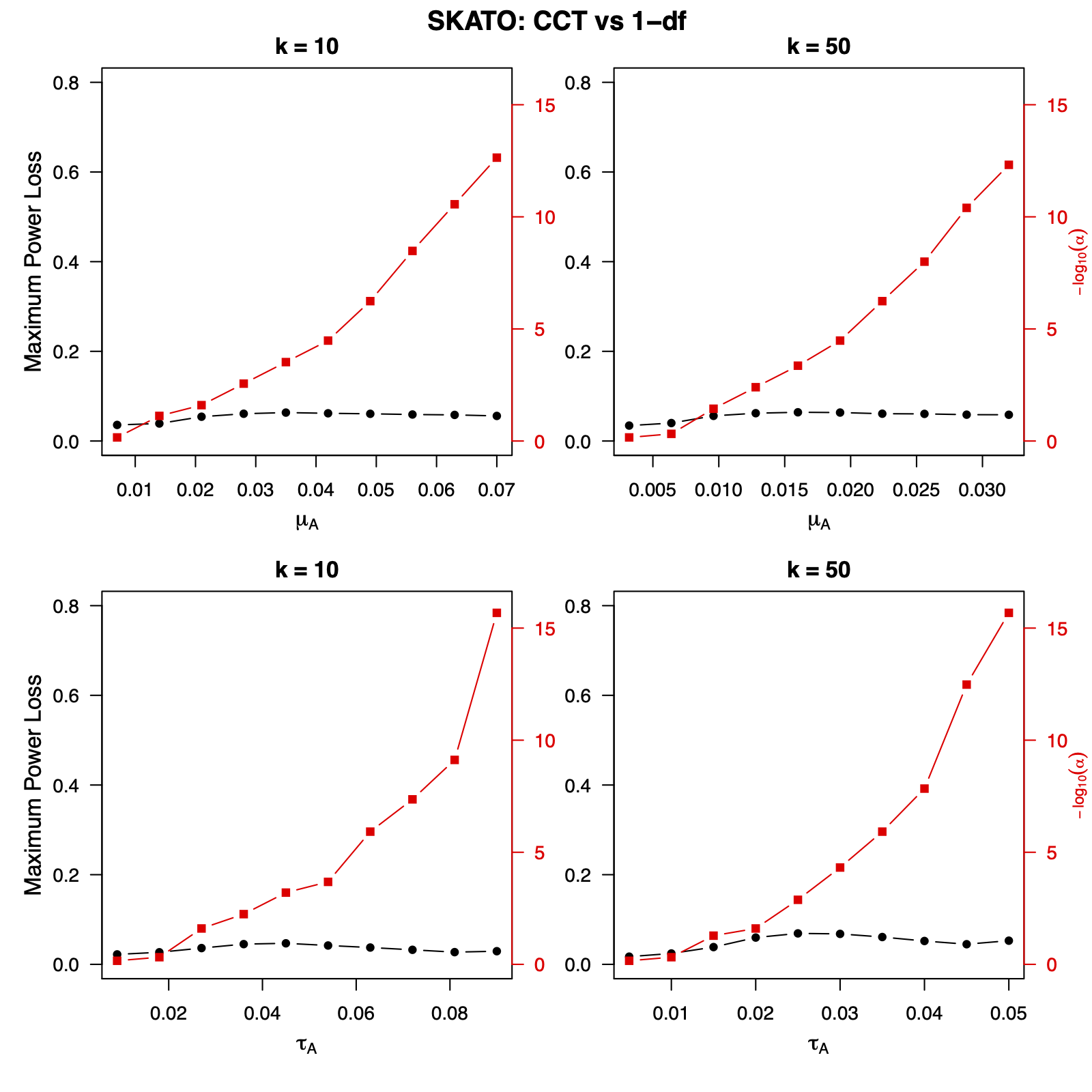}
    \caption{Power Loss Curves of SKATO CCT vs 1-df. In the first row, $k$ variants have fixed effects, \(\tau_A=0, \mu_A>0\), in the second row $k$ variants have random effects \(\tau_A>0,\) setting \(\mu_A = 0\). We compare the power loss due to model misspecification, CCT vs 1-df, as 1-df is the correct model. Black Curve: maximum power loss. Red Curve: values of \(-\text{log}_{10}\alpha\) to reach the maximum power loss. Left column: $k=10$, right column: $k=50$.}
    \label{PowerLoss_SKATO_CCT vs 1−df}
\end{figure}

\begin{figure}[hp]
    \centering
    \includegraphics[width=1\textwidth]{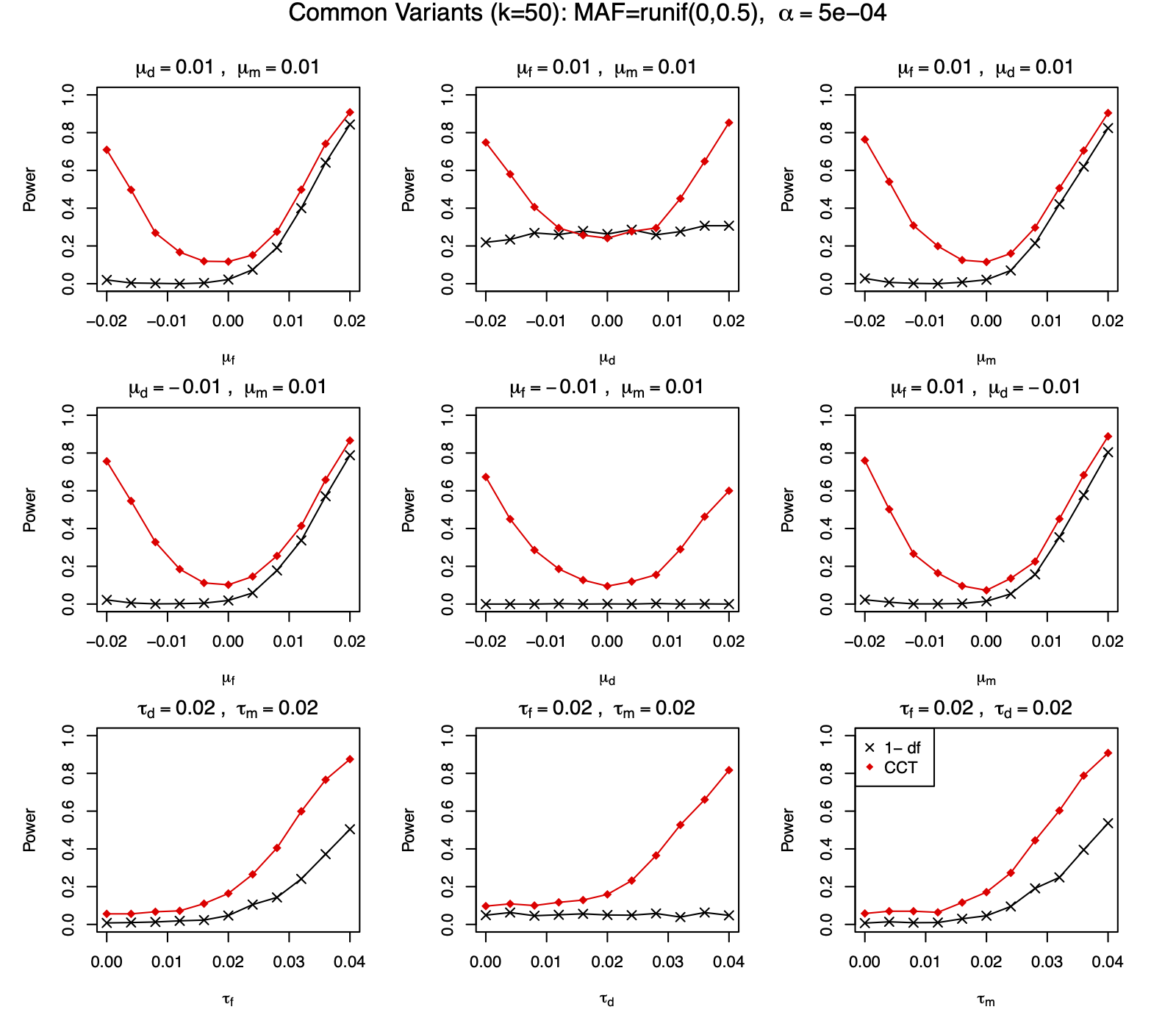}
    \caption{$S_\text{new}$ test power for common variants. First row: two of $\mu_f$, $\mu_d$ and $\mu_m$ are held fixed with same direction while the third is varied, and all \(\tau\) values are set to 0. Second row: two of $\mu_f$, $\mu_d$ and $\mu_m$ are held fixed with opposite directions while the third is varied, and all \(\tau\) values are set to 0. Third row: two of $\tau_f$, $\tau_d$ and $\tau_m$ are held fixed while the third is varied, and all \(\mu\) values are set to 0. Red curve: CCT based multilocus \( S_{\text{new}} \) test power. Black curve: 1-df additive multilocus \( S_{\text{new}} \) test power.}
\end{figure}
\begin{figure}[hp]
    \centering
    \includegraphics[width=1\textwidth]{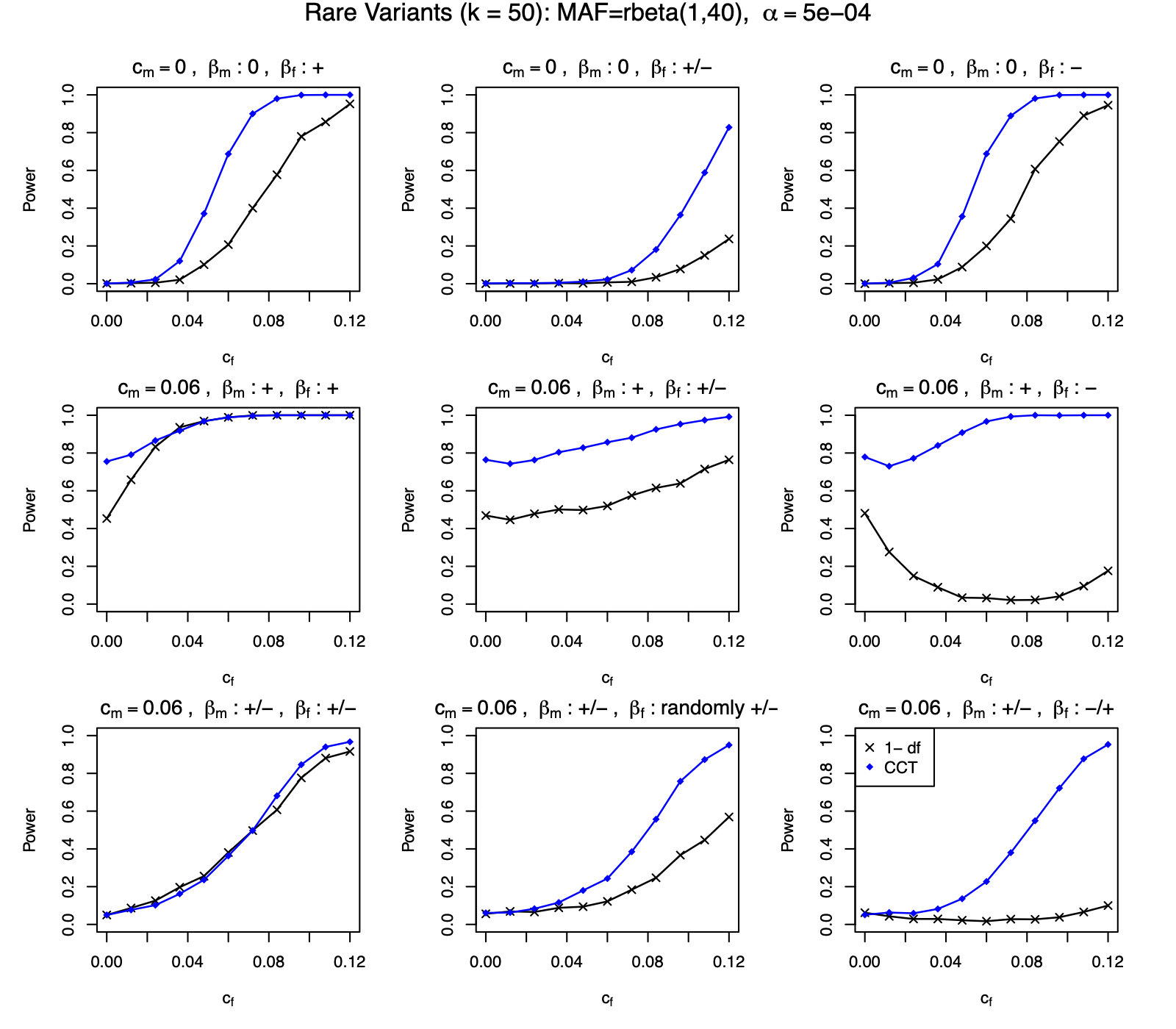}
    \caption{SKATO test power for rare variants. First row: \(\bm{\beta_m = 0}\), and \(\bm{\beta_f}\) takes on values with positive, 50\% positive / 50\% negative, and negative signs. Second row: \(\bm{\beta_m > 0}\), and the directions of \(\bm{\beta_f}\) and \(\bm{\beta_m}\) vary as follows: 100\% concordant, 50\% concordant / 50\% discordant, and 100\% discordant. Third row: the half part of \(\bm{\beta_m}\) is randomly assigned positive values and the other half negative. Correspondingly, \(\bm{\beta_f}\) direction is set as 100\% concordant, randomly half positive, or 100\% discordant. Blue curve: CCT based multilocus SKATO test power. Black curve: 1-df additive multilocus SKATO test power.}
\end{figure}

\newpage
\bibliographystyle{unsrt}  
\bibliography{references}